\documentclass{ws-ijmpa}
\usepackage[super,compress]{cite}
\usepackage{graphicx}
\begin{document}
\markboth{C.E.Gerber, C.Vellidis}
{The Top Quark Chapter}

%
\catchline{}{}{}{}{}
%

\def\pbarp{$p\overline{p}~$}             
\def\ttbar{$t\overline{t}~$}             
\def\etal{{\sl et al.}}                 
\def\ifb{fb$^{-1}$}
\def\ipb{pb$^{-1}$}
\newcommand{\rargap}    {\mbox{ $\rightarrow$ }}
\newcommand{\ppbar}     {\mbox{$p\bar{p}$}}
\def\met{\mbox{${\hbox{$E$\kern-0.6em\lower-.1ex\hbox{/}}}_T$}} 

\title{Review of Physics Results from the Tevatron: Top Quark Physics}

\author{CECILIA E. GERBER}

\address{University of Illinois at Chicago - Department of Physics\\
845 W. Taylor St. M/C 273, Chicago IL 60607-7059, USA\\
gerber@uic.edu}

\author{COSTAS VELLIDIS}

\address{Fermi National Accelerator Laboratory\\
P.\ O.\ Box 500, Batavia, IL, 60510-5011, USA\\
vellidis@fnal.gov}

\maketitle

\begin{history}
\received{Day Month Year}
\revised{Day Month Year}
\end{history}

\begin{abstract}
We present results on top quark physics from the CDF and D0 collaborations at the Fermilab Tevatron \pbarp collider. These include legacy results from Run II that were published or submitted for publication before mid-2014, as well as a summary of Run I results. The historical perspective of the discovery of the top quark in Run I is also described. 
\end{abstract}

\ccode{PACS numbers: 14.65.Ha}

\newpage
\tableofcontents

\newpage
\section{Introduction}
The top quark is the heaviest known elementary particle and 
completes the quark sector of the three-generation structure of the
standard model (SM). It differs from the
other quarks not only by its much larger mass, but also by its
lifetime which is too short to build hadronic bound states. 

The SM predicts that top quarks are created via two 
independent production mechanisms at hadron colliders. 
The primary mode, in which a \ttbar\ pair is produced from a $gtt$ 
vertex via the strong interaction~\footnote{$g$ stands for gluon}, was used by the D0 and CDF 
collaborations to discover the top quark in 
1995~\cite{top-obs-1995-cdf,top-obs-1995-d0}. 
The second production mode of top quarks at hadron colliders is the 
electroweak production of a single top quark from a $Wtb$ vertex. 
The predicted cross section for single top quark production is about half 
that of \ttbar\ pairs~\cite{toppp, singletop-xsec-kidonakis} 
but the signal-to-background ratio is much worse; 
observation of single top quark production 
has therefore until recently been impeded by its low rate and difficult 
background environment compared to the top pair 
production~\cite{PRL-103-092002-2009,PRL-103-092001-2009}.

Since its discovery, all properties of the top quark have been measured at the Tevatron with increasing precision as new data from Run II were coming in. Most attention was focused on its mass, which is a crucial property of this particle: it is not predicted by theory and, together with the $W$ boson mass, it constrains the Higgs boson mass through global electroweak fits. The large value of the top quark mass indicates a strong Yukawa coupling to the Higgs, and could provide special insights in our understanding of electroweak symmetry breaking. Other properties, including the electric charge, decay width and branching ratio, spin correlations, and $V-A$ structure of the $Wtb$ vertex, have also been studied by both CDF and D0. Both collaborations have also measured the production cross sections of both \ttbar and single top processes, and, with the larger data samples available at the end of Run II, differential cross sections as a function of various variables, including the forward-backwards asymmetry $A_{FB}$. In addition, both collaborations have extensively pursued searches for new physics in events with top quarks, including tests of fundamental symmetries in the top-quark sector as well as direct searches for new particles coupled to top quarks.

\section{The Top Quark in Run I}
\subsection{Historical perspective}
The existence of the top quark was expected since the discovery of its partner, the bottom quark, in 1977~\cite{b-discovery}. The absence of flavor-changing neutral currents in $b$ decay,  evidenced by the small branching ratio of the $b \to s e^+ e^-$ decay, indicated that the $b$-quark had isospin $-1/2$, thus requiring a 
$+1/2$ partner to complete the weak-isospin doublet. However, no firm prediction about the mass of the top quark was available. During the 1980's, a series of new lepton colliders searched for the $e^+e^-\to t\overline{t}$ process, increasing the lower bound on the top quark mass from $m_t=23.3\;\rm GeV$ at PETRA to $30.2\;\rm  GeV$ at TRISTAN, and finally to $45.8\;\rm GeV$ at SLC and LEP. The developments of hadron colliders led to searches for the production of $W$-bosons with subsequent decay $W\to t b$. After a false-positive observation of a $m_t=40\pm10\;\rm GeV$ top quark at the CERN $Sp\overline{p}S$~\cite{PLB147-493-1984}, superseded by~\cite{ZPC37-505-1988} and replaced by a new lower limit of $m_t>69\;\rm GeV$~\cite{ZPC46-179-1990}, the focus switched to searching for a top quark that is heavier than the $W$ boson, with the dominant production mechanism of 
$p\overline{p} \to t\overline{t}$, and the subsequent decay $t \to W b$. 
The CDF detector started taking data at the Fermilab Tevatron collider in 1998 (Run 0), eventually setting a lower limit of $m_t>91\;\rm GeV$ in 
1992~\cite{PRL68-447-1992}. A second detector, D0, was being commissioned at the time. 

\subsection{The Discovery}
Run I of the Tevatron, with proton-antiproton collisions at $\sqrt{s}=1.8\;\rm TeV$, started in 1992 and continued until 1995. During this time the two detectors, CDF and D0, raced for the discovery of the top quark. 
In 1994, D0 set a new limit of $m_t=131\;\rm GeV$ using $15\;\rm pb^{-1}$ of data~\cite{PRL72-2138-1994}. Later that year, CDF claimed first evidence for \ttbar production using $19.3\;\rm pb^{-1}$ of data~\cite{PRL73-225-1994}. Using 12 candidate events, CDF measured a cross section of $13.9^{+6.1}_{-4.8}\;\rm pb$ 
(about 2.5 times the one predicted by the SM at the time~\cite{NPB303-607-1988}) and a mass of $174\pm10^{+13}_{-12}\;\rm GeV$. D0 had a similar expected sensitivity of about 2 standard deviations~\cite{ICHEP94-Grannis}, observing 7 candidate events, with an expected background of $3.2 \pm 1.1$ events. Finally, in 1995, the CDF and D0 collaborations jointly announced the discovery of the top quark in the strong \ttbar production~\cite{top-obs-1995-cdf,top-obs-1995-d0}. The top quark discovery is a major legacy of the Tevatron accelerator that opened a new window to our understanding of electroweak symmetry breaking mechanisms with couplings to heavy flavor.

In the discovery paper CDF used $67\;\rm pb^{-1}$ of data and saw a signal inconsistent with the background at the $4.8\sigma$ level~\cite{top-obs-1995-cdf}. They measured the \ttbar production cross section $\sigma_{t\overline{t}}=6.8^{+3.6}_{-2.4}\;\rm pb$ and the top quark mass $m_t=176\pm8\pm10\;\rm GeV$. D0 used 
$50\;\rm pb^{-1}$ of data and saw a signal inconsistent with the background at the $4.6\sigma$ level~\cite{top-obs-1995-d0}. They measured the \ttbar production cross section $\sigma_{t\overline{t}}=6.2\pm2.2\;\rm pb$ and the top quark mass $m_t=199^{+19}_{-21}\pm22\;\rm GeV$. Figures~\ref{fig:CDFobs} 
and~\ref{fig:D0obs} show the fitted masses  for CDF and D0, respectively. Both collaborations worked with two candidate samples with different purity. At CDF, the difference was the application or not of algorithms to identify jets originating from the decay of long-lived B hadrons (b-tagging) that relied on the resolution of the SVX detector. 
At D0, the loose sample was obtained by relaxing topological cuts. 
\begin{figure}[h,t,b,p]
\includegraphics[scale=0.45]{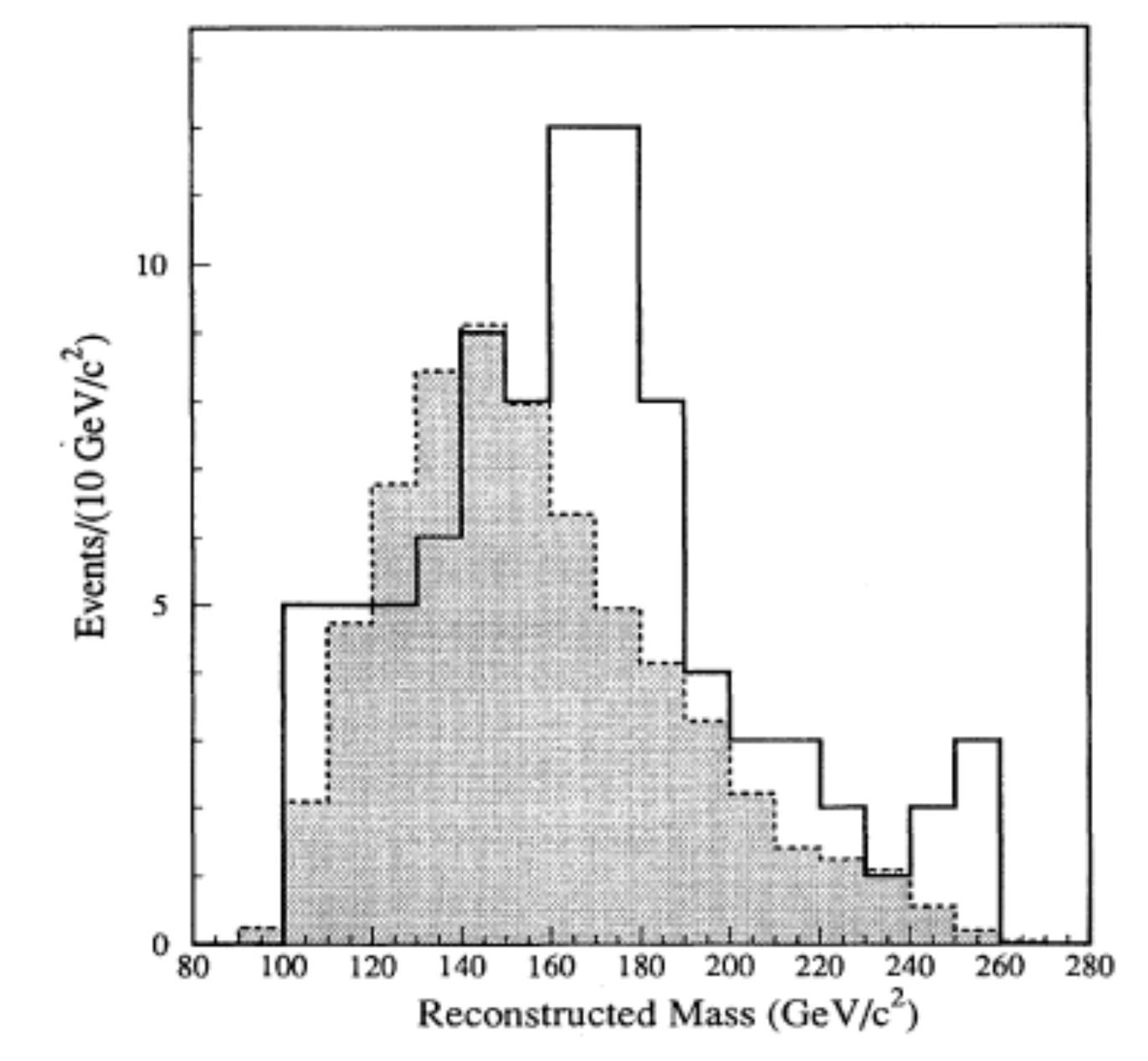}
\includegraphics[scale=0.45]{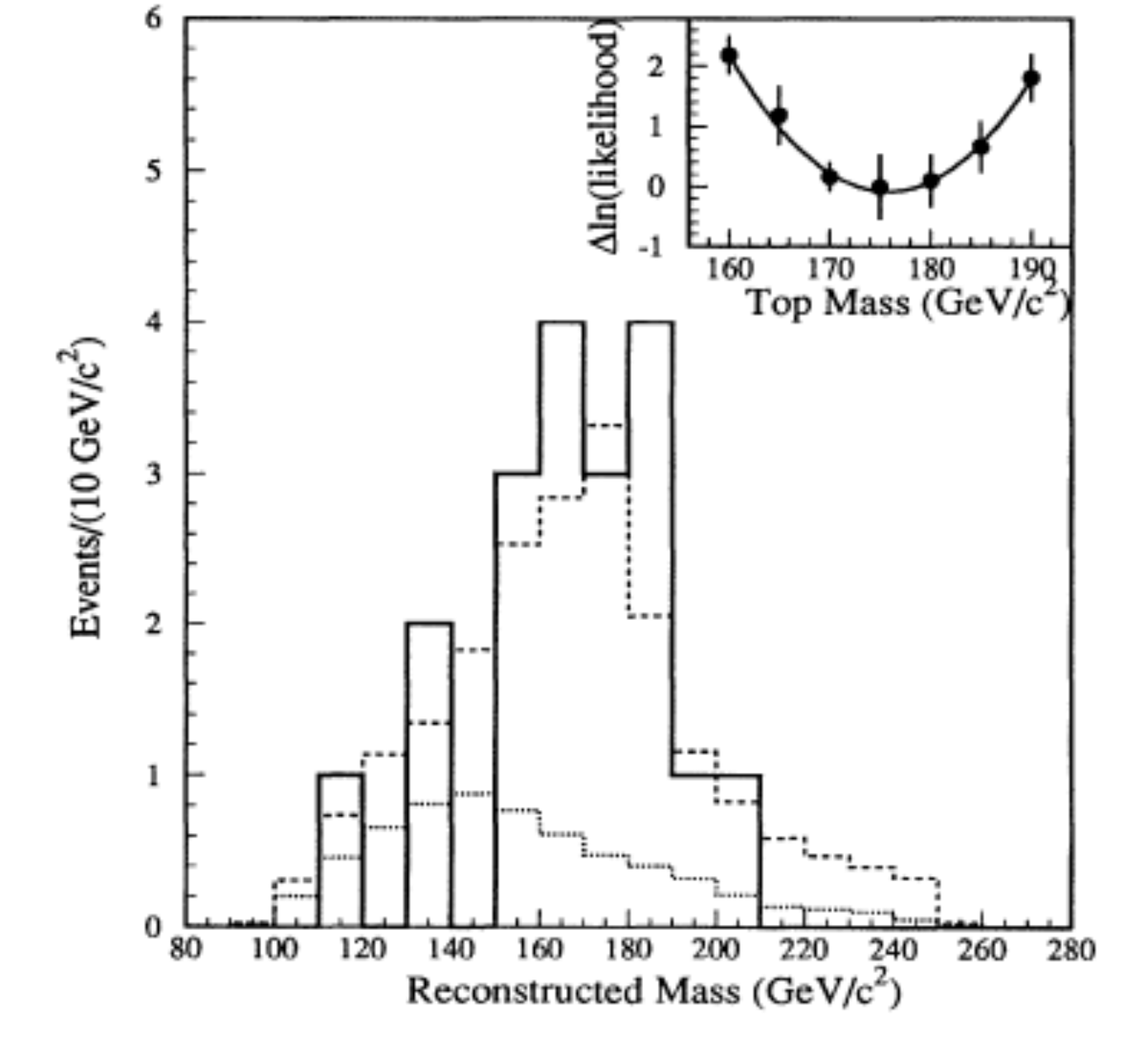}
\caption{Mass distributions from CDF's top discovery paper~\cite{top-obs-1995-cdf}. 
Left: Reconstructed mass distribution for the $W+$ 4-jet
sample prior to b tagging (solid). Also shown is the background
distribution (shaded) with the normalization constrained
to the calculated value. Right: Reconstructed mass distribution for the b-tagged $W+$4-jet events (solid). Also shown are the background
shape (dotted) and the sum of background plus \ttbar Monte
Carlo simulations for $m_t = 175\;\rm GeV$ (dashed), with the
background constrained to the calculated value.
The inset shows the likelihood fit used to determine the top
mass.}
\label{fig:CDFobs}
\end{figure}
\begin{figure}[h,t,b,p]
\centerline{\includegraphics[width=15cm]{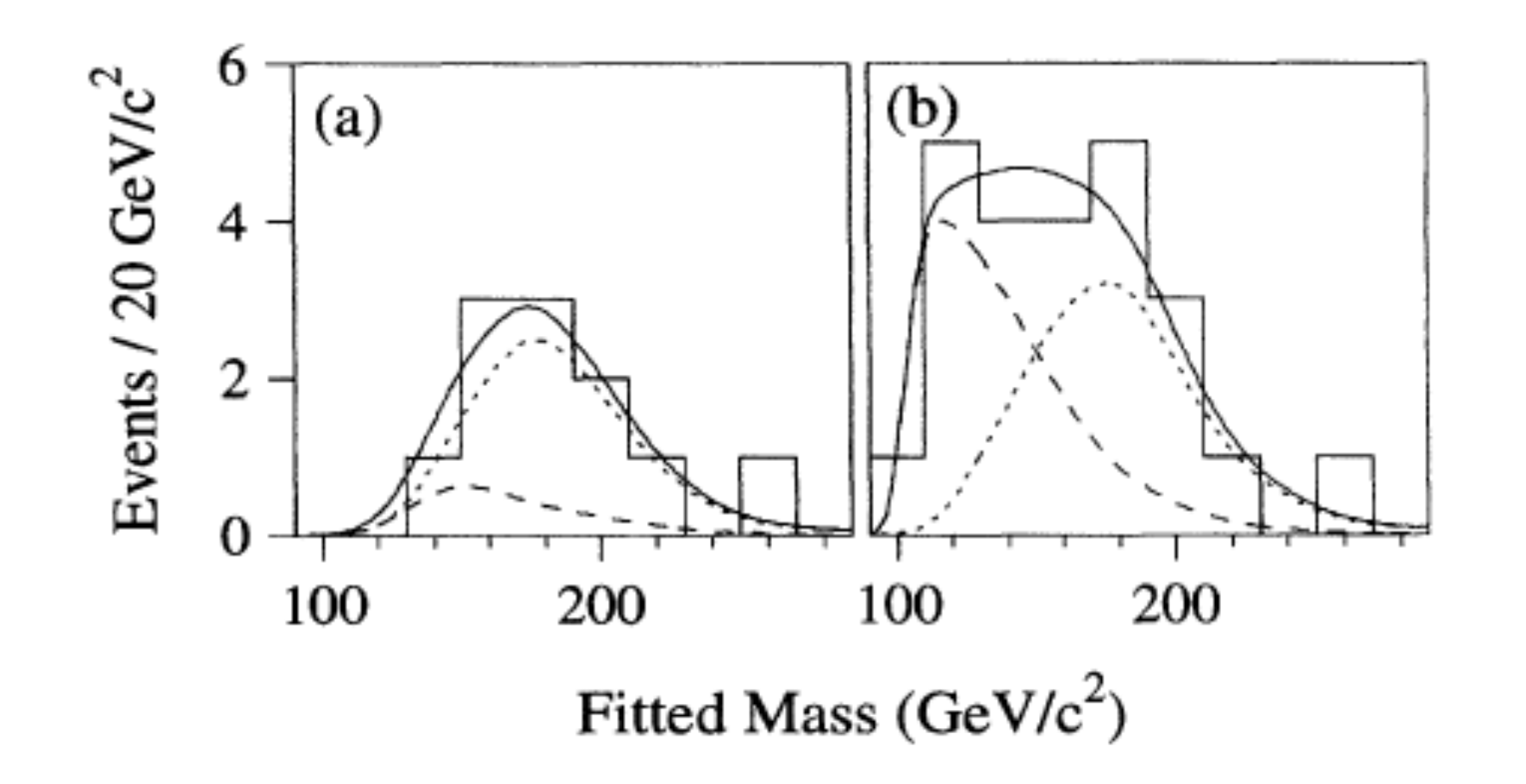}}
\caption{Mass distributions from D0's top discovery paper~\cite{top-obs-1995-d0}. 
Fitted mass distribution for candidate events (histogram) with the expected mass distribution for $199\;\rm GeV$ top quark events (dotted curve), background (dashed 
curve), and the sum of top and background (solid curve) for (a) standard and (b) loose events selection. 
\label{fig:D0obs}}
\end{figure}

\subsection{Final Results from Run I}
The entire Run I dataset of $109\;\rm pb^{-1}$ for CDF and $125\;\rm pb^{-1}$ for D0 roughly doubled the amount of luminosity used for the observation. Using those datasets, CDF and D0 produced final results for both the top quark mass and the \ttbar production cross section at $\sqrt{s}=1.8\;\rm TeV$. 
CDF combined data from all decay channels except those including a hadronically decaying tau lepton and measured~\cite{PRD64-032002-2001} a \ttbar production cross section of 
$\sigma_{t\overline{t}}=6.5^{+1.7}_{1.4}\;\rm pb$ for $m_t=175\;\rm GeV$. The result included measurements that relied on identifying $b$ quarks both by reconstructing secondary vertices (SECVTX) and the presence of soft leptons within the jets (SLT). All individual results were in agreement with each other. D0 based their result~\cite{PRD67-012004-2013} on the same channels, however, as it had no silicon vertex detector, the ability to reconstruct secondary vertices was not available. D0 thus utilized a series of topological variables designed to separate the \ttbar sample from the backgrounds. All results were also in agreement with each other, as can be observed in Fig.~\ref{fig:D0Run1xsec}. Their combination yield 
$\sigma_{t\overline{t}}=5.69\pm 1.21(stat) \pm 1.04(syst)\;\rm pb$ for $m_t=172.1\;\rm GeV$. The measurements from both collaborations were in good agreement with SM expectations. At the end of Run I, the uncertainty on the theoretical prediction for the \ttbar cross section was about 20\% of the experimental uncertainty. 
\begin{figure}[h,t,b,p]
\centerline{\includegraphics[width=10cm]{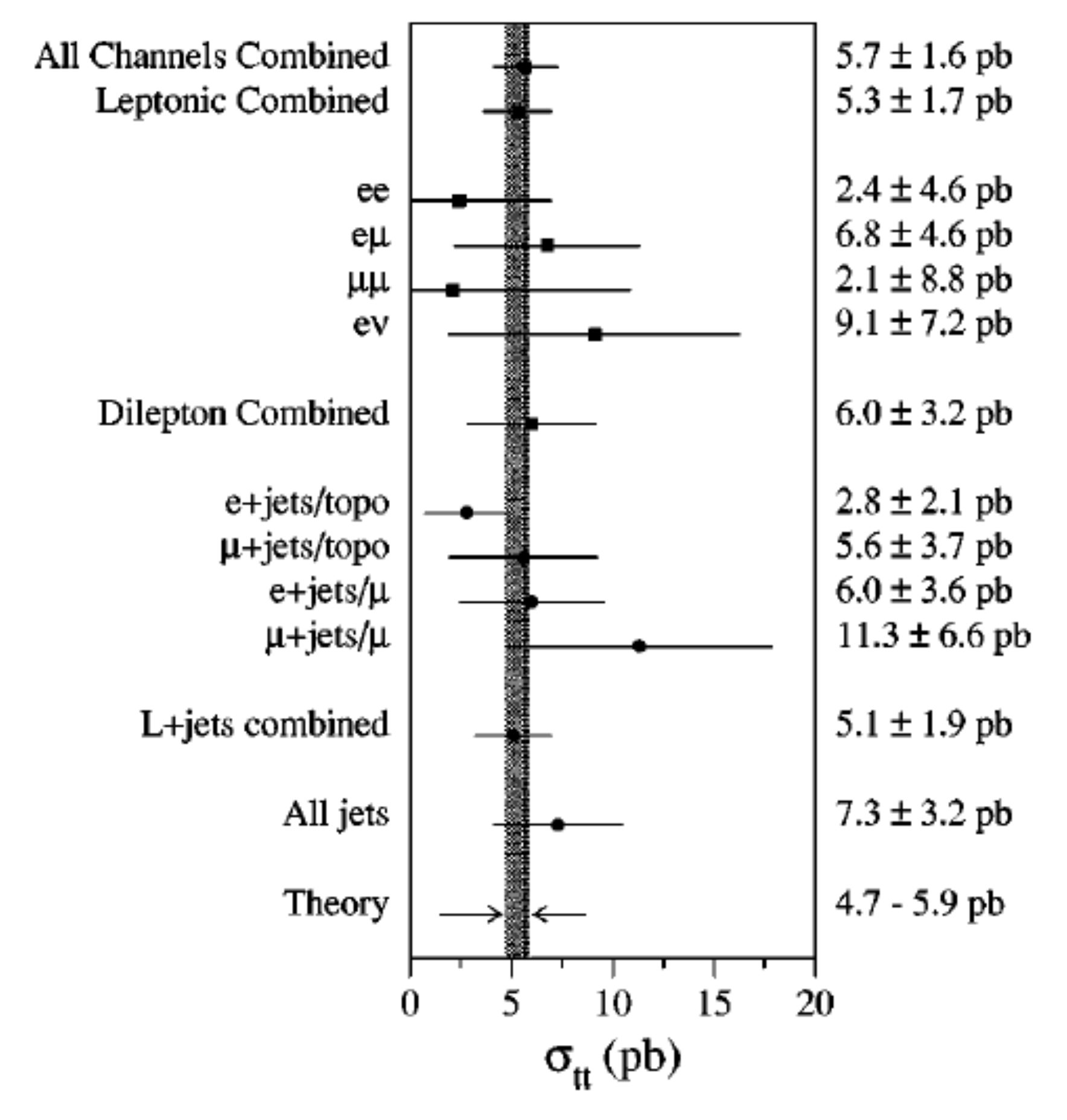}}
\caption{Measured \ttbar production cross section values for all
channels used by the D0 analyses, assuming a top quark mass of $172.1\;\rm GeV$. The vertical
line corresponds to the cross section for all channels combined and
the shaded band shows the range of theoretical predictions.
\label{fig:D0Run1xsec}}
\end{figure}

The most precise measurements of the top quark mass were obtained by both collaborations using data in final states containing an isolated electron or muon, large missing energy and at least 4 jets~\cite{PRL80-2767-1998,PRD58-052001-1998}. The mass is extracted using a maximum-likelihood method and templates extracted from simulated samples generated at different masses that have been reconstructed according to a \ttbar hypothesis. Figure~\ref{fig:CDFRun1mass} shows the reconstructed mass distribution for the four subsamples used in the CDF analysis and their combination. The combined measured mass was found to be 
$m_t=175.9\pm 4.8 (stat) \pm 4.9 (sys)\;\rm GeV$. The systematic uncertainty was dominated by the calibration of the jet energy and the signal modeling. The corresponding result from D0 was $m_t=173.3\pm 5.6 (stat) \pm 5.5 (sys)\;\rm GeV$. A reanalysis of the same dataset was published in 
2004~\cite{Nature429-638-2004} and yielded a final 
result of $m_t=180.1\pm 3.6 (stat) \pm 3.9 (sys)\;\rm GeV$, consistent with the previous result but with a precision comparable to all previous top mass measurements combined. The new method used leading-order matrix elements to calculate an event weight that was then used to assign more importance to events that were well measured and thus are more likely to correspond to the \ttbar signal. This method became the method of choice of the D0 collaboration for top quark mass measurements in Run II.

\begin{figure}[h,t,b,p]
\includegraphics[scale=0.3]{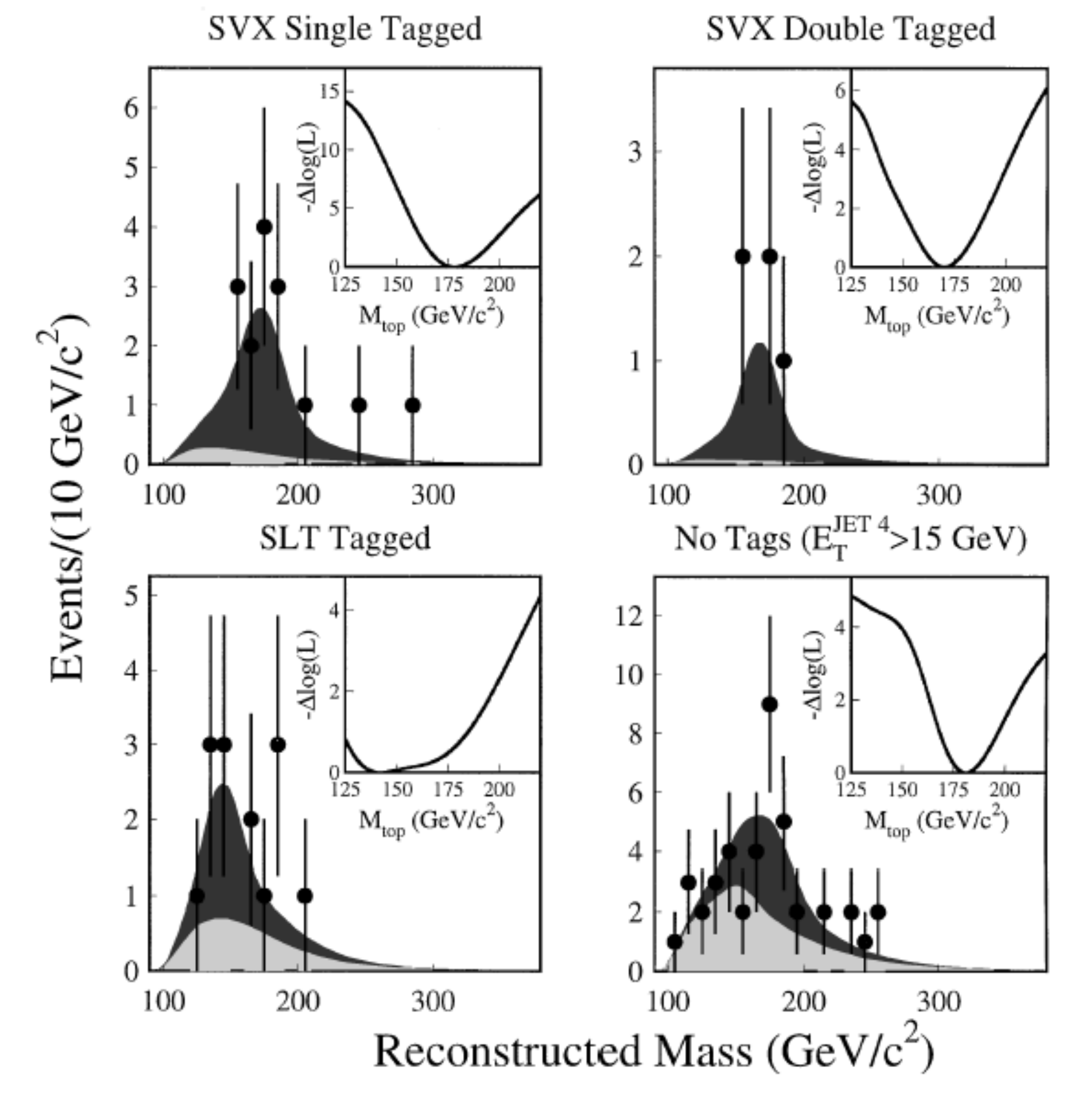}
\includegraphics[scale=0.3]{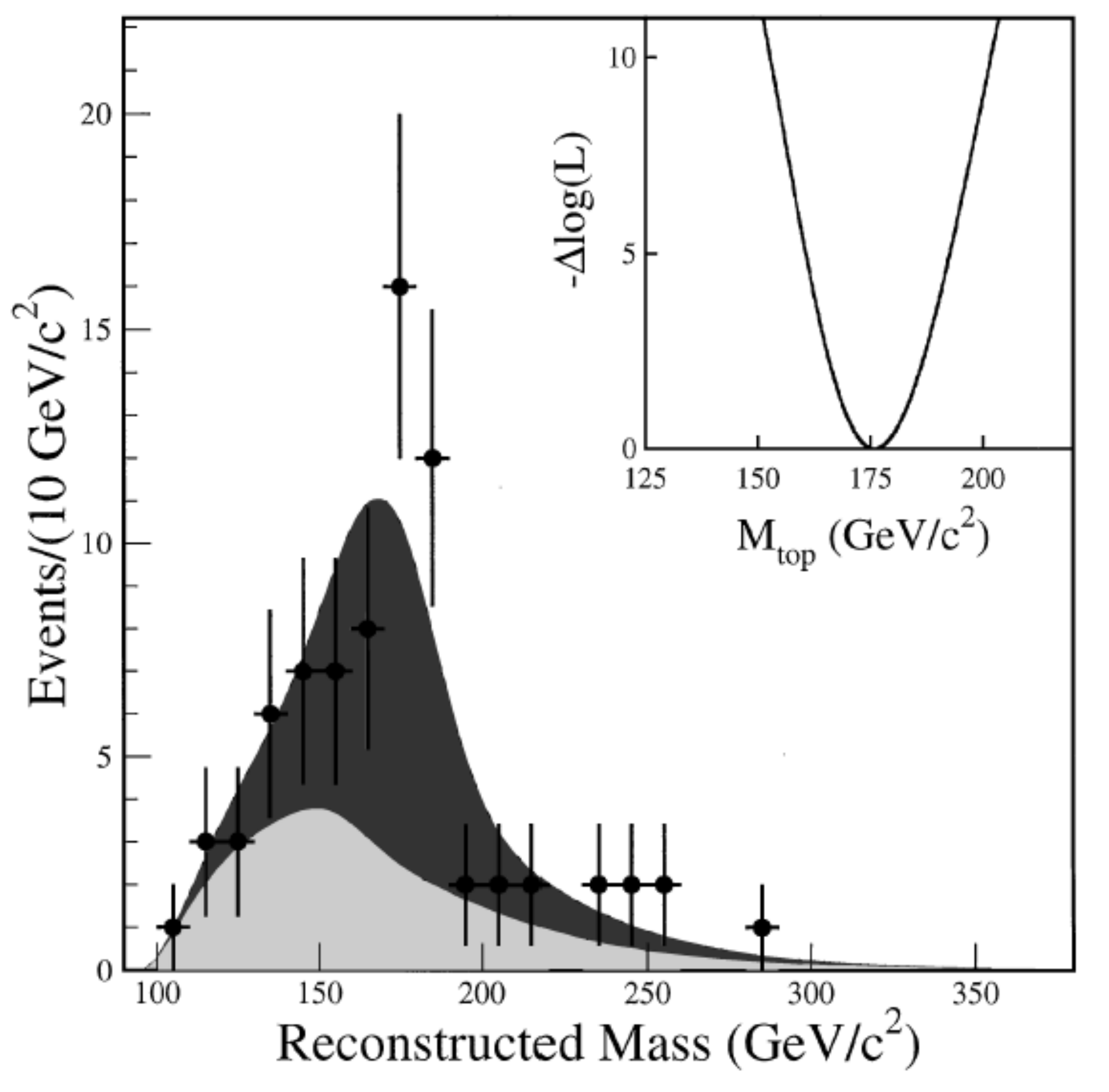}
\caption{Reconstructed mass distributions from the CDF experiment in (left) each of the four mass subsamples and (right) all subsamples combined. Each plot shows the data (points), the result
of the combined fit to top + background (dark shading), and
the background component of the fit (light shading). The insets
show the variation of the negative log-likelihoods with Mtop.
\label{fig:CDFRun1mass}}
\end{figure}
 
Both collaborations also measured the top quark mass using samples containing two isolated leptons and at least 2 jets~\cite{PRL82-271-1999,PRD60-052001-1999} and CDF also measured the top quark mass in events with six jets and no reconstructed isolated leptons~\cite{PRL79-1992-1997}. All results were in agreement with each other, giving further support to the hypothesis that the events originated from \ttbar production.  

The CDF collaboration used the \ttbar sample to directly measure the $t \to W b$ branching ratio from the number of events with 0, 1 and 2 $b$-tags. Assuming that top decays 100\% to $W b$ and that only three generations of fermions exist, the CKM matrix element~\cite{top-ckm} was 
extracted~\cite{PRL86-3233-2001} $|V_{tb}|=0.97^{+0.16}_{-0.12}$ or $|V_{tb}|>0.75$ at 95\% C.L. 

Both collaborations set upper limits at the 95\% C.L. on the electroweak production of single tops~\cite{PRD65-091102-2002,PLB517-282-2001} that were roughly six times larger than the SM prediction, an indication that far more data would be required before single top quark production could be observed. Both collaborations also searched for non-SM top decays, in particular, charged Higgs~\cite{PRL79-357-1997,PRL82-4975-1999} and flavor changing neutral currents~\cite{PRL80-2525-1998} (CDF), and measured other \ttbar properties, namely $W$ helicity~\cite{PRL84-216-2000} (CDF) and \ttbar spin correlations~\cite{PRL85-256-2000} (D0). In all instances, the results agreed with SM expectations. The publications pioneered analysis methods that would be applied to the much larger \ttbar samples in Run II and also to the data collected at the Large Hadron Collider by the ATLAS and CMS collaborations.

\section{Studies of Top Quark Pair Production}	
\subsection {\ttbar Production and Decay}
At Tevatron energies, top quarks are produced predominantly as a \ttbar pair. 
The total \ttbar production cross section for a hard
scattering process initiated by a $p\overline{p}$ collision at a center of mass energy of $1.96\;\rm TeV$ 
is a function of the top quark mass $m_t$. 
For a top quark mass of $172.5\;\rm GeV$, the predicted
SM \ttbar production cross section calculated with the  TOP++ program~\cite{toppp} at full NNLO+NNLL is 
$7.35^{+0.11}_{-0.21}\;\rm pb$, where the MSTW2008nnlo68cl PDF set was used.
Deviations of the measured cross section from the theoretical
prediction could indicate effects beyond quantum chromodynamics (QCD) perturbation theory.
Explanations might include substantial non-perturbative effects, new
production mechanisms, or additional top quark decay modes beyond the SM.

Within the SM, the top quark decays almost exclusively
into a $W$ boson and a $b$ quark, resulting in two $W$ bosons and 2 
$b$ jets in each \ttbar\ pair event. 
The $W$ boson itself decays into one lepton and its 
associated neutrino, or hadronically. 
 \ttbar\ pair decay channels are generally classified as follows:
the dilepton channel where both $W$ bosons decay leptonically 
into an electron or a muon ($ee$, $\mu\mu$, $e\mu$), 
the lepton + jets channel where one of the $W$
bosons decays leptonically and the other hadronically ($e$+jets,
$\mu$+jets), and the all-jets channel where both $W$ bosons decay
hadronically.  While the all-jets channel has the largest branching fraction ($\approx 46\%$), it also suffers from large backgrounds from multi jet production. The dilepton channels, with a branching ratio of $\approx 4\%$, results in the cleanest sample with minimal contamination from Drell-Yan and diboson production. The lepton + jets channel, with a branching ratio of $\approx 35\%$, has a background dominated by $W+$ jets production at a level significantly smaller than the all-jets channel,  making
it the channel of choice for the measurement of top quark properties. 
In many analyses, the identification of jets originating from the decay of a bottom quark ($b$-tagging) is used for the rejection of backgrounds. 

\subsection{Inclusive \ttbar Cross Sections}
\ttbar production cross sections $\sigma_{t\overline{t}}$ have been measured in all decay channels, and good agreement with the theoretical prediction has been found. Figure~\ref{fig:xsec-sum} summarizes the latest $\sigma_{t\overline{t}}$ measurements from CDF and D0, as well as their 
combination~\cite{PRD89-072001-2014}. 
In this publication, based on data samples with integrated luminosity between $2.9\;\rm fb^{-1}$ and $8.8\;\rm fb^{-1}$, results from six analyses are combined, reducing the uncertainty on the measured  $\sigma_{t\overline{t}}$ to $\approx 5\%$. The following results were included in the combination: 
\begin{itemize}
\item CDF dilepton uses the entire $8.8\;\rm fb^{-1}$ dataset and is based on counting the number of events with at least one $b$-tag~\cite{PRD88-091103-2013}. 
\item Two CDF lepton + jets analyses  based on  $4.6\;\rm fb^{-1}$ of data~\cite{PRL105-012001-2010}. While the first result relies on $b$-tagging to separate the signal from the backgrounds, the second one uses an artificial neural network (NN) and no $b$-tagging to discriminate between \ttbar signal and $W$+jets background. Both analyses reduce the uncertainty arising from the luminosity by concurrently measuring the ratio of the \ttbar to the $Z/\gamma^{*}$ cross sections. This technique 
replaces the $6\%$ uncertainty on the luminosity with a $2\%$ combined theoretical and experimental uncertainty on the $Z/\gamma^{*}$ cross section, resulting in the most precise single $\sigma_{t\overline{t}}$ measurement. 
\item CDF all-jets, based on $2.9\;\rm fb^{-1}$ of data, requires the presence of six to eight jets that satisfy a series of kinematical requirements imposed by means of a NN~\cite{PRD81-052011-2010}. At least one of these jets must be $b$-tagged. 
\item D0 dilepton, based on $5.4\;\rm fb^{-1}$ of data, uses a likelihood fit to a discriminant based on a NN $b$-tag algorithm~\cite{PLB704-403-2011}. 
\item D0 lepton + jets, based on $5.3\;\rm fb^{-1}$ of data, uses events with at least three jets 
which are split into subsamples according to the total
number of jets and $b$-tags~\cite{PRD84-012008-2011}. A random forest multivariate discriminant is used in background-dominated subsamples to separate the signal from the background. The cross section is extracted by simultaneously fitting the number of events in the signal dominated subsamples (3 jets with 2 $b$-tags and 4 jets with at least 1 $b$-tag) 
and the random forest discriminant in the background dominated samples.
\end{itemize}
\begin{figure}[h,t,b,p]
\centerline{\includegraphics[width=14cm]{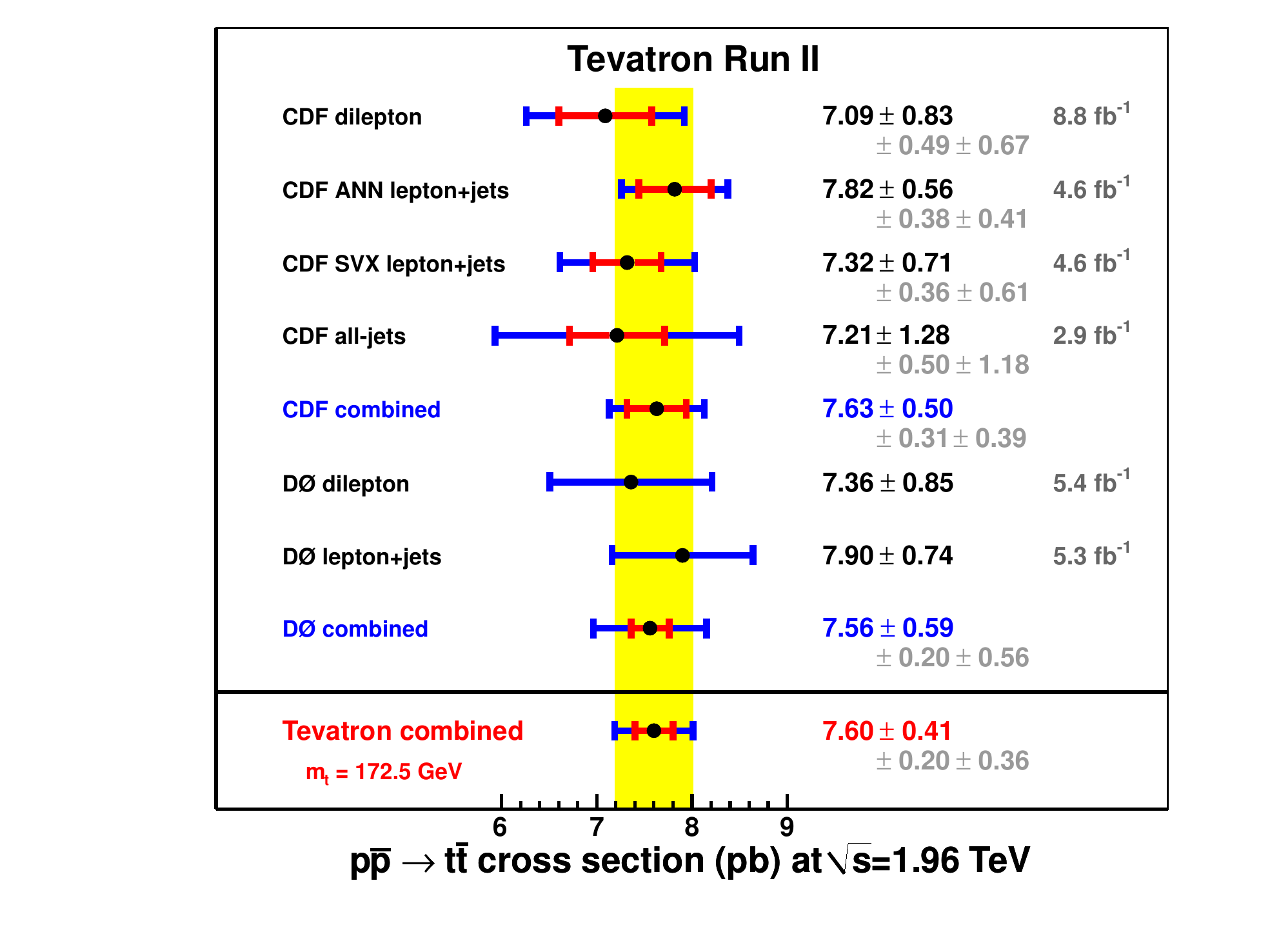}}
\caption{Measured \ttbar cross sections from D0 and CDF together with their combinations~\cite{PRD89-072001-2014}. 
The outer (blue) error bars show the total uncertainties on each measurement. The inner (red) error bars indicate the contribution from the 
statistical uncertainties for those measurements where this quantity is available. 
\label{fig:xsec-sum}}
\end{figure}

Channels with a $\tau$-lepton decaying into an electron or a muon are included in the corresponding dilepton or lepton + jets channels. Events where the $\tau$-lepton decays hadronically are treated separately. CDF measured $\sigma_{t\overline{t}}$  in the $\tau+$ lepton (electron or muon) channel~\cite{arXiv:1402.6728} using the complete data set corresponding to $9.0\;\rm fb^{-1}$ of data. 
D0 measured the \ttbar cross section in the $\tau$+ jets channel~\cite{PRD82-071102-2010} using  $1.0\;\rm fb^{-1}$ of data. Similarly, CDF used $0.3\;\rm fb^{-1}$ of data and events with an inclusive signature of large missing transverse energy and jets~\cite{PRL96-202002-2006} to maintain the sensitivity to hadronic $\tau$ decays. 
These measurement are in good agreement with results from other channels and theoretical predictions. 

Because beyond the standard model (BSM) processes can affect different \ttbar decay modes differently, it has been a priority of the Tevatron experiments to measure $\sigma_{t\overline{t}}$ in different channels and with different techniques, in particular with and without the assumption of the presence of two $b$-quarks in the final state. Even though the early measurements suffered from rather large uncertainties ($\approx 30\%$), recent individual results have reached precisions as small as 
$7\%$~\cite{PRL105-012001-2010}, allowing for stringent tests of QCD 
predictions~\cite{PRL109-132001-2012,PRD80-054009-2009,PRD78-074005-2008,JHEP09-127-2008}. 

\subsection{Differential \ttbar Cross Sections and Forward-Backward \ttbar Asymmetry}
The larger \ttbar samples available in Run II also allowed for differential \ttbar cross section measurements. A result from D0 uses the entire dataset of $9.7\;\rm fb^{-1}$ and the lepton + jets channel to measure the \ttbar production cross section as a function of the transverse momentum and absolute rapidity of the top quarks as well as of the invariant mass of the \ttbar pair~\cite{ arXiv:1401.5785}. The data are corrected for detector efficiency, acceptance and bin migration by means of a regularized
unfolding procedure. In all cases, the
differential cross sections agree well with QCD MC generators and predictions at approximate NNLO~\cite{JHEP1009-097-2010}. 
Figure~\ref{fig:xsec-diff} shows, as an example, the unfolded differential cross section as a function of the \ttbar pair invariant mass compared to predictions. 
\begin{figure}[h,t,b,p]
\centerline{\includegraphics[width=12cm]{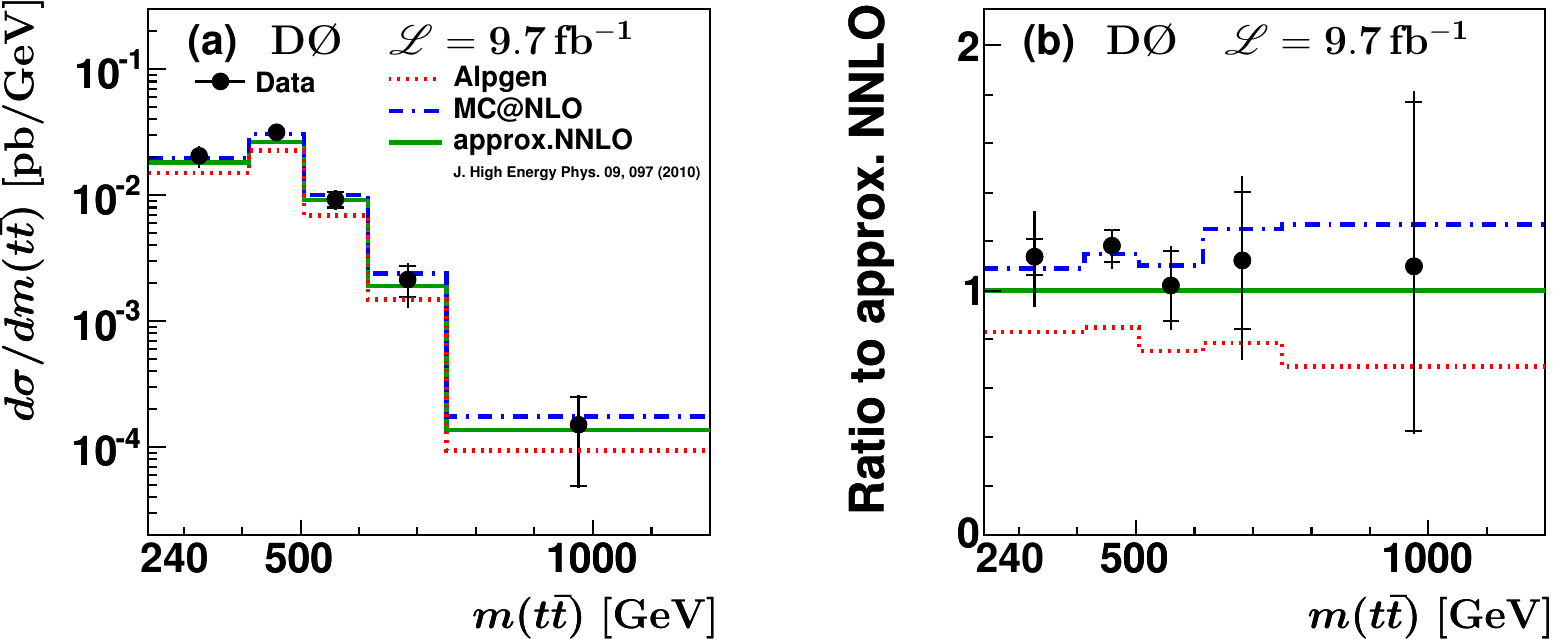}}
\caption{(a) The differential cross section as a function of the invariant mass of the \ttbar pair for data compared to several QCD
predictions. (b) The ratio of cross section to the QCD prediction at approximate NNLO~\cite{JHEP1009-097-2010}.  In both cases, the inner error bars correspond to the statistical uncertainties and the outer error bars to the systematic uncertainties.
\label{fig:xsec-diff}}
\end{figure}

The CDF and D0 collaborations also investigated other properties of the \ttbar production mechanism in search of deviation from the SM predictions. Some 
deviation from SM predictions was found on the mass and rapidity dependent forward-backward \ttbar asymmetry reported by CDF using $5.3\;\rm fb^{-1}$  of lepton+jets 
data~\cite{PRD83-112003-2011}. The corresponding D0 analysis~\cite{PRD84-112005-2011}, based on $5.4\;\rm fb^{-1}$  of lepton+jets 
data, also finds a disagreement between data and SM predictions, however, it finds no statistically significant enhancement of the asymmetry neither as a function of the \ttbar mass nor for the rapidity. A follow up analysis from CDF uses the entire Run II data set~\cite{PRD87-092002-2013} and observes a linear dependence on both the rapidity difference and the \ttbar mass, with higher slopes than the NLO prediction. 
The D0 analysis using the entire dataset~\cite{arXiv:1405.0421}  measures an inclusive forward-backward asymmetry of 
$A_{FB}=(10.6\pm3.0)\%$, in agreement with SM predictions which range from $5\%$ at NLO to $9\%$ once electroweak effects are taken into 
account~\cite{PRD86-034026-2012}. 
The measured dependences of the asymmetry on rapidity and mass are also in agreement with the SM predictions, but do not disagree with the larger asymmetries observed by previous analyses, as can be seen in Fig.~\ref{fig:d0CDFAfb}.
\begin{figure}[h,t,b,p]
\centerline{\includegraphics[width=6.5cm]{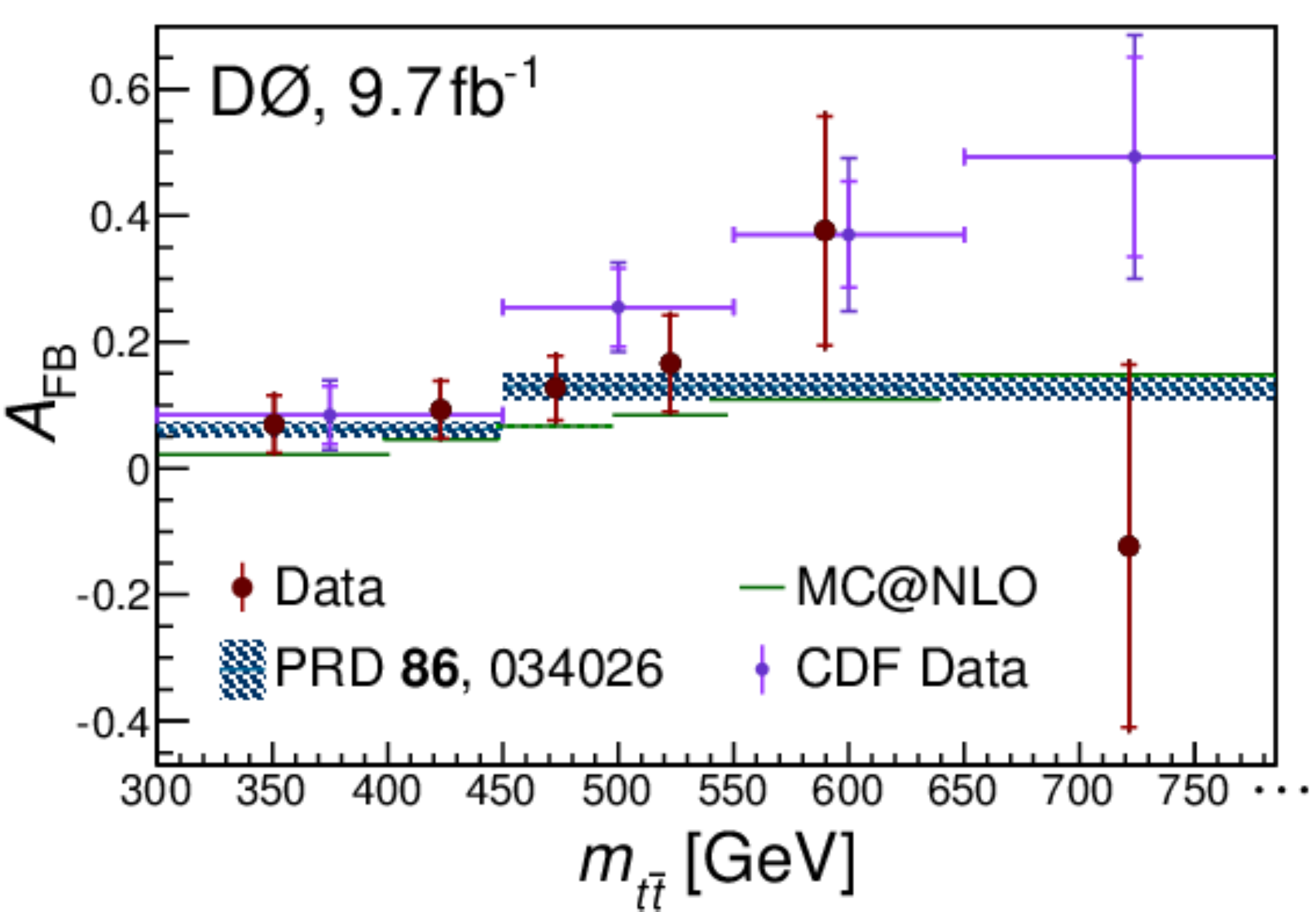}\includegraphics[width=6.3cm]{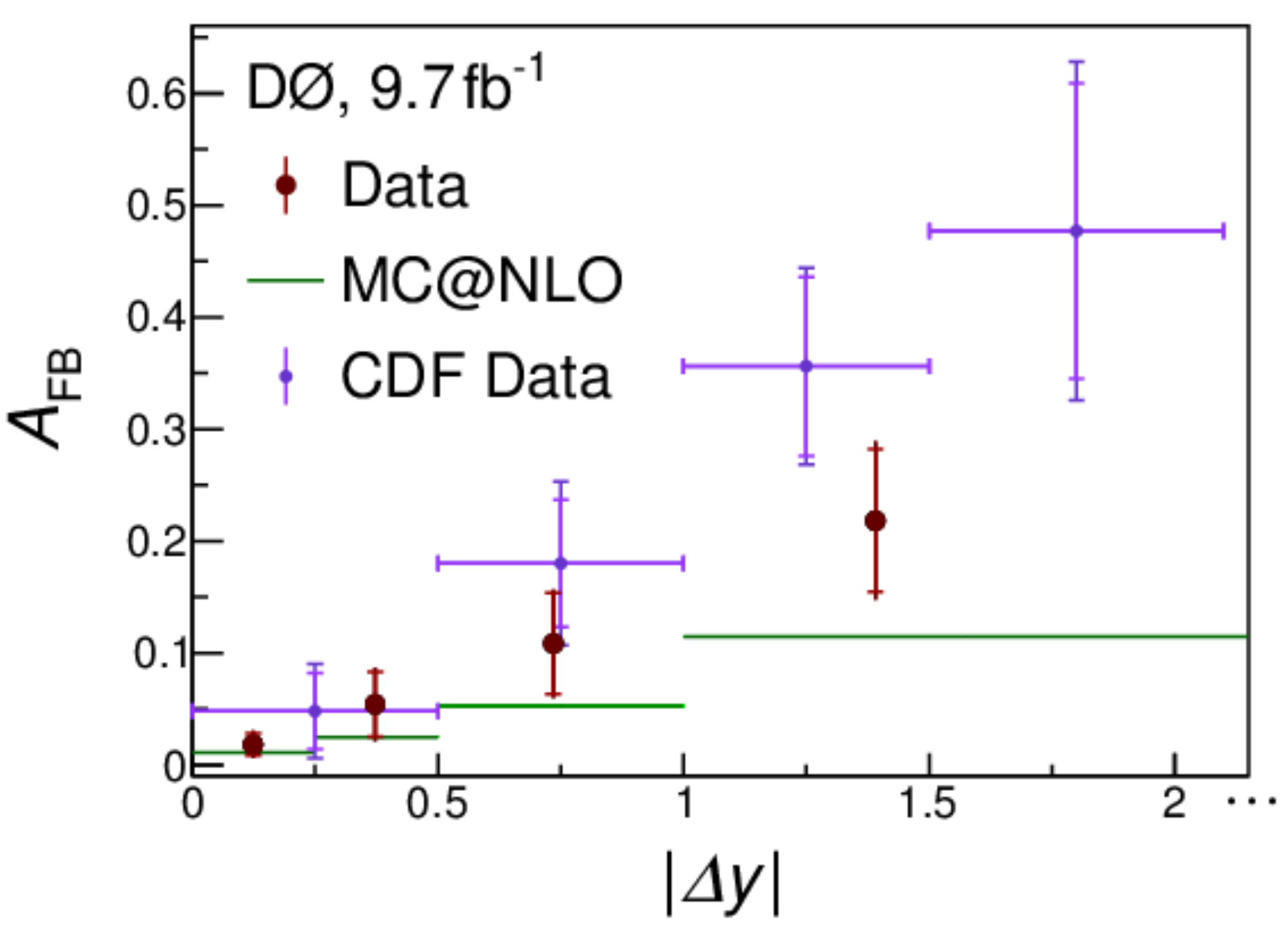}}
\caption{The dependence of the forward-backward asymmetry on the \ttbar invariant mass (left) and 
the difference in rapidities of top quark and antiquark (right). 
The measurements from CDF and D0 are compared to theoretical predictions.
\label{fig:d0CDFAfb}}
\end{figure}

It is important to note that the theoretical predictions for differential distributions, including $A_{FB}$, are only available at order  $\alpha_s^3$, where 
$\alpha_s$ is the strong coupling constant. Since the SM \ttbar forward-backward asymmetry only appears at this order, no full higher order prediction exists. The relative uncertainty on the 
 $\alpha_s^3$ calculation due to higher order corrections could therefore be significant. 

Alternatively, both collaborations also measure the  asymmetry in the charge-weighted rapidity of the lepton, which does not require the reconstruction
of the kinematic properties of the full \ttbar system .  
The D0 collaboration combined $9.7\;\rm fb^{-1}$  of dilepton~\cite{PRD88-112002-2013} and  lepton+jets~\cite{arXiv:1403.1294} data and measures the asymmetry at
production level  $A_{FB}^l=(4.2\pm2.4)\%$ for $|y_l| \le 1.5$, in agreement with the prediction of 2.0\% from the NLO QCD generator MC@NLO.
These two measurements are individually extrapolated to cover the full
phase space (using the MC@NLO simulation), and combined. The extrapolated result of 
$A_{FB}^l ({\rm ex})=(4.7\pm2.3 ({\rm stat}) \pm 1.5 ({\rm syst}))\%$ facilitates the comparison with theoretical calculations and the 
extrapolated result from CDF. 

The CDF collaboration uses  
$9.4\;\rm fb^{-1}$  of lepton+jets data~\cite{PRD88-072003-2013} and finds the extrapolated value of 
$A_{FB}^l=(9.4\pm^{3.2}_{2.9})\%$, to be compared with the prediction of $(3.8\pm0.3)\%$~\cite{PRD86-034026-2012}. A corresponding measurement in the 
dilepton channel $A_{FB}^l=(7.2\pm6.0)\%$~\cite{arXiv:1404.3698}, is consistent with predictions and the D0 results.
Figure~\ref{fig:D0asym} summarizes the measured lepton asymmetries for the various samples used in the analyses. 
\begin{figure}[h,t,b,p]
\centerline{\includegraphics[width=9.4cm]{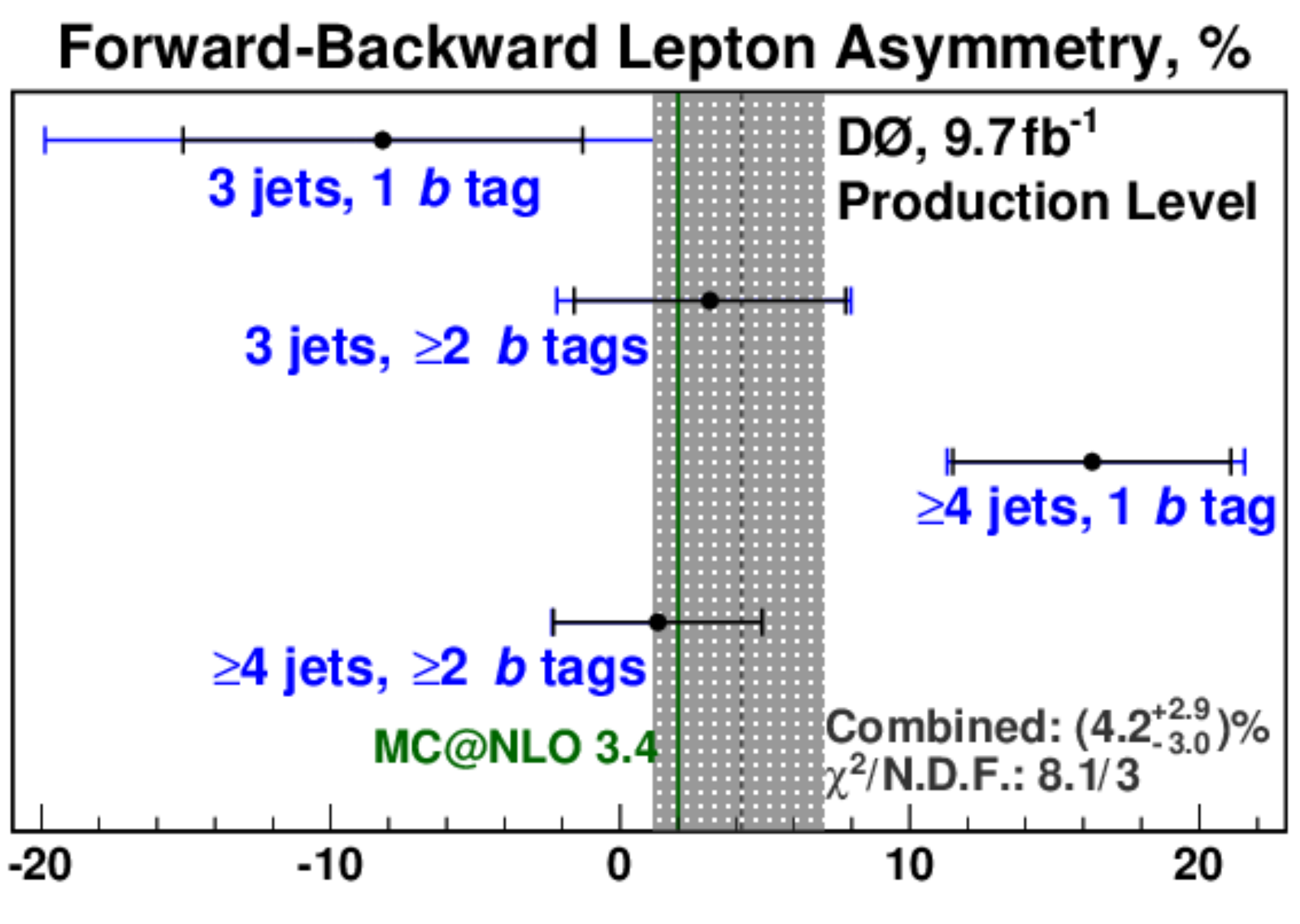}}
\centerline{\includegraphics[width=10cm]{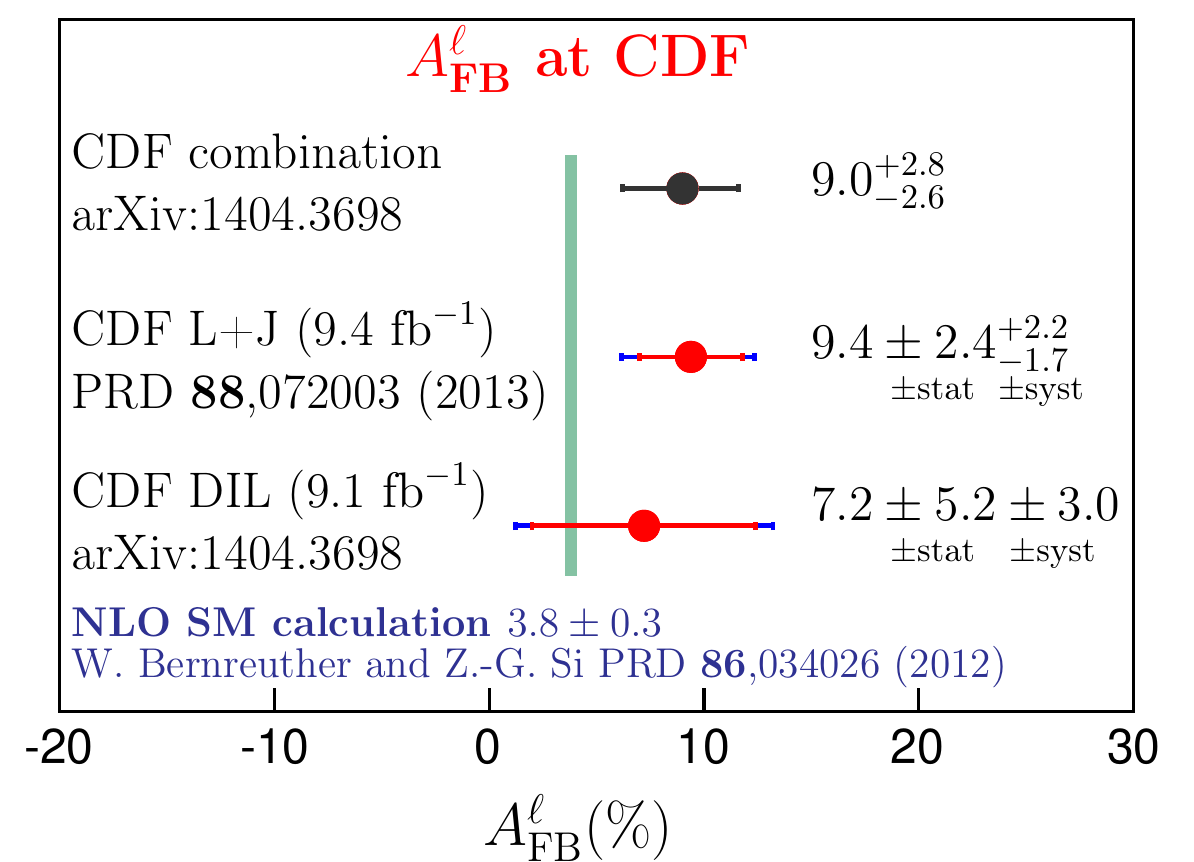}}
\caption{Measured production-level $A_{FB}^l$ by analysis channel compared to the theoretical prediction for D0 (top) and CDF (bottom). 
Statistical (total) uncertainties are indicated by the
inner (outer) vertical lines.
\label{fig:D0asym}}
\end{figure}

The CDF collaboration also used the entire dataset of  $9.4\;\rm fb^{-1}$ and the lepton + jets channel to measure the \ttbar production cross section as a function of 
the production angle of the top quark with respect to the incoming proton momentum as measured in the \ttbar center-of-mass frame 
($\theta_t$)~\cite{PRL111-182002-2013}. 
The inclusive measurements of $A_{FB}$
are equivalent to a two-bin measurement of this normalized differential
cross section, with one bin forward ( $cos\theta_t > 0)$ and one
bin backward ( $cos\theta_t < 0)$. The full shape of the differential
cross section thus provides additional information and has
the potential to identify which aspects of the shape of $d\sigma/d(cos\theta_t)$ affect the forward-backward asymmetry.
The shape of $d\sigma/d(cos\theta_t)$ is characterized by Legendre polynomials with Legendre moments $a_1$ to $a_8$. 
A mild excess is observed in the differential cross section in the linear term $a_1$ in $cos\theta_t$, as can be observed in Fig.~\ref{fig:CDFasym}. All other terms agree well with NLO SM predictions. 
\begin{figure}[h,t,b,p]
\centerline{\includegraphics[width=10cm]{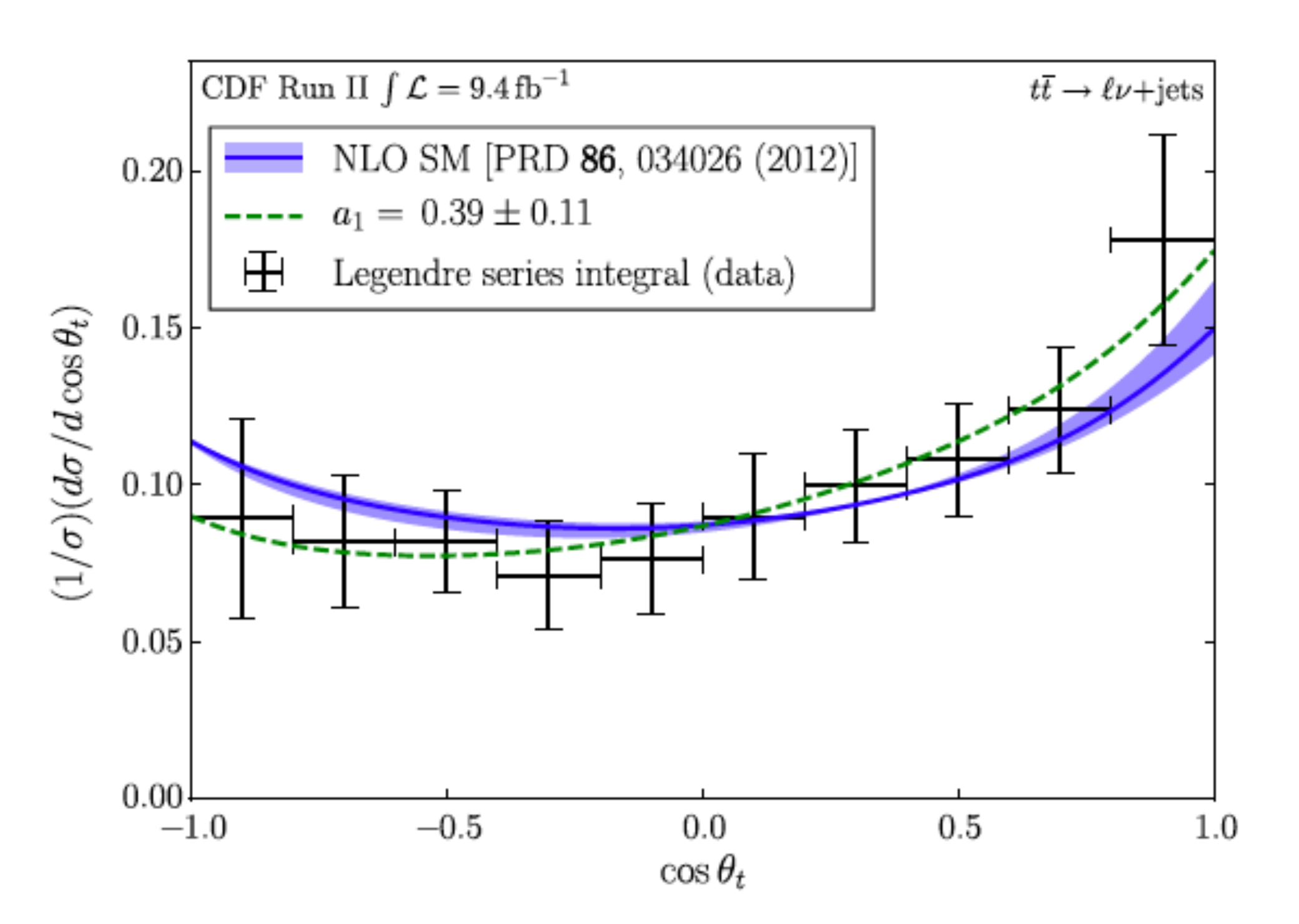}}
\caption{Fraction of cross section accruing in 10
bins of $cos\theta_t$, obtained by integrating the series of Legendre
polynomials over the width of each bin.
\label{fig:CDFasym}}
\end{figure}

These differential measurements, possible for the first time due to the high statistic samples available in Run II, constitute detailed probes into the \ttbar system modeling and  serve as a means to better understand higher-order corrections to SM predictions. 
A precise modeling is vital in many searches for new phenomena, where differential top quark cross sections are used to set constraints on beyond the SM processes. 
A detailed modeling is also needed in searches for rare processes that involve new particles that decay to a \ttbar pair, where particles are produced in 
association with a \ttbar pair, or where \ttbar is among the dominant backgrounds. 

\subsection{Searches for Anomalous \ttbar Production}
Both collaborations have also searched for anomalous \ttbar production. 
Several beyond the SM theories predict resonant 
production of \ttbar\ pairs~\cite{PRD49-4454-1999}. Examples include topcolor models~\cite{PLB266-419-1991} and
models with extra dimensions, such as Kaluza--Klein (KK) excitations of gluons or gravitons
in various extensions of the Randall--Sundrum model~\cite{PRL83-3370-1999}.
Using $9.5\;\rm fb^{-1}$ of data, CDF has studied the
invariant mass distribution in lepton + jets events~\cite{PRL110-121802-2012}. 
The observed spectrum is consistent with 
SM expectations, showing no evidence for additional resonant production 
mechanisms. Consequently, the data is used to 
set upper limits on $\sigma \times B(X \rightarrow t\overline{t})$ 
for different hypothesized 
resonance masses. Similar results had been obtained by the D0 collaboration using  $5.3\;\rm fb^{-1}$ of data~\cite{PRD85-051101-2012}. 

Various axigluon models with different couplings~\cite{arXiv:1401.2443} that enhance $A_{FB}$ also modify the differential cross section distributions, in particular 
for the \ttbar invariant mass. The D0 collaboration has extended its differential cross section analysis~\cite{arXiv:1401.5785} to search for axigluons by comparing various models to the unfolded cross section data. Several of these models were constructed to account for the larger than predicted forward-backward asymmetries 
observed at the Tevatron. Models including heavy masses are disfavored by the Tevatron and LHC data, with the Tevatron data being especially sensitive to the lower 
mass regions. Figure~\ref{fig:axigluons} shows the ratio of the different models to data, as a function of the \ttbar invariant mass and the top transverse momentum. Even though the invariant mass is the most sensitive variable, the top transverse momentum and rapidity shows sensitivity that allows to disentangle
the various models. 
\begin{figure}[h,t,b,p]
\centerline{\includegraphics[width=14cm]{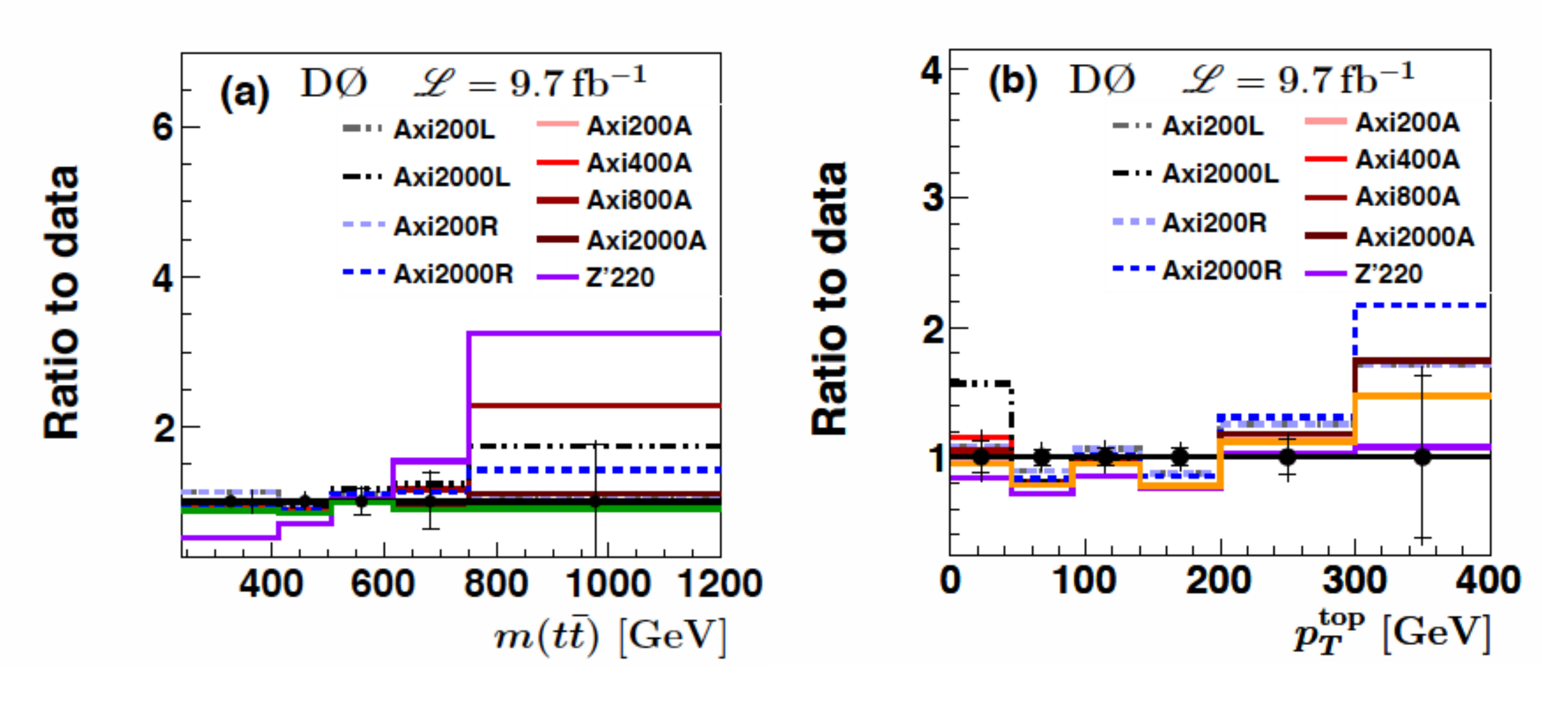}}
\caption{Differential cross section distributions as a function of (a) invariant mass and (b) top transverse momentum
comparing the D0 data to various benchmark models of axigluon contributions to the \ttbar production cross section. 
\label{fig:axigluons}}
\end{figure}
Alternatively, the D0 collaboration has also compared the forward backward asymmetries in the lepton+jets and 
dilepton channels to SM NLO predictions and axigluon models~\cite{PRD88-112002-2013}. The ratio of the measured asymmetry in the lepton+jets and dilepton samples is consistent at the level of 2 standard deviations with the SM prediction, and sensitive to various axigluon models, as can be seen in Figure~\ref{fig:axi2} 
\begin{figure}[h,t,b,p]
\centerline{\includegraphics[width=8cm]{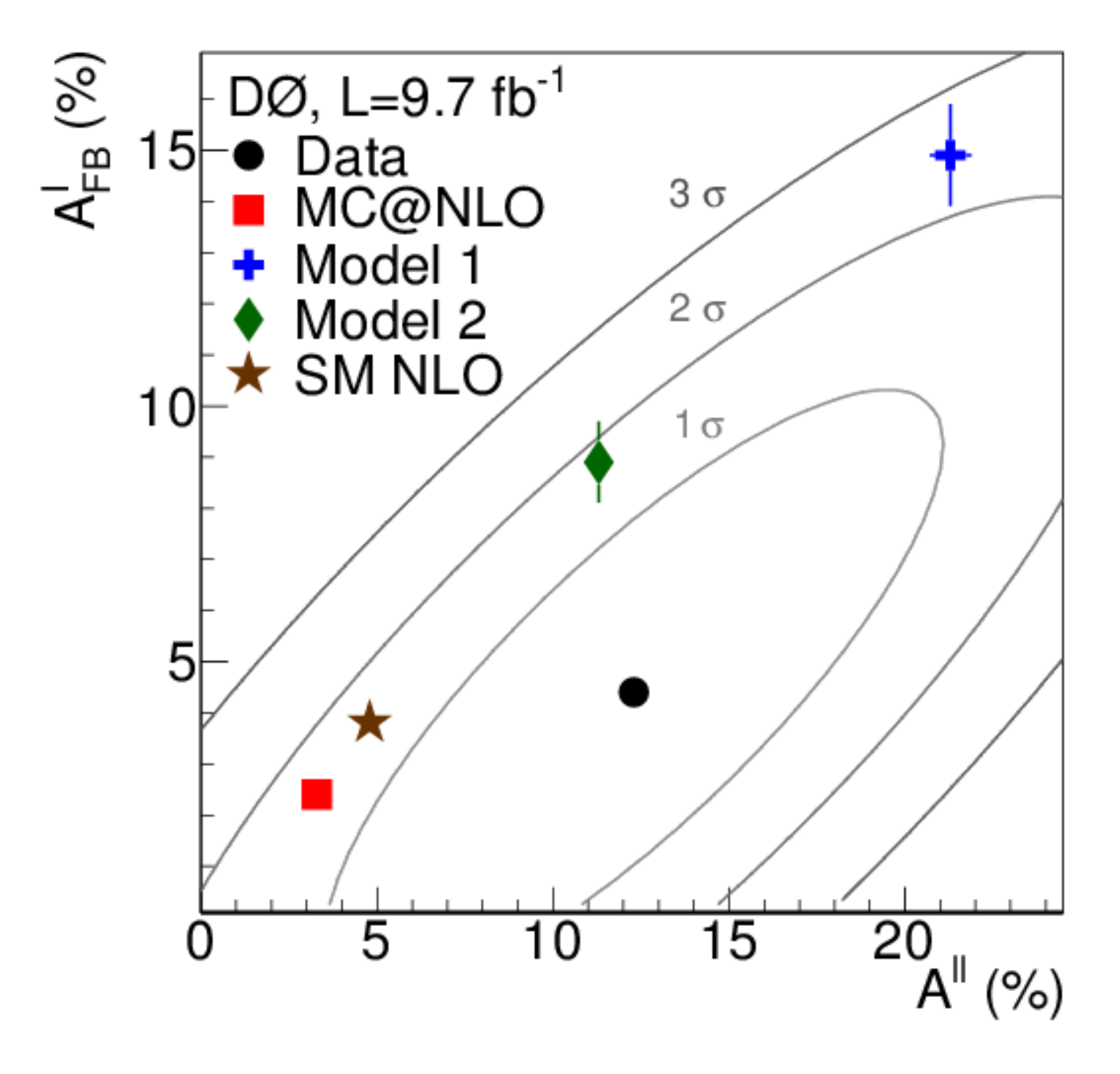}}
\caption{Extrapolated forward-backward asymmetry in the lepton+jets data vs dilepton \ttbar data as measured by D0. The measurement is compared to SM predictions and various axigluon models. The ellipses represent contours of total uncertainty at 1, 2, and 3 SD on the measured
result. All values are given in \%. Predicted asymmetries are
shown with their statistical uncertainties.
\label{fig:axi2}}
\end{figure}

The CDF collaboration has also searched for a heavy vector boson that decays to two gluons (chromophilic $Z^{\prime}$~\cite{PRD85-115011-2012}), in the case where the off-shell gluon converts to a \ttbar pair. Using $8.7\;\rm fb^{-1}$ of data, and selecting events with one charged lepton, large $\met$ and at least five jets, upper limits between 300 and 40~fb (at the 95\% C.L.) are set for $Z^{\prime}$ masses from 400 to 1000~GeV. 

\section{Observation and Studies of Single Top Quark Production}
\subsection{Single Top Quark Production}
In the SM, single top production at hadron colliders provides an 
opportunity to study the charged-current weak-interaction of the top quark. 
At the Tevatron, the dominant production mode is the exchange of a space-like virtual $W$ boson between a light quark and a bottom quark in the $t$-channel. The second mode is the decay of a time-like virtual $W$ boson in the $s$-channel. A third process, usually called ``associated production'' or $Wt$, has negligible cross section at the Tevatron, but has been recently observed at the LHC~\cite{PRL112-231802-2014}. Figure~\ref{fig:stop} shows the lowest level Feynman
diagrams for single top quark production at the Tevatron.
\begin{figure}[h,t,b,p]
\centerline{\includegraphics[width=5cm]{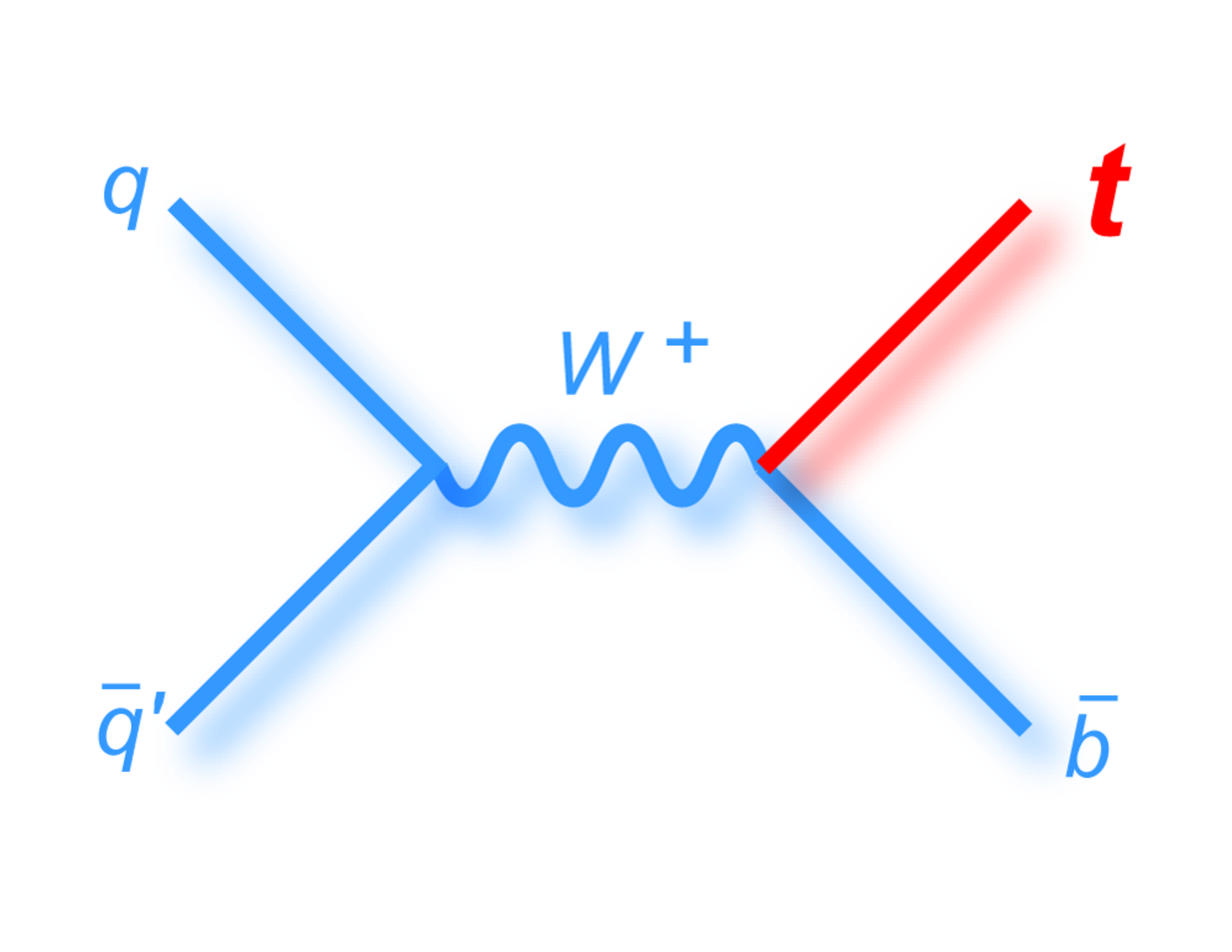}\includegraphics[width=5cm]{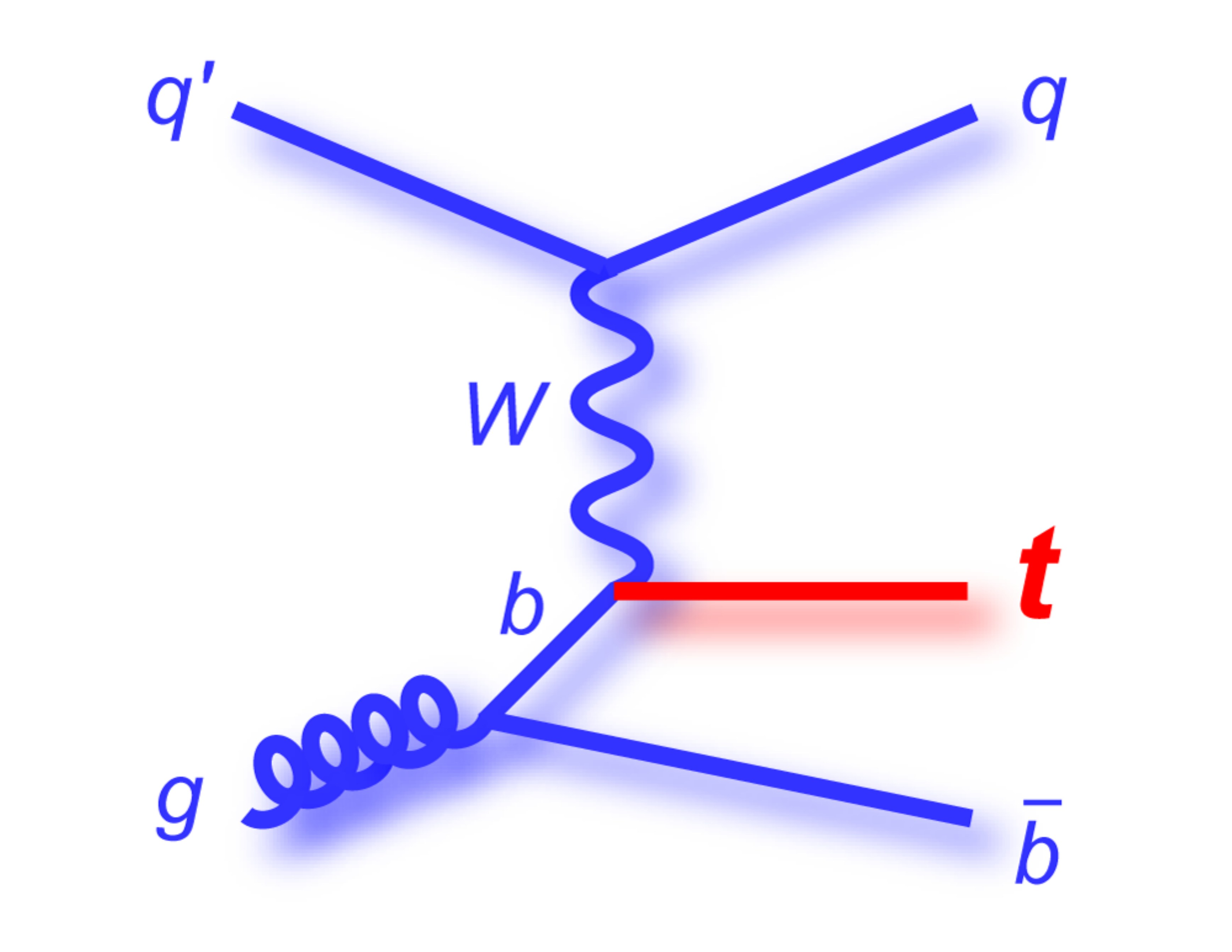}\includegraphics[width=4cm]{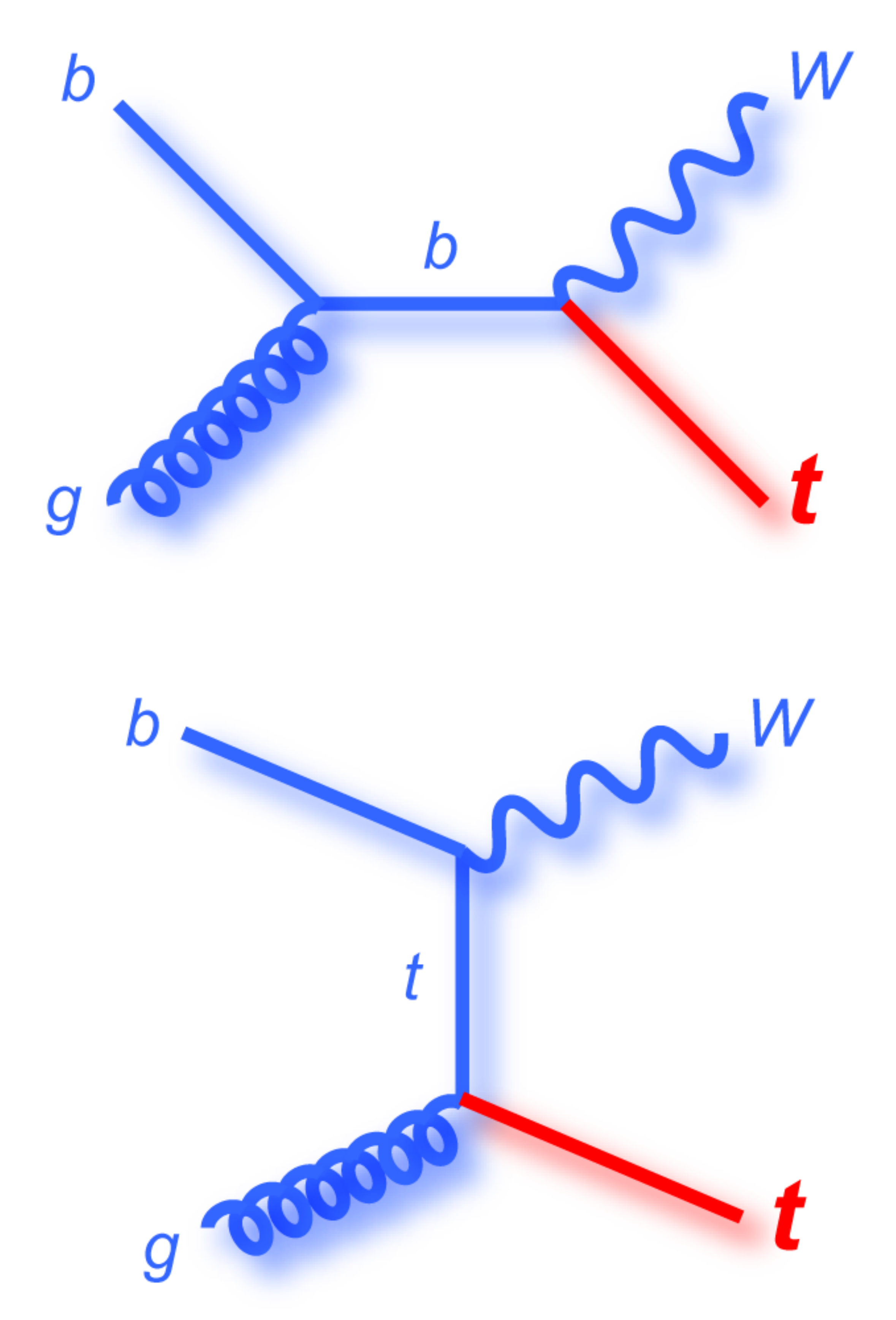}}
\caption{Lowest level Feynman
diagrams for (left) $s$-channel and (center) $t$-channel and (right) associated single top quark production at the Tevatron.
\label{fig:stop}}
\end{figure}

For a top mass of $172.5\;\rm GeV$ the predicted cross section, calculated at NLO+NNLL, is $2.10\pm0.13\;\rm pb$ for the $t$-channel and $1.05\pm0.06\;\rm pb$ for the $s$-channel~\cite{PRD-83-091503-2011,PRD81-054028-2010}, about half the rate as the one for \ttbar production. Naively one would expect the production rate
via the electroweak force to be much lower, however, the strong interaction cannot  change the flavor of the particles, which means a top quark produced by
it must be accompanied by a top antiquark. The weak interaction can change one kind of particle
into another, and thus it may produce one top quark at a time. The strong force is the stronger
one, but the requirement of enough energy to produce two top quarks suppresses the production
cross section.

\subsection{Single Top Quark Discovery}
In the search for single top production, both collaborations select events containing an  isolated electron or muon and large missing transverse energy ($\met$) from the decay of the $W$ boson originating from the top quark. Events are also required to have at least 2 jets, with at least one of the jets being $b$-tagged. This topology is referred to as  ``lepton+jets". The CDF collaboration also selects events containing jets and $\met$, but no reconstructed leptons, referred to as ``\met+jets''. The D0 collaboration also searched for 
single top production in the $\tau$+jets channel, using $5.4\;\rm fb^{-1}$ of data, and selecting events containing an isolated hadronically decaying $\tau$ lepton and at least two jets, with at least one being $b$-tagged~\cite{PLB690-5-2010}. 
In all cases, the resulting samples are dominated by backgrounds, and the expected amount of signal is 
smaller than the uncertainties on those backgrounds. Both collaborations thus developed multivariate analysis techniques (MVA) to separate the single top signal from the overwhelming backgrounds, as a simple counting experiment is not possible. 
In most cases, multiple MVAs were used on the same dataset, each defining a discriminant that was then used to constrain the 
uncertainties on the backgrounds and extract the signal contribution. The correlation between the outputs of the individual methods was typically found to be 
$\approx 70\%$. An increase in sensitivity can therefore be obtained by using their outputs as inputs to a ``superdiscriminant", a method employed by both collaborations. The cross section measurements were in all cases obtained using a Bayesian statistical analysis of the superdiscriminant output, where the data is compared to the sum of the predictions for signal and background processes. 

Single top quark production for the combined $s$ and $t$ channel was first observed by the CDF and D0 collaborations in 2009~\cite{PRL-103-092002-2009, PRD-82-112005-2010, PRL-103-092001-2009} using $3.2\;\rm fb^{-1}$ and $2.3\;\rm fb^{-1}$ of integrated luminosity, respectively. The measured cross sections for the 
combined $s$ and $t$-channel production were $2.32^{+0.6}_{-0.5}\;\rm pb$ and $3.94\pm 0.88\;\rm pb$, respectively. For these analyses, the CDF collaboration combined their lepton+jets and their \met+jets samples, while the D0 collaboration relied on the combination of their lepton+jets samples selected depending on the 
number of jets and the number of b-jets in the events. 
Figure~\ref{fig:stop-discovery} shows the output of the superdiscriminant for the CDF and the D0 analyses.
\begin{figure}[h,t,b,p]
\centerline{\includegraphics[width=7.5cm]{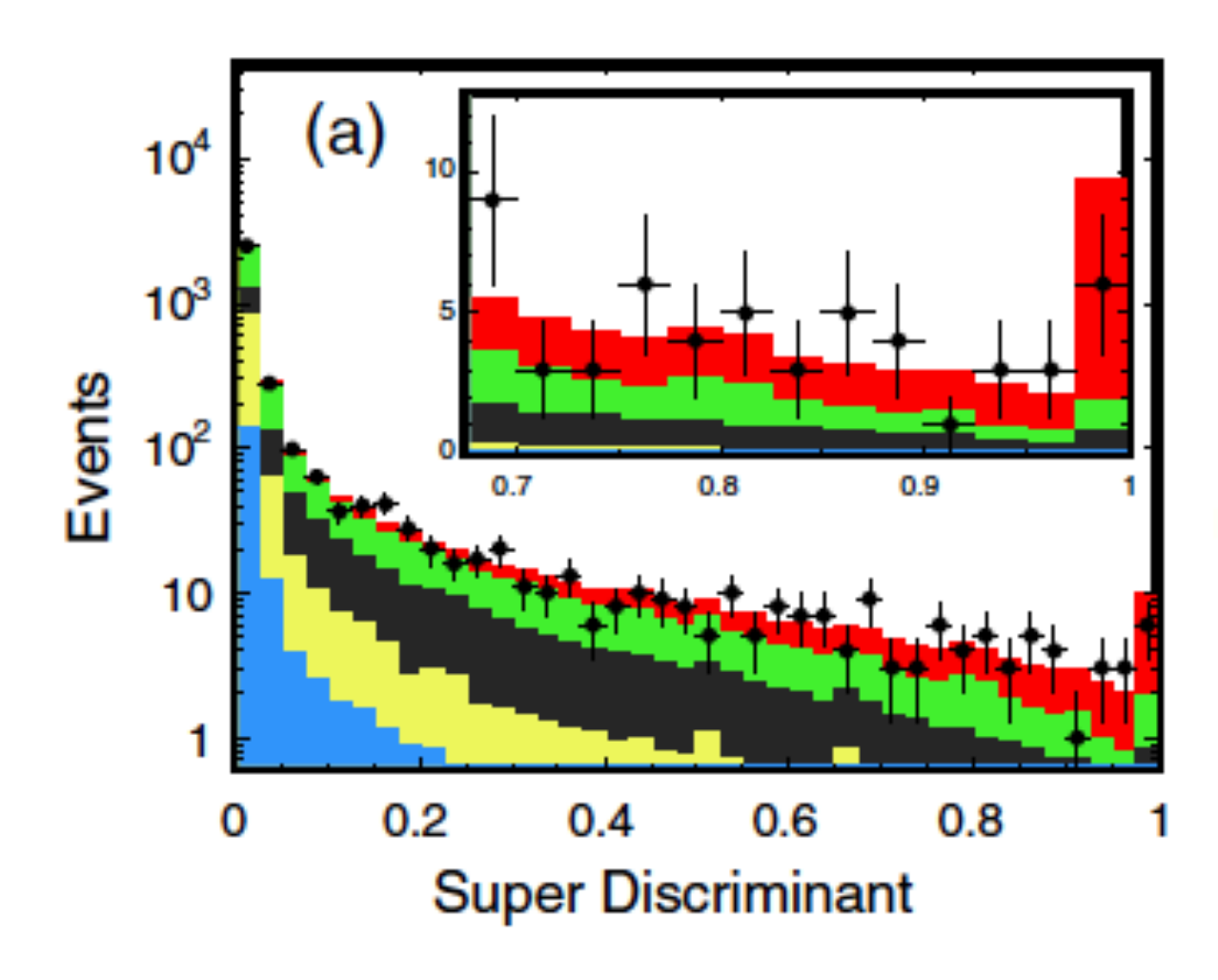}\includegraphics[width=6cm]{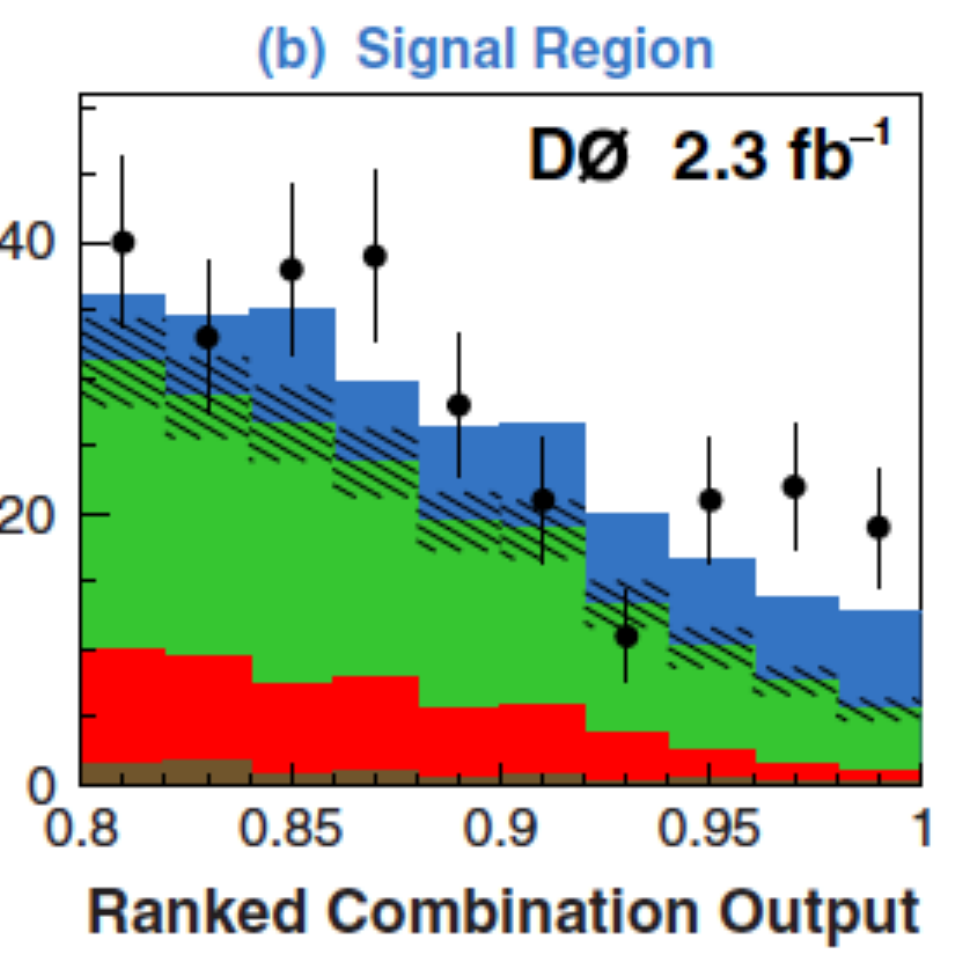}}
\caption{Output of the superdiscriminant for the CDF (Left) and  D0 (Right). The single top signal is shown in red in the CDF figure and in blue in the D0 figure.
\label{fig:stop-discovery}}
\end{figure}

\subsection{Final results from Run II}
\label{sec:stop-Run2}
Subsequent measurements by the D0 
collaboration~\cite{PRD-84-112001-2011,PLB-705-313-2011,PLB-726-656-2013 }  used larger datasets of $5.4\;\rm fb^{-1}$ and 
$9.7\;\rm fb^{-1}$ and reported both the $s$+$t$ cross section, as well as the individual contributions, and include the individual observation of the $t$-channel production  with a measured cross section of $\sigma_t=2.9\pm0.59\;\rm pb$, as well as evidence for $s$-channel production. The CDF collaboration later also announced evidence for $s$-channel production using the entire Run II dataset~\cite{arXiv:1402.0484, arXiv:1402.3756}. 
All results are in good agreement with each other and with SM expectations, as can be observed in Figure~\ref{fig:singletop-sum}. 
\begin{figure}[h,t,b,p]
\centerline{\includegraphics[width=10cm]{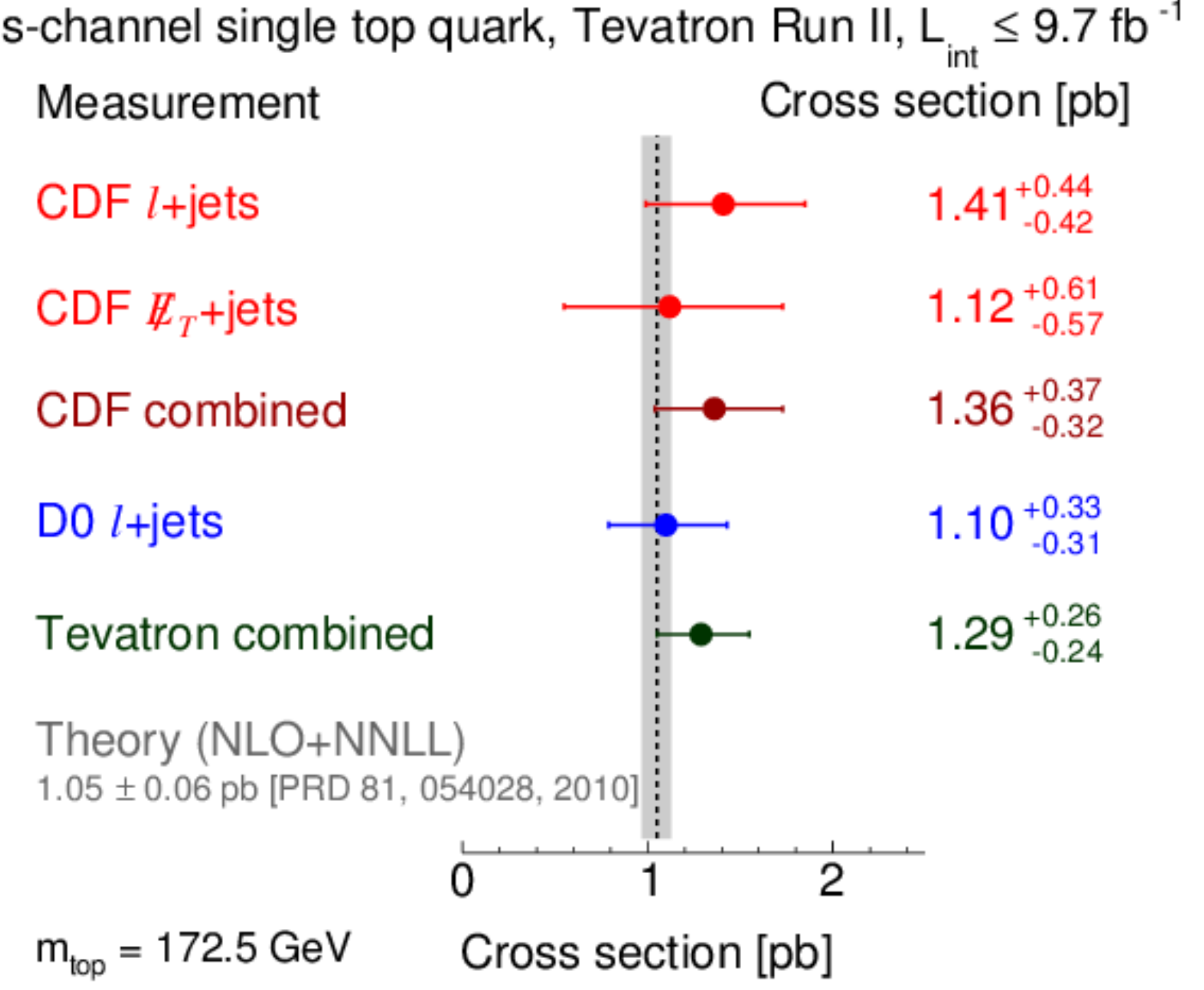}}
\caption{Measured $s$-channel single top quark cross sections from D0 and CDF together with their combinations, taken from~\cite{arXiv:1402.5126}. 
The combined result is in good agreement with SM expectations. 
\label{fig:singletop-sum}}
\end{figure}

The D0 and CDF results were combined~\cite{arXiv:1402.5126} and resulted in the observation of $s$-channel production with a 
significance of  6.3 standard deviations and a measured cross section of $\sigma_s=1.29^{+0.26}_{-0.24}\;\rm pb$. Figure~\ref{fig:tb-discovery} shows the output 
discriminant for the combined result. The combination of the $t$-channel measurements is forthcoming but was not available at the time of submission of this review.
\begin{figure}[h,t,b,p]
\centerline{\includegraphics[width=10cm]{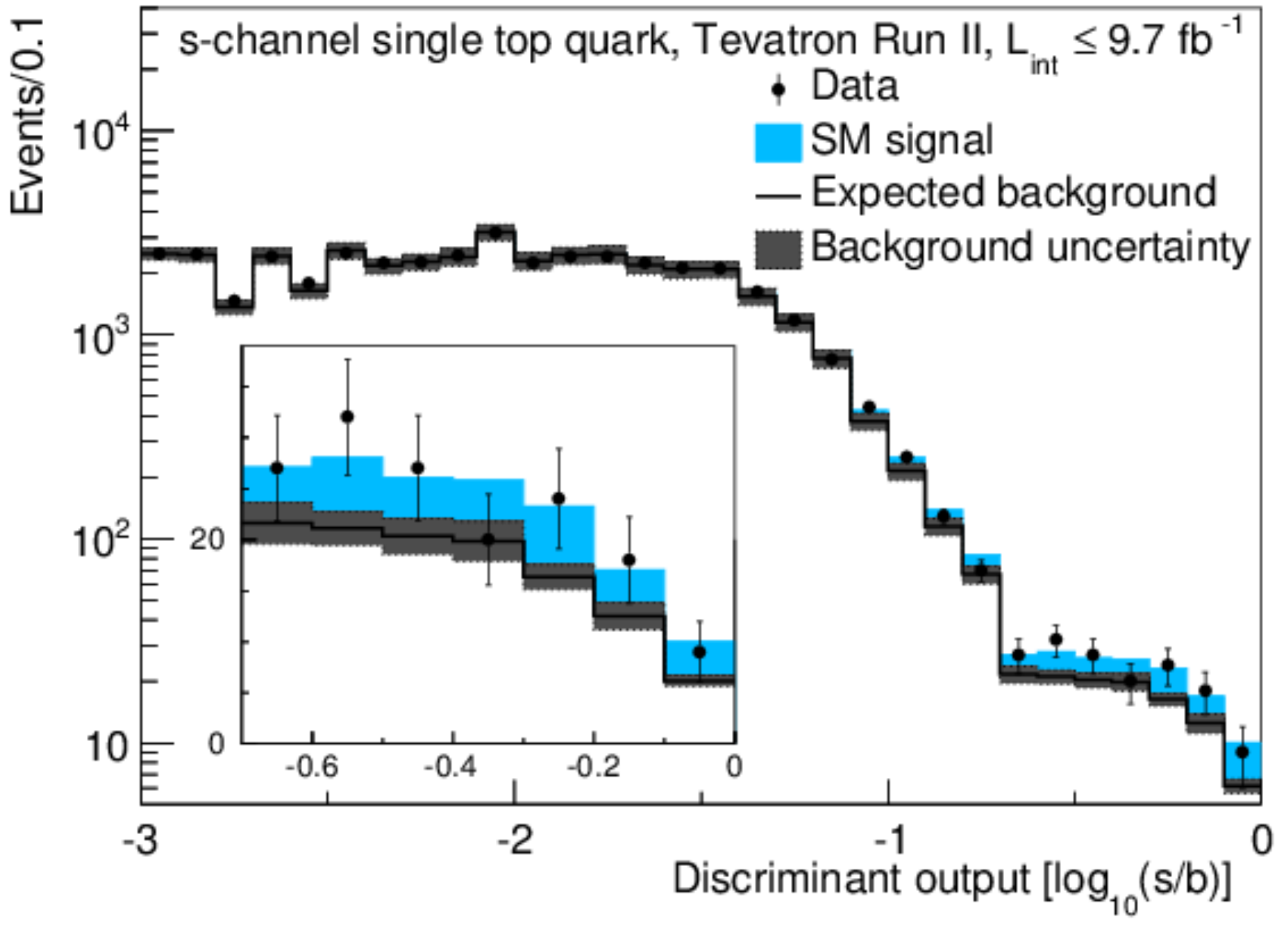}}
\caption{Distribution of the output discriminant, 
summed for bins with similar signal-to-background
ratio (s/b). The expected sum of the backgrounds is shown
by the unfilled histogram, and the uncertainty of the background
is represented by the grey shaded band. The expected
s-channel signal contribution is shown by a filled blue histogram.
\label{fig:tb-discovery}}
\end{figure}

The SM predicts that the top quark decays almost exclusively to a 
$W$ boson and a bottom quark with $B(t \to Wb) \approx 1$. The rate for the process 
leads to a firm prediction for the top quark decay width $\Gamma_{t}$. 
If there are only three generations, the unitarity 
constraint of the CKM matrix implies that $|V_{tb}|$ is very close to unity. But  
the presence of a heavy fourth generation quark with a large CKM coupling to the 
top quark could be consistent with large values of $B(t \to Wb)$, while resulting in 
an almost entirely unconstrained value for $|V_{tb}|$. A direct measurement of 
$|V_{tb}|$, which is possible through the measurement of the single top production cross section, 
can therefore explore the possibility of a fourth generation or an additional heavy quark singlet that mixes with the top quark.
The most stringent limit on the value of $|V_{tb}|$ comes from the full Run II dataset analysis by the D0 collaboration~\cite{PLB-726-656-2013} that used 
data corresponding to $9.7\;\rm fb^{-1}$ of integrated luminosity. 
Using the lepton+jets channel, the $s$+$t$ cross section $\sigma_{s+t}=4.11^{+0.60}_{-0.49}\;\rm pb$ translates in a lower limit on 
$|V_{tb}|>0.92$ at the 95\% C.L., for a top mass of $172.5\;\rm GeV$. CDF used $7.5\;\rm fb^{-1}$ and the same channel and obtained 
$\sigma_{s+t}=3.04^{+0.57}_{-0.53}\;\rm pb$ which translates into $|V_{tb}|>0.78$ at the 95\% C.L.~\cite{ arXiv:1407.4031}, for a top mass of $172.5\;\rm GeV$. 


\subsection{Searches for Anomalous Single Top Quark Production}
The single top quark production is also expected to be sensitive to several models of physics
beyond the SM~\cite{PRD63-014018-2001}, in particular, the $t$-channel is most sensitive to those in which flavor changing
neutral current (FCNC) couplings between a gluon, a top quark,
and up or charm quarks may be large, and the $s$-channel is most sensitive to additional particles that contribute to the single top production, such as a $W^{\prime}$ vector boson that couples to top and bottom quarks. Both collaborations searched for these 
processes~\cite{PRL102-151801-2008, PLB693-81-2010, PRL103-041801-2009,PLB699-145-2011} and find no indication of BSM processes. 
Using $7.7\;\rm fb^{-1}$ of data, the CDF collaboration also searched for a dark-matter candidate produced in association with a single top quark that decays hadronically~\cite{PRL108-201802-2012}. The final state considered consists of three jets and $\met$. Figure~\ref{fig:anomaly} shows the $\met$ distribution for data, which is in agreement with the prediction from the SM and is used to set limits of $\approx 0.5\;\rm pb$ for a dark-matter particle with mass below $150\;\rm GeV$. 

Using $9.7\;\rm fb^{-1}$ of data, the D0 collaboration constructed a two-dimensional 
(2D) posterior probability density as a function of the $s$ and $t$-channel production cross sections~\cite{PLB-726-656-2013}, with the position of the maximum defining the value of the cross sections, and the width of
the distribution in the minimal region that encompasses
68\% of the entire area defining the uncertainty (statistical 
and systematic components combined). This 2D posterior is shown in Fig.~\ref{fig:anomaly}, which also shows the sensitivity to some models of beyond the SM physics that would change the $s$- or $t$-channel cross sections. 
 \begin{figure}[h,t,b,p]
\centerline{\includegraphics[width=7cm]{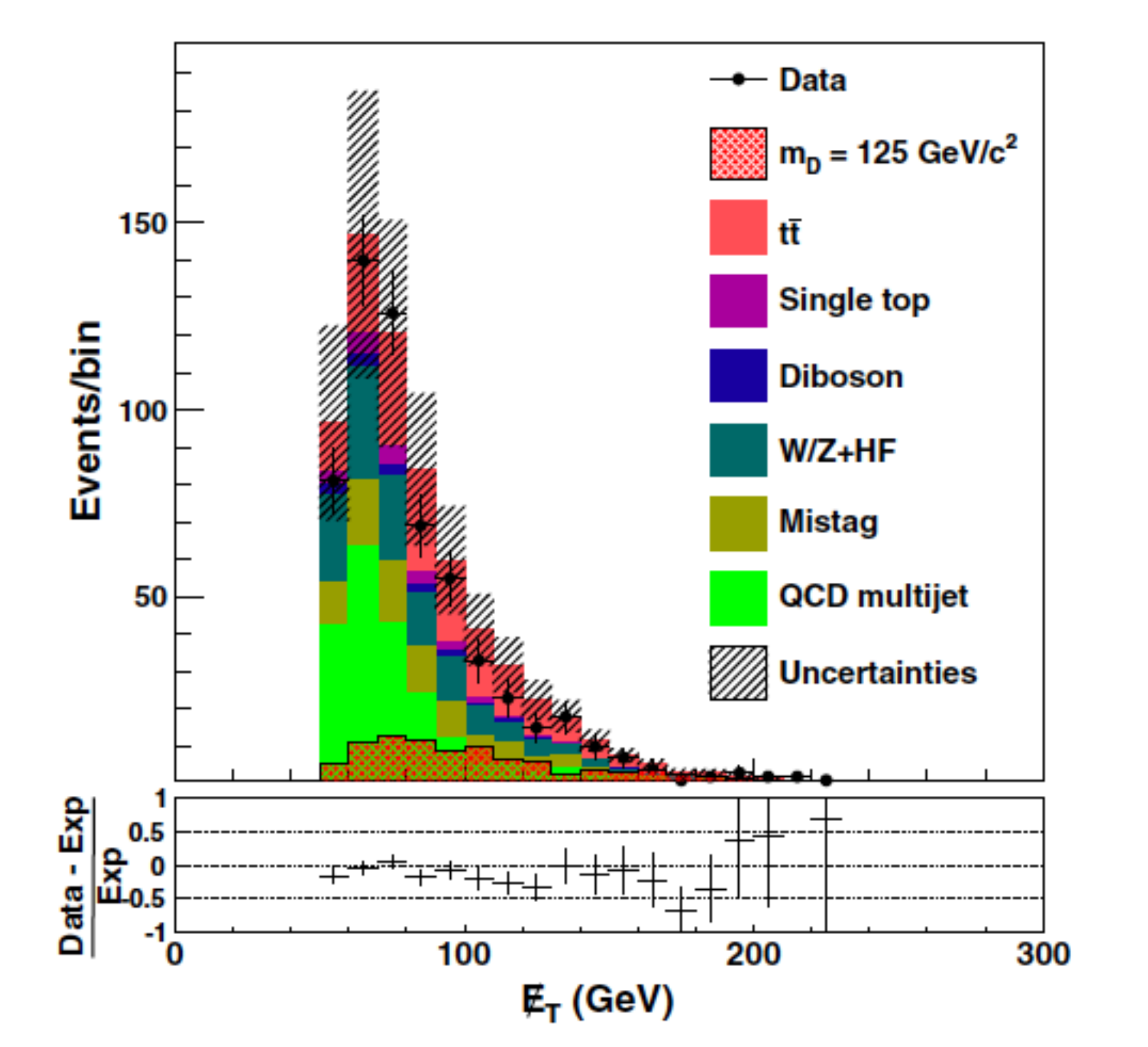}\includegraphics[width=7cm]{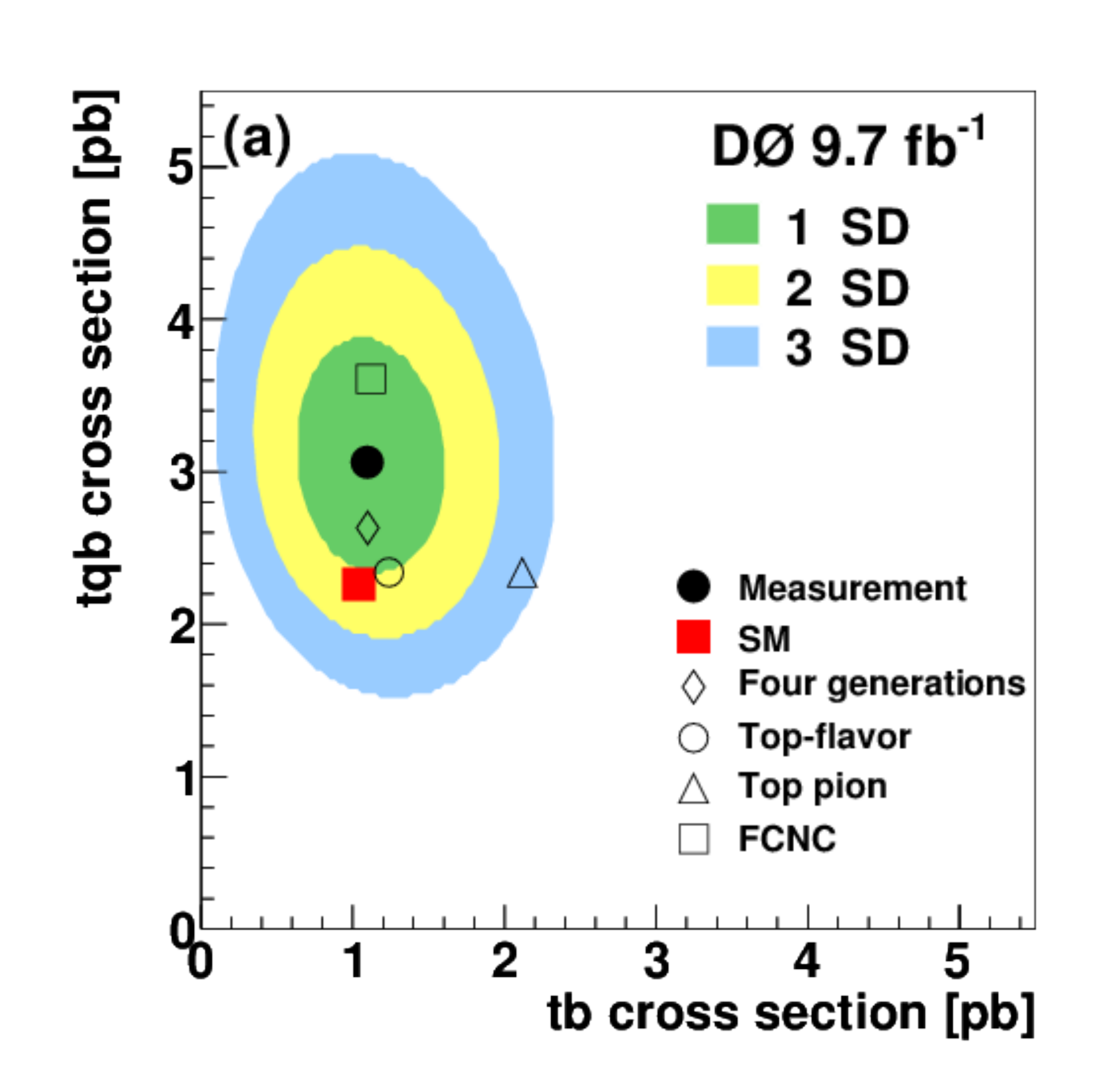}}
\caption{Left (CDF): The $\met$ distribution in the signal region. 
The data are compared to the sum of the SM contributions. A dark matter signal of mass $125\;\rm GeV$ is also shown.
Right (D0):2D posterior density distribution with SM prediction and predictions from various beyond the SM theories that would modify the $s$ (tb) or $t$ (tqb) cross section.
\label{fig:anomaly}}
\end{figure}

\section{Top Quark Mass}
\subsection{Introduction}
One of the fundamental properties of an elementary particle is its mass. In the SM, fermions acquire mass through interactions with the Higgs field~\cite{higgs}, but actual values of these masses are not predicted by the SM. In theoretical calculations, the mass of a particle can be defined in more than one way, and it depends on how higher-order terms in perturbative QCD calculations are renormalized. In the modified minimal subtraction scheme (${\rm\overline{MS}}$), for example, the mass definition reflects short-distance effects, whereas in the pole-mass scheme the mass definition reflects long-distance effects~\cite{msbar}. The concept of the pole mass is not well defined since color confinement does not provide S-matrix poles at $m=m_t$~\cite{pole}. Direct mass measurements rely on Monte Carlo (MC) generators to extract $m_t$. Hence the measured mass corresponds to the mass parameter in the MC. Work is proceeding to address the exact difference between the measured mass and the pole mass, as presented, for example, in Appendix C of Ref.~\cite{scheme}. One alternative way to address this problem is to extract $m_t$ from a measurement of the $t\bar t$ cross section~\cite{mtop-xsec}. The D0 Collaboration has shown that the directly measured mass of the top quark is closer to the pole mass extracted from a measurement of the $t\bar t$ cross section than to an ${\rm\overline{MS}}$ mass extracted in a similar way~\cite{mtop-xsec}. Hence, within the precision of theory and data, the directly measured $m_t$ is best interpreted as the top-quark pole mass.

Before 1995, global fits to electroweak data from the CERN and SLAC $e^+e^-$ colliders (LEP and SLC) and from other experiments produced estimates of $m_t$ that ranged from $\approx 90\;\rm GeV$ to $\approx 190\;\rm GeV$~\cite{mtop-pre95}. At the time of the first observation of the top quark in 1995, the fits indicated a mass close to the current Tevatron value of $m_t$, but with an uncertainty of $\approx\pm$10\% and an assumption of 300 GeV for the mass of the Higgs boson~\cite{mtop-at95}. 
Since then, the CDF and D0 Collaborations have developed many novel measurement techniques and published about 50 journal papers on their measurements of $m_t$~\cite{mtop-review}.

\subsection{Methods for Measuring the Top Mass}
At the Tevatron, the mean transverse momentum ($p_T$ ) of the $t\bar t$ system at parton level is $\approx$20 GeV, which is attributed to initial-state radiation (i.e., gluon emission). The mean transverse momentum of the top quarks at parton level is $\approx$95 GeV~\cite{PLB693-81-2010}. 
Top quarks have a lifetime of $\approx$$0.3\times 10^{-24}$ s~\cite{lifetime-th}, which is an order of magnitude smaller than the time scale for parton evolution and hadronization. Hence, when top quarks decay, they transfer their kinematic characteristics to the $W$ boson and $b$ quark, and the measured energy-momentum four-vectors of the final-state particles can be used to reconstruct the mass of the top quark, except for the presence of initial or final-state radiation.

The four-vector of every jet emerging from quarks can be reconstructed, but neutrinos emitted in semileptonic decays of $b$ quarks and jet energy resolution effects will lead to energy mismeasurements. The momentum of the neutrino from the $W \to l\nu_l$ decay is not detected, but the transverse component can be inferred from the negative of the vector sum of all transverse momenta of particles detected in the calorimeter and muon detectors. We estimate the longitudinal momentum of $\nu_l$ by constraining the mass of the charged lepton and neutrino system to the world average value of $M_W$~\cite{wmass}. We also use $M_W$ to choose the two light jets from $W\to q{\bar q}^\prime$ decay, and we use that information for an {\it in situ} calibration of jet energies. In dilepton events, the analysis is more complicated because there are two final-state neutrinos from the leptonic decays of both $W$ bosons. Therefore, the longitudinal and transverse-momentum components of the neutrinos cannot be determined without the application of more sophisticated tools. These involve assuming a value for $m_t$ to solve the event kinematics and assigning a weight to each $m_t$ hypothesis to determine the most likely value of $m_t$ consistent with the hypothesis that the event is a $t\bar t$ event.

A major issue in $t\bar t$ final-state reconstruction is the correct mapping of the reconstructed objects to the partons from the decays of the top quark and $W$ boson. The problem arises because often the jet charge and flavor cannot be uniquely determined. This creates combinatorial ambiguities in the $t\bar t$ event reconstruction that vary from 90 possible jet-to-parton assignments for the all-jets final state to 2 in the dilepton channel. In the lepton$+$jets and dilepton final states, additional ambiguities may arise from multiple kinematical solutions for the longitudinal component of the neutrino momentum.

Two methods are used to measure the value of $m_t$. In the first method, the reconstructed mass distribution in data or a variable correlated with $m_t$, such as the decay length of the $B$ hadron or the transverse momentum of a lepton, is compared to template distributions composed of contributions from background and simulation of $t\bar t$ events. One template is used to represent background and another for each putative value of $m_t$. The second method uses event probabilities based on the LO matrix element for the production of $t\bar t$. For each event, a probability is calculated as a function of $m_t$ that this event is from $t\bar t$ production, as based on the corresponding production and decay matrix element. Detector resolution is taken into account in the calculation of these probabilities through transfer functions that correlate parton-level energies and their measured values. The value of $m_t$ is then extracted from the joint probability calculated for all selected events, based on the probability for signal and background (also defined through its matrix element). This method produces the most accurate results, but the computations are time consuming.

\begin{figure}[h,t,b,p]
\centerline{\includegraphics[width=4.2cm]{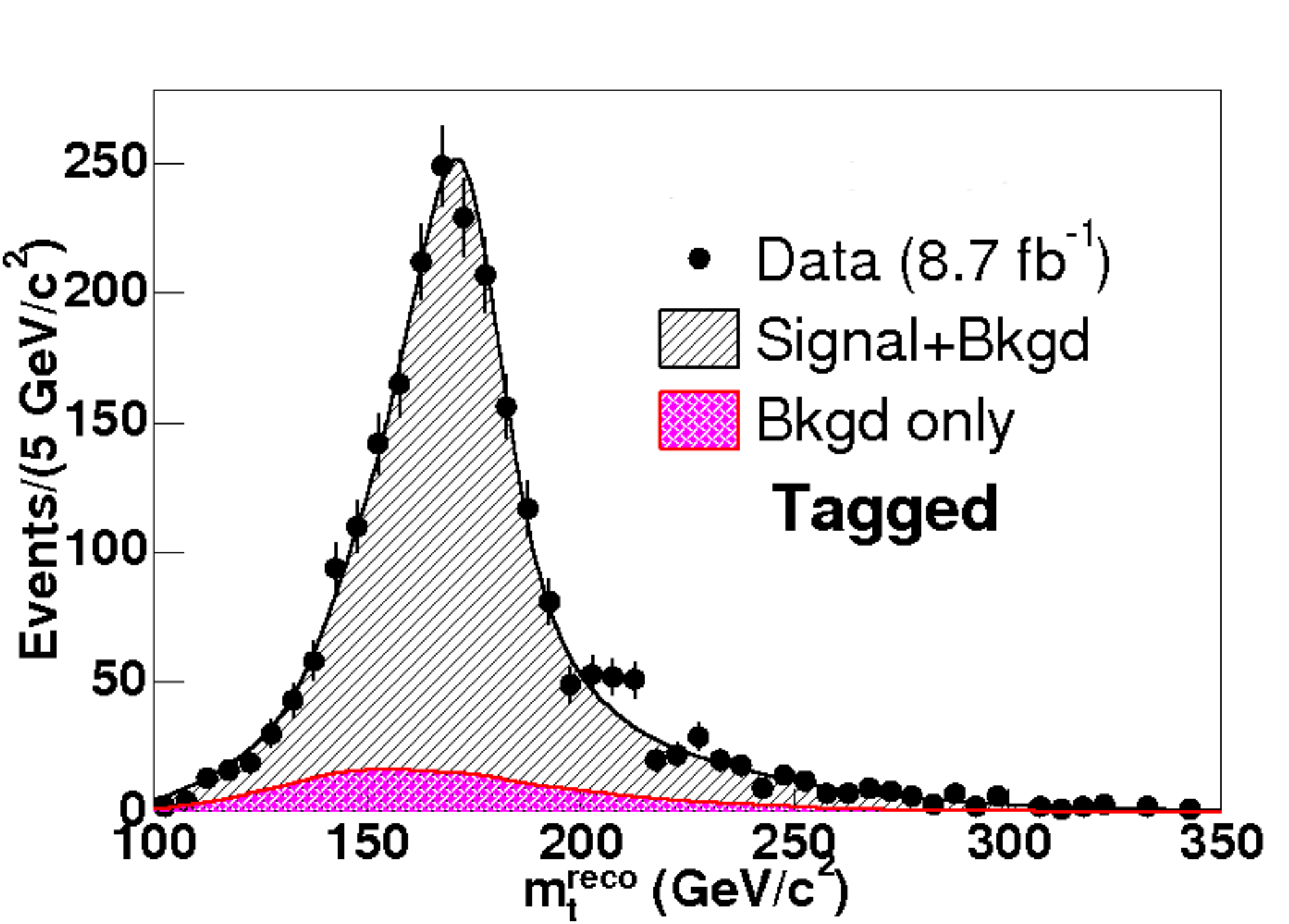}\includegraphics[width=4.2cm]{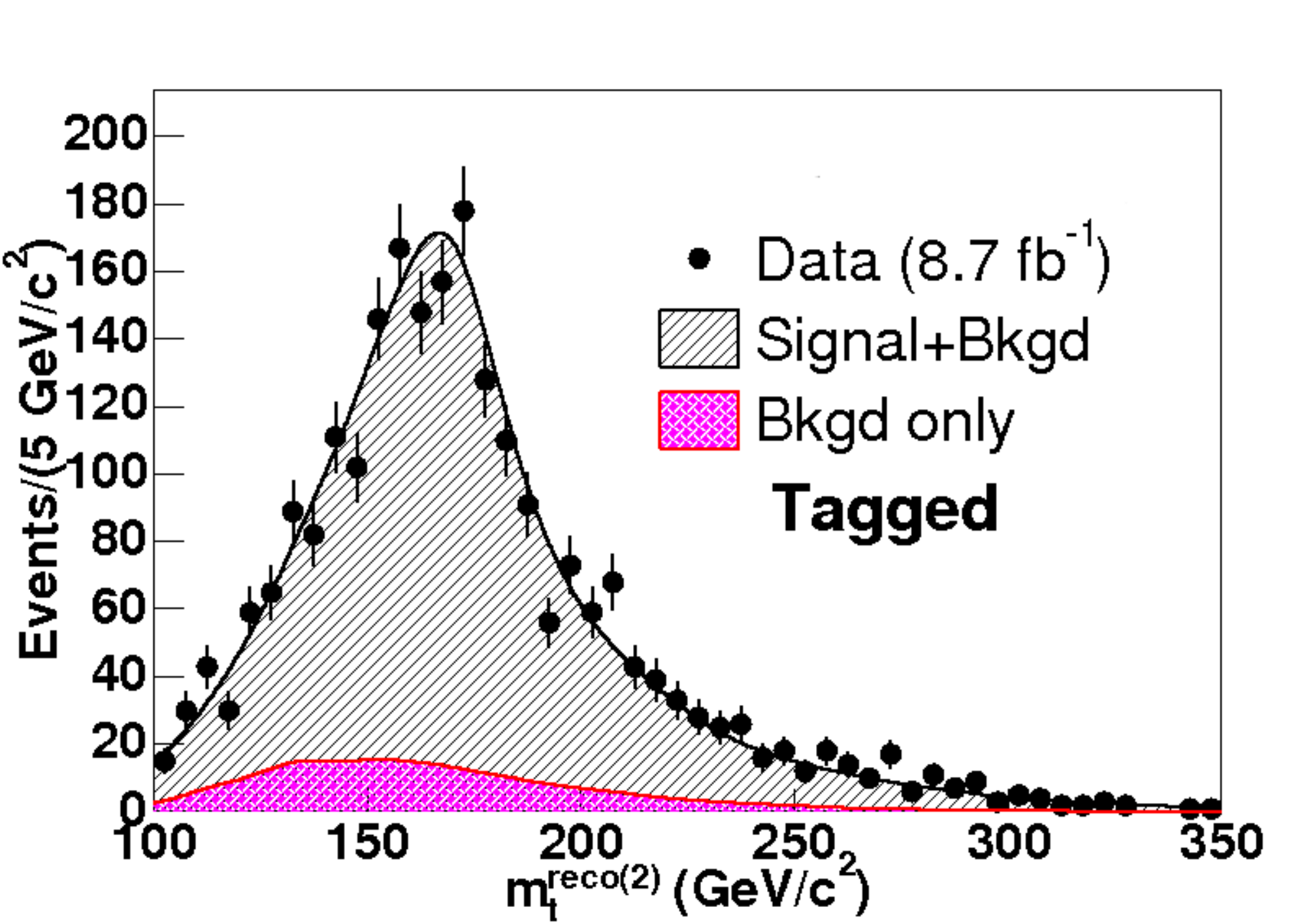}\includegraphics[width=4.2cm]{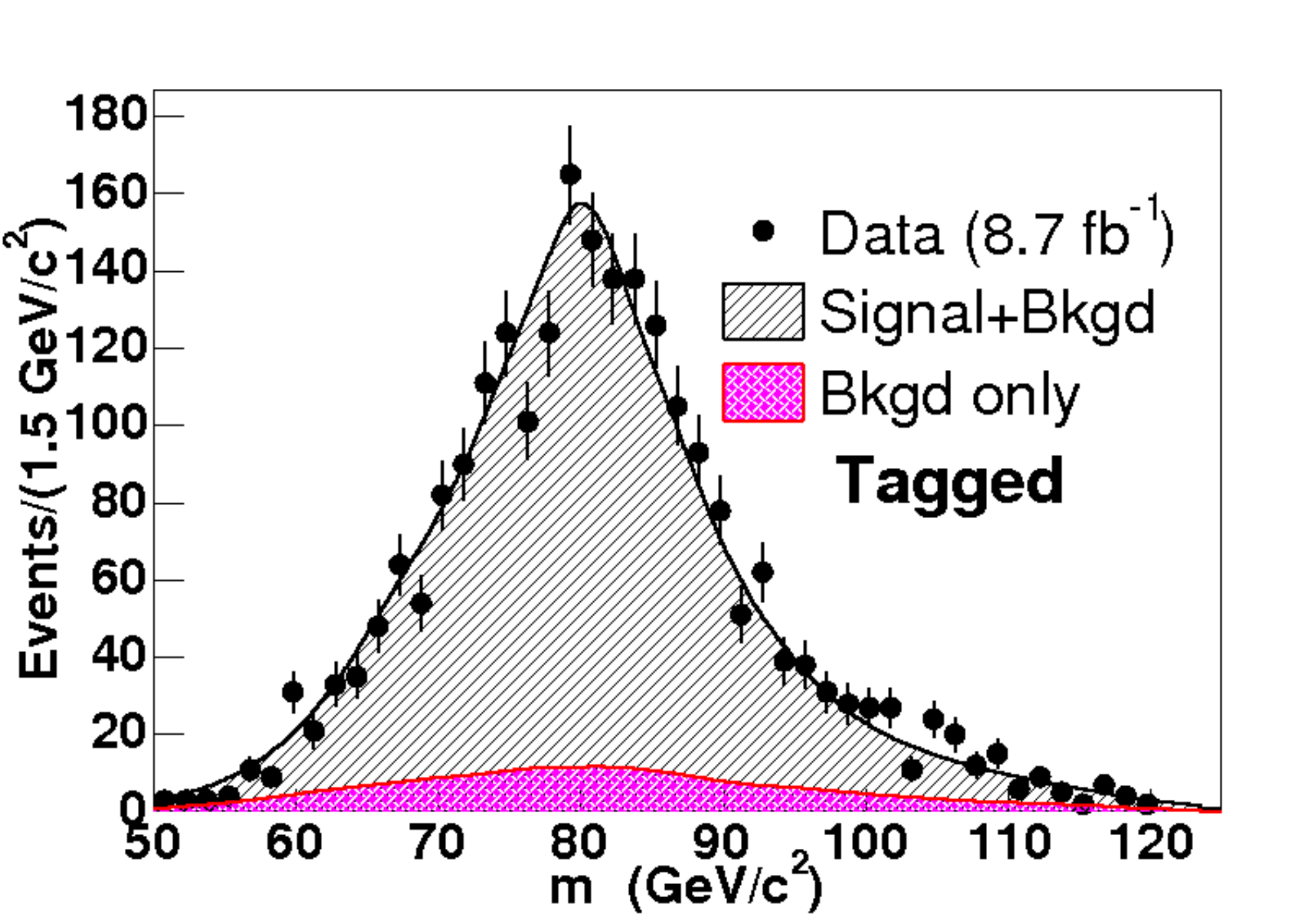}}
\caption{Distributions of the three variables used to measure $m_t$ for the $b$-tagged events in a CDF data sample corresponding to 8.7 fb$^{-1}$ of integrated luminosity. From left to right: Best fit reconstructed $m_t$, next-to-best fit reconstructed $m_t$, and dijet mass from hadronic $W$ decays. The data are overlaid with predictions from the signal model with input $m_t=173$ GeV and the full background model.
\label{fig:cdf-lj-full-lumi}}
\end{figure}

\subsection{Mass Measurement in lepton+jet events}
CDF measured $m_t$ in the lepton$+$jets channel using a template method and the full luminosity of 8.7 fb$^{-1}$~\cite{cdf-lj-full-lumi}. In this analysis, for each selected $t\bar t$ candidate event, a $\chi^2$ minimization is performed to reconstruct $m_t$. The fit is based on the hypothesis that the event is $t\bar t$: it uses $W$-mass constraints on the hadronic and leptonic side and requires the two top-quark masses in the event to be equal. Only the four leading jets are assigned to the four quark daughters from the $t\bar t$ decay. The jet-parton assignment that yields the lowest $\chi^2$ after minimization is used for the analysis and the corresponding $m_t$ value is used for the template construction. To include additional information corresponding to the true top-quark mass, the second best reconstructed $m_t$ with different jet-parton assignment from the best $\chi^2$ fit is also used to construct a separate template. The jet energy scale (JES) is constrained by constructing a dijet mass $m_{jj}$ template from the $W$ boson decaying hadronically. No fit is used to obtain $m_{jj}$, although events failing the $\chi^2$ cut in the $m_t$ reconstruction are also not used to constrain JES. In events with two $b$-tags, there is only one $m_{jj}$ construction among the leading four jets consistent with $b$-tagging (i.e., not tagged). In events with zero and one $b$-tag, there are 12 and 3 $m_{jj}$ constructions, respectively, consistent with $b$-tagging. The single $m_{jj}$ construction closest to the well-known $W$ boson mass is taken as the single $m_{jj}$ value per event. The three templates, two for $m_t$ and one for $m_{jj}$, thus constructed from signal ($t\bar t$ MC) and background samples are used to define a likelihood for various values of MC input $m_t$ and JES, which is then maximized on the data events. Figure~\ref{fig:cdf-lj-full-lumi} shows the three templates constructed with MC input $m_t=173$ GeV for events with one or two $b$-tags, together with the total background template, overlaid with the data. The result of $m_t=172.85\pm 0.71({\rm stat})\pm 0.85({\rm syst})=172.85\pm 1.11$ GeV from the likelihood maximization is the single most-precise top-quark mass measurement from CDF.

A complementary measurement from CDF to the one described above used only $\met$ originating from the undetected neutrino to identify the leptonically decaying $W$ boson~\cite{cdf-vj-full-lumi}. In this $\met$ $+$jets topology, events with charged leptons are vetoed in order to select a sample orthogonal to the lepton$+$jets sample used in the previous measurement. The technique is identical to the one with the lepton$+$jets topology, using the same three variables for the top-quark mass and the {\it in situ} JES calibration in a maximum likelihood fit, but the resolution is worse because the $t\bar t$ event reconstruction is missing the kinematic information carried by the lepton. As a result, this measurement is less accurate than the previous one. Using again the full luminosity of 8.7 fb$^{-1}$, the result is $m_t=173.93\pm 1.64({\rm stat})\pm 0.87({\rm syst})=173.93\pm 1.85$ GeV.

\begin{figure}[h,t,b,p]
\centerline{\includegraphics[width=6.5cm]{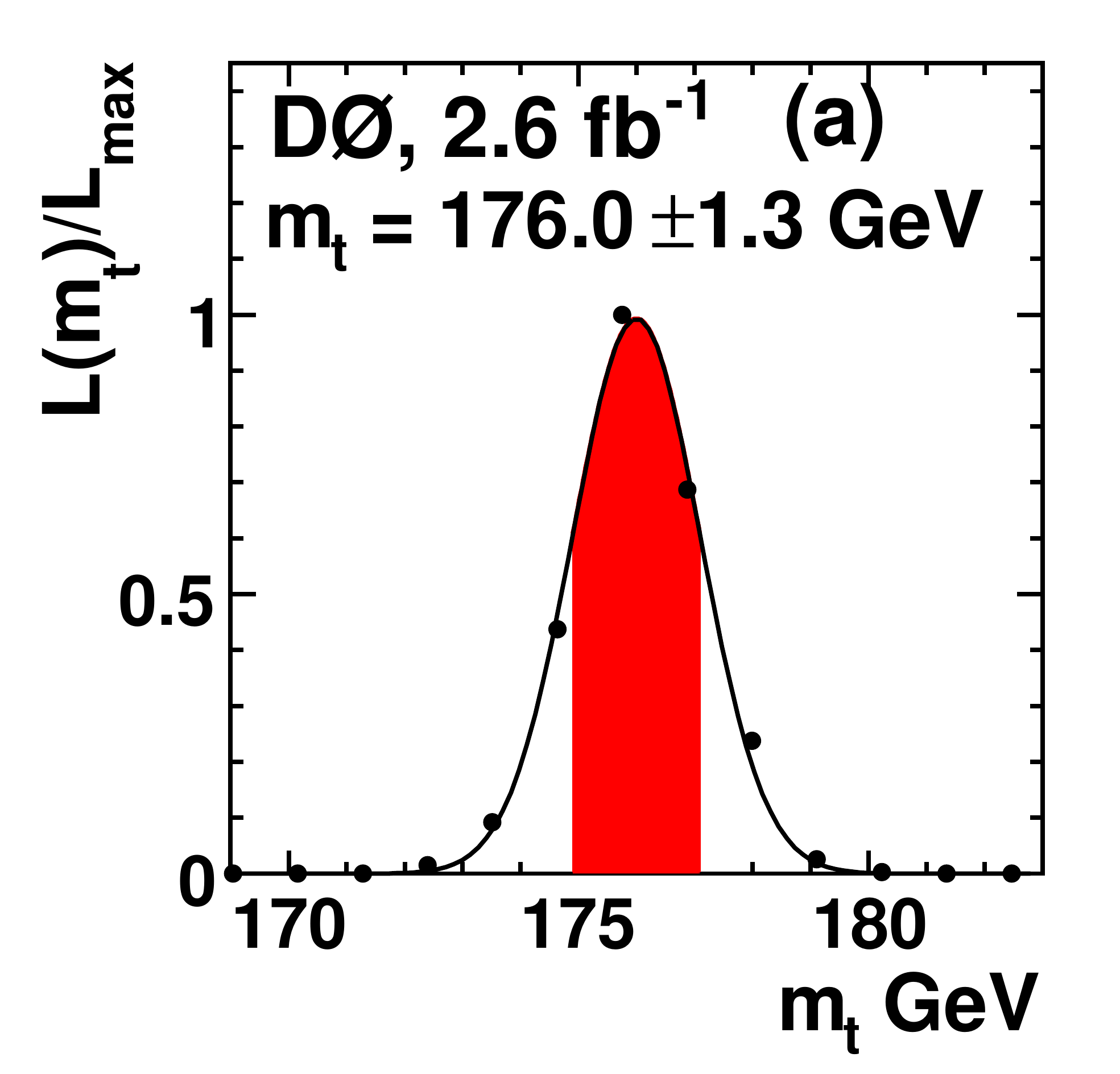}\includegraphics[width=6.5cm]{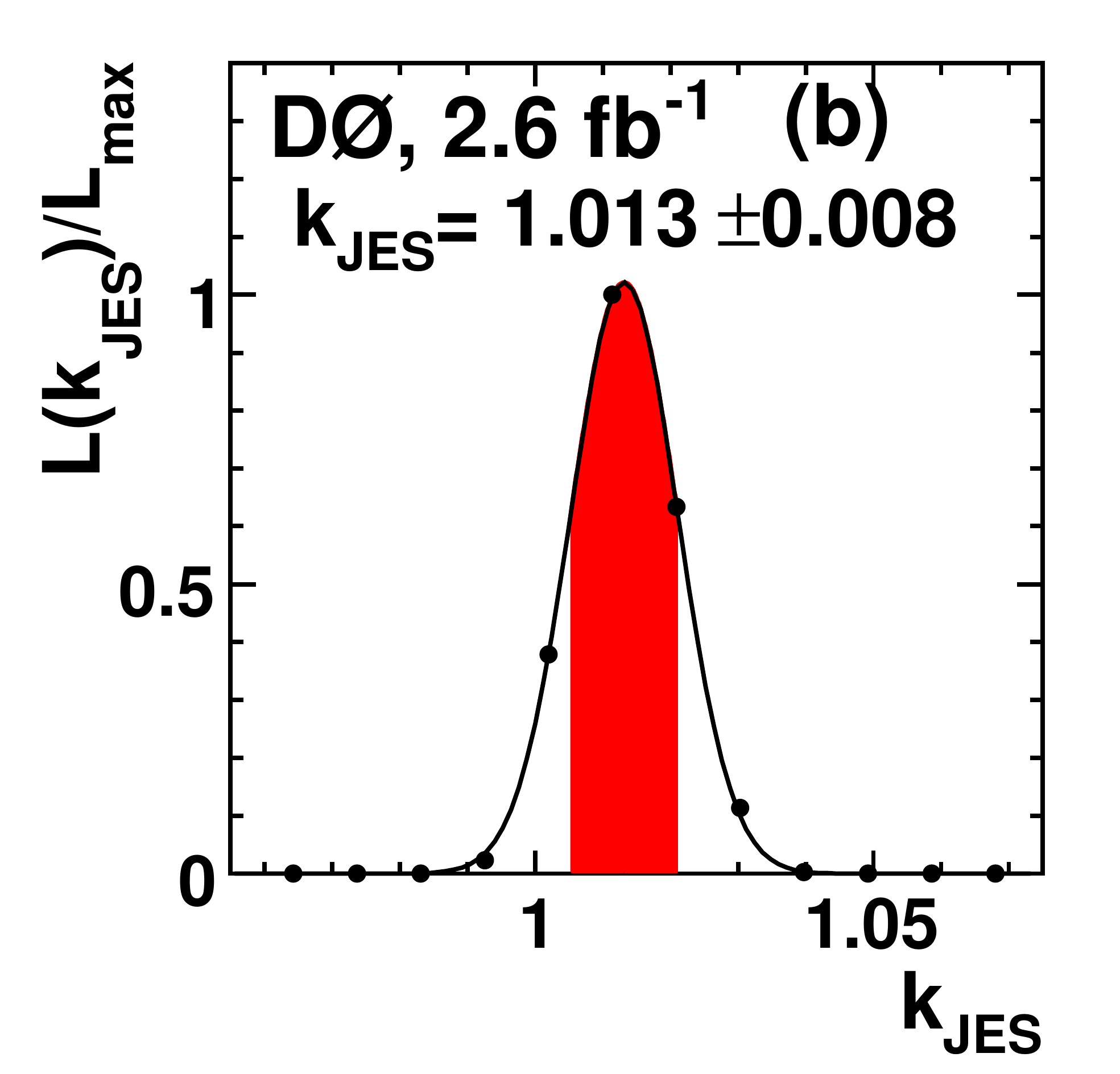}}
\caption{Projections of the data likelihood onto the (a)  $m_t$ and (b) JES axes for lepton$+$jets events selected from a D0 data sample corresponding to 2.6 fb$^{-1}$ of integrated luminosity. The shaded areas indicate the 68\% C.L. regions of the fit.
\label{fig:d0-lj}}
\end{figure}

D0  measured $m_t$ in the lepton$+$jets channel using a matrix element method~\cite{d0-lj}. For each $t\bar t$ candidate event, an event probability is calculated assuming that it is either a signal event or a background event with a matrix element of the dominant background process in the channel, which is $W$$+$4 jets production. The transfer functions, accounting for hadronization smearing and detector resolution effects, are approximated by gaussian functions with mean values equal to the measured kinematic quantities of the event and standard deviations equal to the estimated average resolutions. To reduce the dimensions of 4-momentum integrations of the matrix element and transfer functions over the phase space of the observed final-state particles in the event probability calculation, the angles of all particles are assumed to be measured with infinite precision and overall 4-momentum conservation is enforced, thus reducing the integration dimensions from 24 to 10. The event probability is summed over the 24 possible jet permutations, each weighted by a weight constructed from the average tagging efficiency for heavy quarks, light quarks, and gluons, and over all possible solutions (up to eight) for the neutrino kinematics. The event probabilities of all events in the sample are used to define a likelihood for the sample. An {\it in situ} calibration is performed by applying the $W$-mass constraint on the light-flavored jets consistent with $b$-tagging and allowing both $m_t$ and JES to vary simultaneously in the maximum likelihood fit to the data. Two data samples are used separately with the same method, one corresponding to the first 1 fb$^{-1}$ and the other corresponding to the next 2.6 fb$^{-1}$ of luminosity. The results from the fits of the two samples are combined to yield $m_t=174.94\pm 0.83({\rm stat})\pm 0.78({\rm JES})\pm 0.96({\rm syst})=174.94\pm 1.49$ GeV. Figure~\ref{fig:d0-lj} shows projections of the data likelihood fit onto the $m_t$ and JES axes with 68\% confideince level (C.L.) regions indicated by the shaded areas.

Very recently, D0 updated the measurement in the lepton$+$jets channel using the same matrix element technique, but with data corresponding to the full luminosity of 9.7 fb$^{-1}$~\cite{d0-lj-new}. Other changes include the use of an improved detector calibration, in particular the $b$-jet energy scale corrections, and a new 
integration method that reduced the integration speed by two orders of magnitude, significantly augmenting the number of MC events available for the analysis.  
These changes allowed to reduce substantially systematic uncertainties previously dominated either by limited MC statistics or by poor detector resolution. Figure~\ref{fig:d0-mtop-ljets} shows a comparison of the SM prediction to data for the preferred values of 
the top quark mass and the energy calibration for the reconstructed $W$ boson and invariant mass of the \ttbar system.
The new result of $m_t=174.98\pm 0.58({\rm stat+JES})\pm 0.49({\rm syst})=174.98\pm 0.76$ GeV is the single most-precise top-quark mass measurement from the Tevatron, comparable in precision to the latest measurements from the LHC. 
\begin{figure}[h,t,b,p]
\centerline{\includegraphics[width=6.5cm]{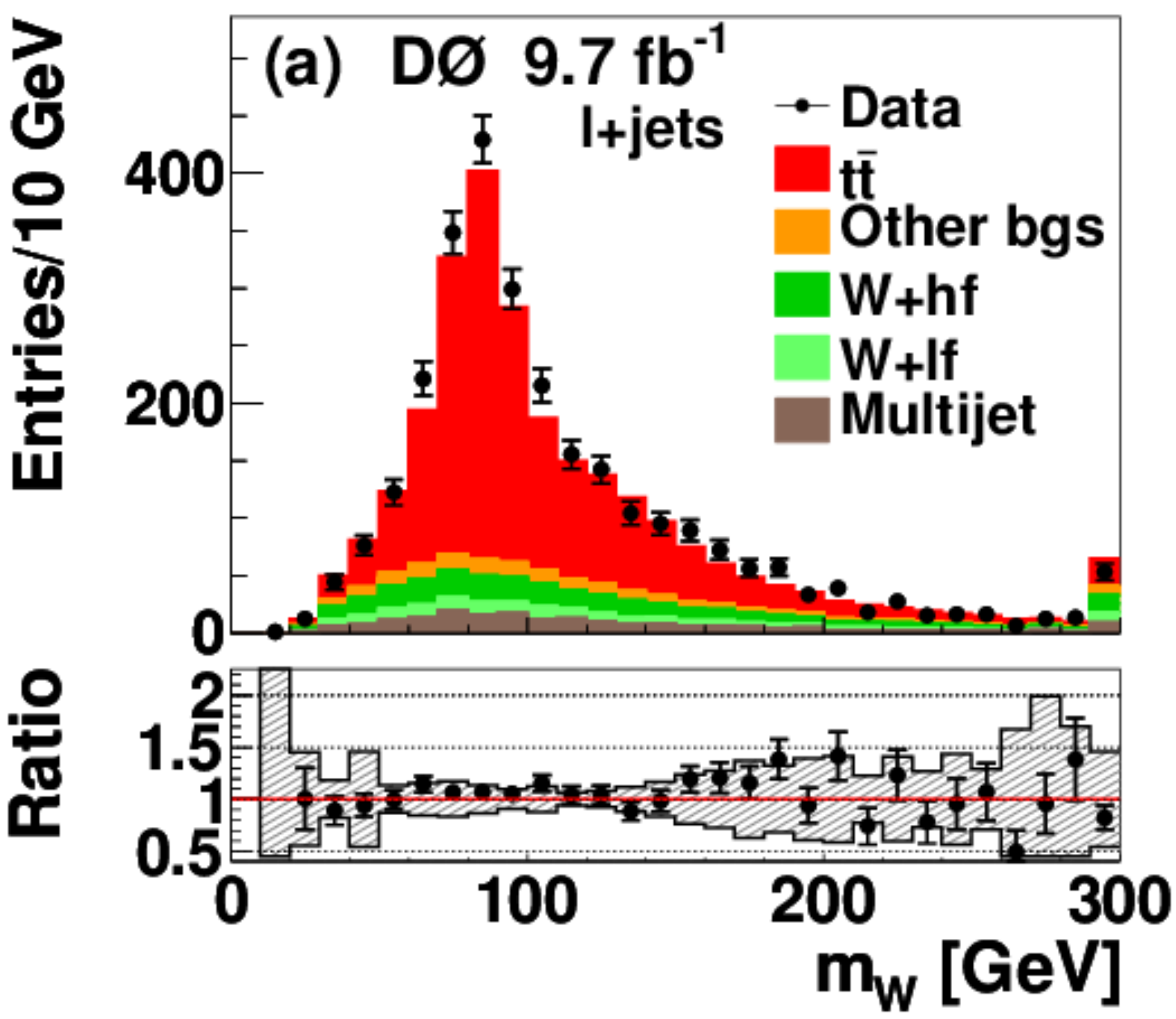}\includegraphics[width=6.5cm]{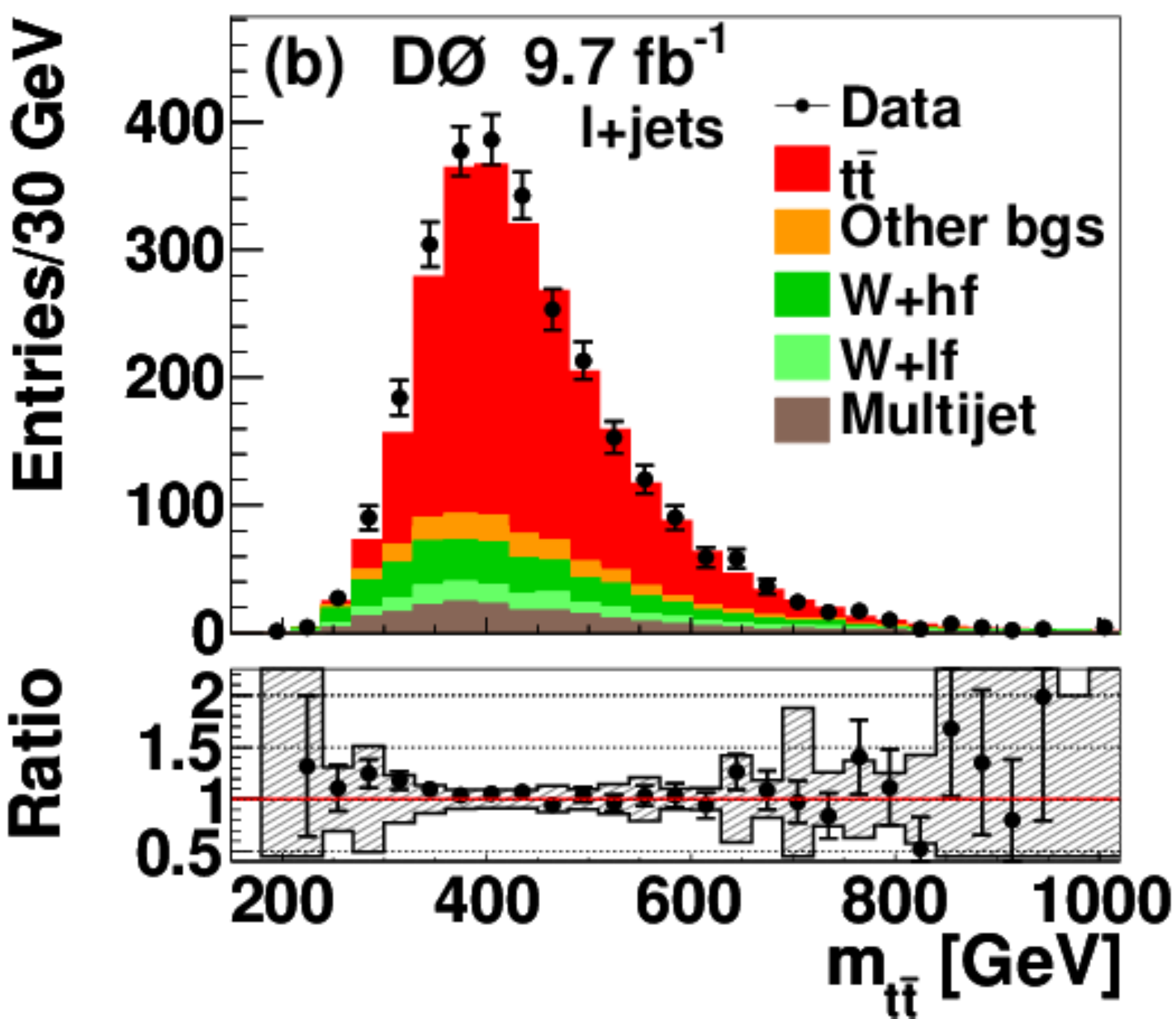}}
\caption{Data/MC comparison for the D0 lepton+jets data corresponding to 9.7 fb$^{-1}$ of integrated luminosity. (a) Invariant mass of the jet pair
matched to one of the $W$ bosons. (b) Invariant mass of the
\ttbar system. In the ratio of data to SM prediction, the total
systematic uncertainty is shown as a shaded band.
\label{fig:d0-mtop-ljets}}
\end{figure}

CDF developed methods to measure $m_t$ without using information from the hadronic calorimeter, i.e. jet energies, which is always the major source of systematic uncertainty. In the first such analysis, using lepton$+$jets events selected from data corresponding to 1.9 fb$^{-1}$ of luminosity, two measurements were performed and combined together~\cite{cdf-mtop-tracking}. One used the mean transverse decay length from the primary to the secondary vertex of $b$-tagged jets, which is sensitive to the boost of the $b$-quarks from $m_t$ and thus can be calibrated with MC against the input top-quark mass to determine $m_t$ from the data. This measurement yielded $m_t=166.9^{+9.5}_{-8.5}({\rm stat})\pm 2.9({\rm syst})$ GeV. The other used in the same way the mean transverse momentum ($p_T$) of the leptons (electrons or muons) from the $W$ boson decays, which is also sensitive to the boost of the $b$-quarks from $m_t$, and yielded $m_t=173.5^{+8.8}_{-8.9}({\rm stat})\pm 3.8({\rm syst})$ GeV. The combination of the two measurements resulted in a top-quark mass of $170.7\pm 6.3({\rm stat})\pm 2.6({\rm syst})=170.7\pm 6.8$ GeV. In the second analysis, the shape of the full lepton $p_T$ distribution was used in the context of the template method~\cite{cdf-mtop-leptons}. Figure~\ref{fig:cdf-mtop-leptons} shows the maximum likelihood fit of the signal and background templates to the lepton $p_T$ distribution of the data. The measurement used lepton$+$jets events selected from a sample corresponding to 2.7 fb$^{-1}$ of luminosity. A dedicated calibration of the lepton $p_T$ was performed, using data from leptonic $W$ and $Z$ boson decays, in order to minimize uncertainties related with absolute calibration of the tracker (for muons) and the electromagnetic calorimeter (for electrons), which is the dominant source of systematic uncertainty in this case. The result of this measurement is $176.9\pm 8.0({\rm stat})\pm 2.7({\rm syst})=170.7\pm 8.4$ GeV.
\begin{figure}[h,t,b,p]
\centerline{\includegraphics[width=12cm]{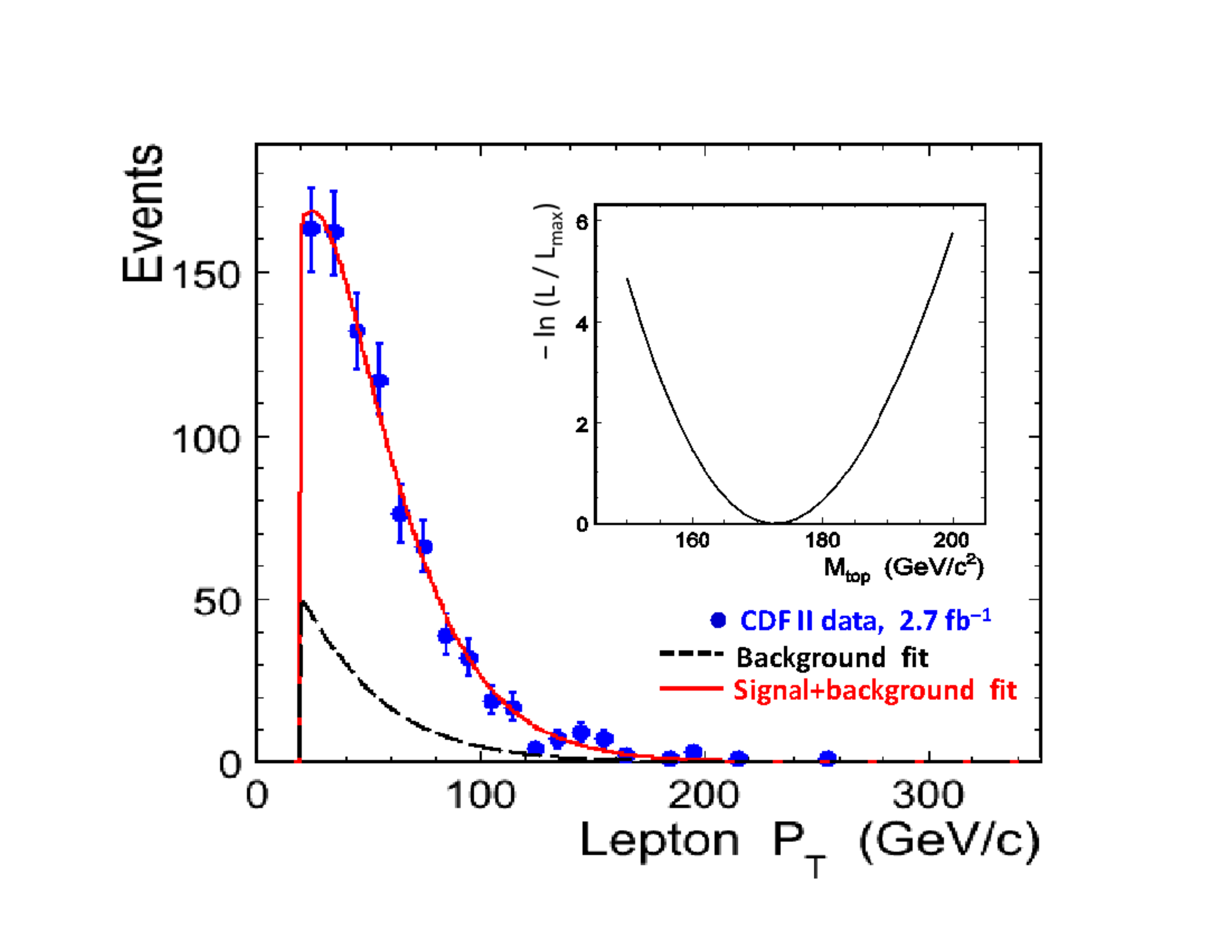}}
\caption{Fit of the lepton transverse momentum templates to the data in the lepton$+$jets channel. The inset shows the $m_t$-dependent negative log-likelihood function of the fit.
\label{fig:cdf-mtop-leptons}}
\end{figure}

\subsection{Mass Measurement in all-jet events}
CDF measured $m_t$ in the all-jets channel using 5.8 fb$^{-1}$ of luminosity~\cite{cdf-allj}. Events with six to eight jets are selected by a neural network algorithm and by the requirement that at least one of the jets is tagged as a $b$-quark jet. As in the lepton$+$jets and $\met$ $+$jets channels, the kinematics of the selected $t\bar t$ candidate events are reconstructed by $\chi^2$ minimization using $W$-mass constraints and assuming the two top-quark masses in the event to be equal. $m_t$ and $M_W$ are reconstructed per event and the corresponding templates are constructed from their distributions. Figure~\ref{fig:cdf-allj} shows the two templates constructed from signal and background corresponding to the measured $m_t$ and JES. The measurement is performed with a maximum likelihood fit, which simultaneously determines $m_t$ and the JES calibration. The result is $m_t=172.5\pm 1.4({\rm stat})\pm 1.0({\rm JES})\pm 1.1({\rm syst})=172.5\pm 2.0$ GeV.
\begin{figure}[h,t,b,p]
\centerline{\includegraphics[width=6.5cm]{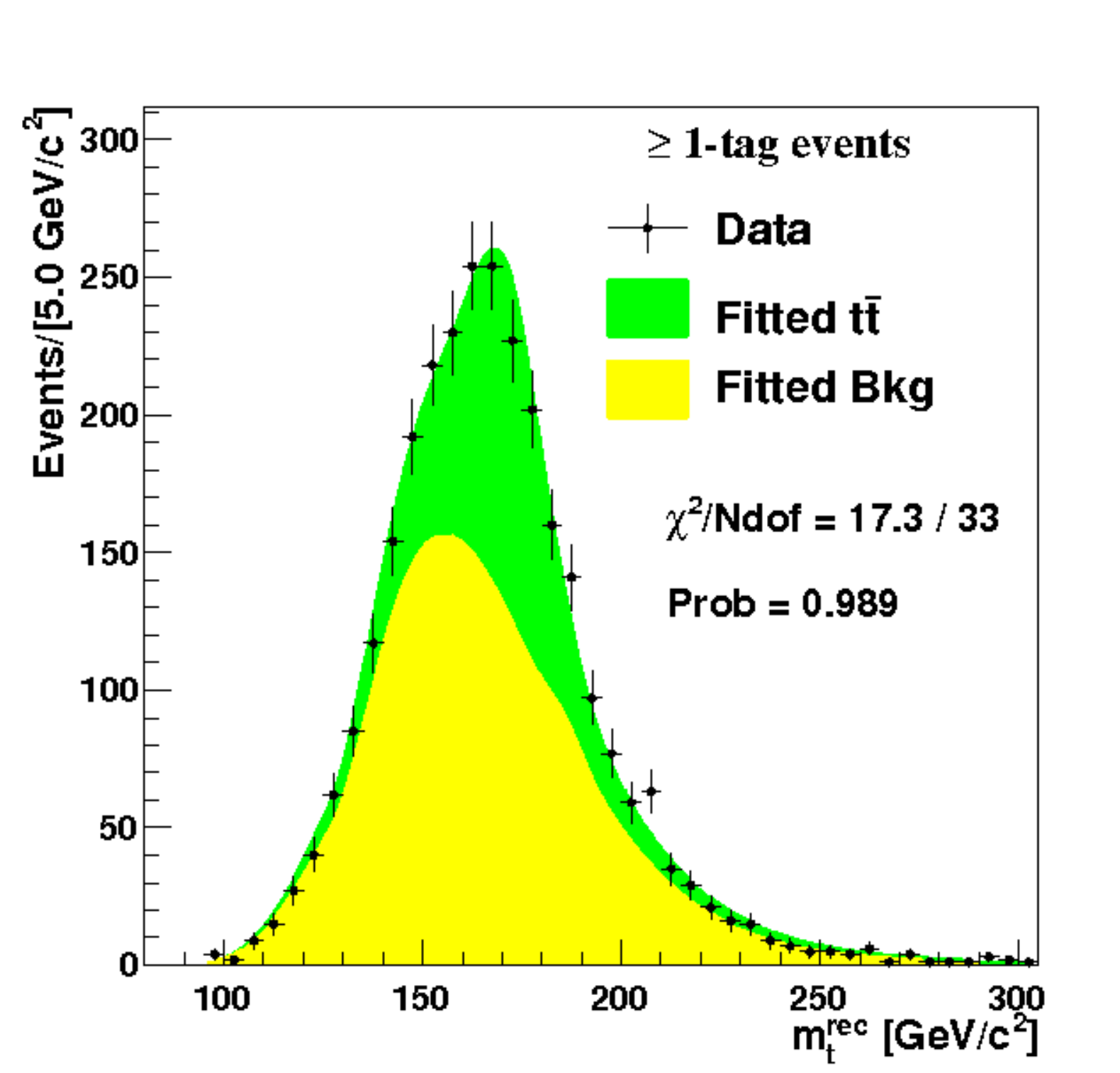}\includegraphics[width=6.5cm]{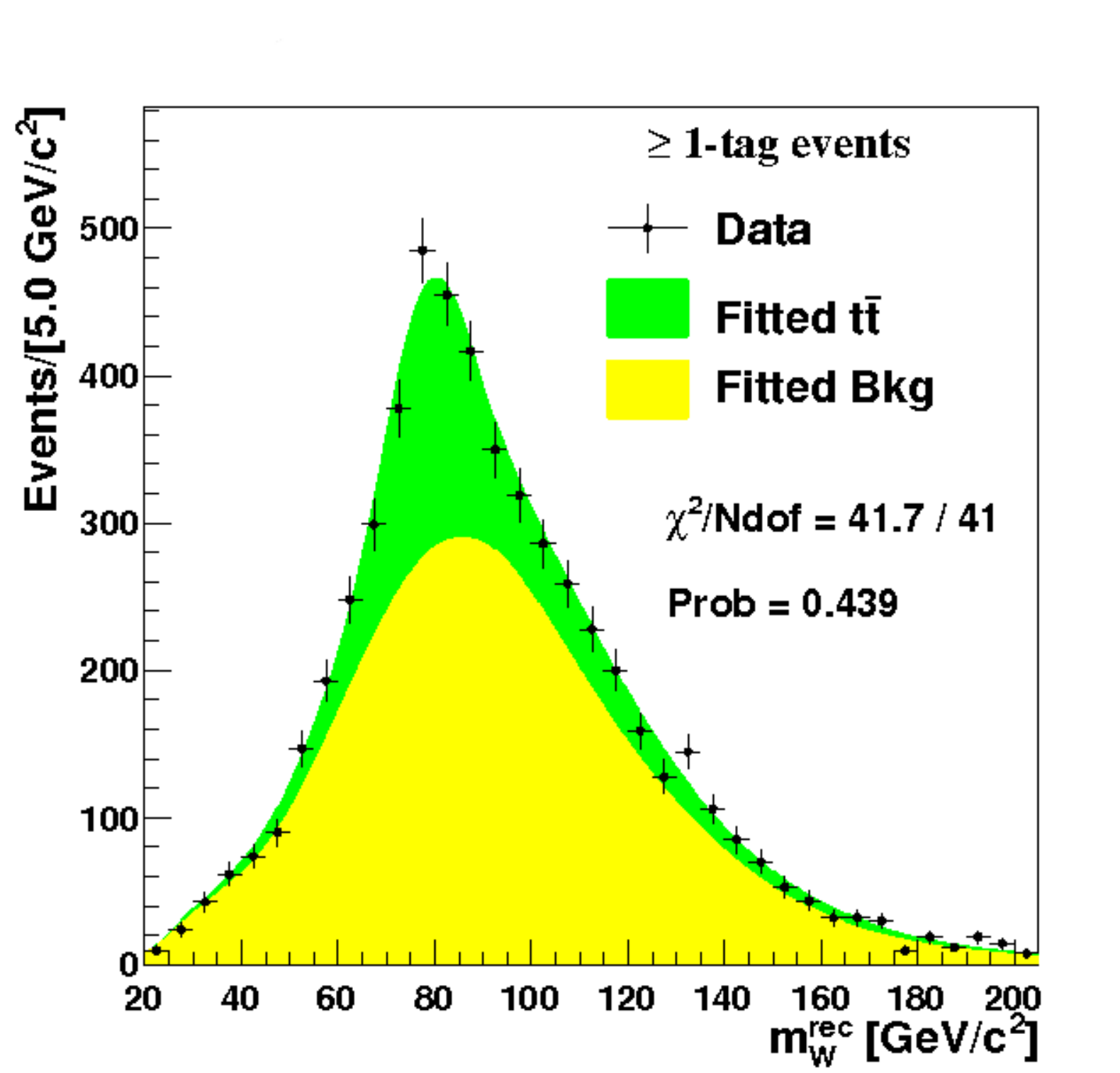}}
\caption{Distributions of $m_t$ (left plot) and $m_W$ (right plot) reconstructed from the all-jets events selected from a CDF data sample corresponding to 5.8 fb$^{-1}$ of integrated luminosity. The data (points) are compared to the signal and background expectations corresponding to the measured $m_t$ and JES. The expected distributions are normalized to the best fit yields.
\label{fig:cdf-allj}}
\end{figure}

\subsection{Mass Measurement in dilepton events}
As discussed previously, the dilepton channel is the most difficult one to measure $m_t$ with good precision because of the low statistics, the poor kinematic reconstruction of $t\bar t$ candidate events, and the lack of possibility to perform {\it in situ} JES calibration. CDF measured $m_t$ from a sample of 328 events with a charged electron or muon and an isolated track, corresponding to luminosity of 2.9 fb$^{-1}$, selected as $t\bar t$ candidates~\cite{cdf-dil}. In this analysis, to account for the unconstrained event kinematics, the phase space of the azimuthal angles $(\phi_{\nu_1},\phi_{\nu_2})$ of neutrinos is scanned and $m_t$ is reconstructed for each $\phi_{\nu_1},\phi_{\nu_2}$ pair by minimizing a $\chi^2$ function in the $t\bar t$ dilepton hypothesis. $\chi^2$-dependent weights are assigned to the solutions in order to build a preferred mass for each event. Preferred mass distributions (templates) are built from simulated $t\bar t$ and background events, and parametrized in order to provide continuous probability density functions. A maximum likelihood fit to the mass distribution in data as a weighted sum of signal and background probability density functions (p.d.f.) is performed, with the background normalization constrained to be Gaussian-distributed with mean equal to the background expectation and standard deviation equal to the uncertainty of the expectation. The fit is illustrated in Figure~\ref{fig:cdf-dil} and gives a top-quark mass of $165.5^{+3.4}_{-3.3}({\rm stat})\pm 3.1({\rm syst})=165.5^{+4.6}_{-4.5}$ GeV.
\begin{figure}[h,t,b,p]
\centerline{\includegraphics[width=10cm]{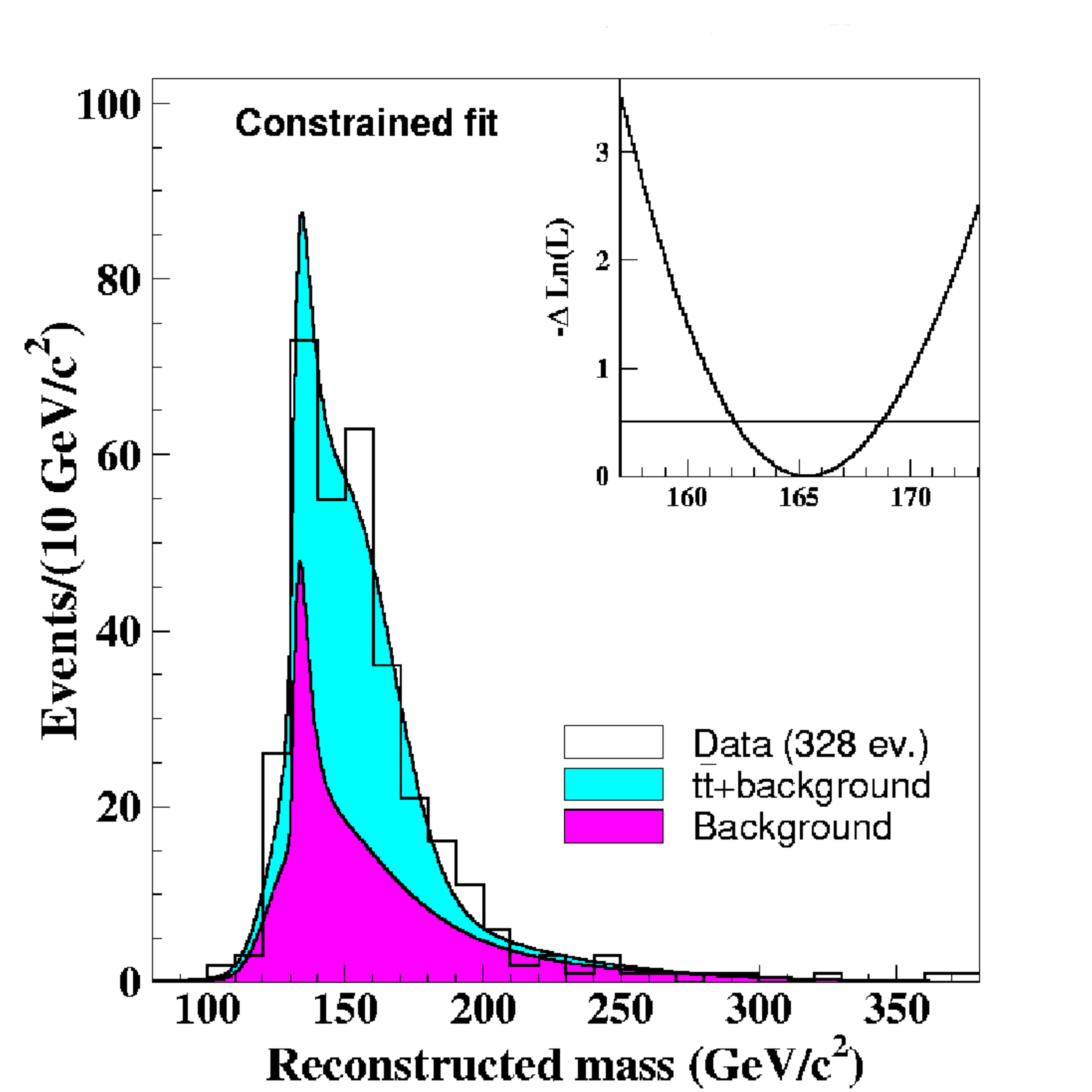}}
\caption{Fit of $m_t$ to the 328 lepton$+$track events selected from a CDF data sample corresponding to 2.9 fb$^{-1}$ of integrated luminosity. Background and signal$+$background probability density functions, normalized according to the yields determined by the fit, are superimposed to the $m_t$ distribution reconstructed from the data. The inset shows the $m_t$-dependent negative log-likelihood of the fit.
\label{fig:cdf-dil}}
\end{figure}

D0 measured $m_t$ in the dilepton channel using 5.4 fb$^{-1}$ of luminosity and a matrix element method similar to the one used in the lepton$+$jets analysis described above~\cite{d0-dil}. The background event probability is calculated from a matrix element corresponding to the dominant background process in this channel, which is $Z$$+$2 jets production. In this case, the event probability is summed over two possible jet assignments, up to two real solutions for the neutrino energies, and integrated over 7 to 9 kinematic variables for a number of 0 to 2 muons in the event, respectively, assuming that the electron energies are measured with infinite precision but the muon momenta are not, for kinematics consistent with $t\bar t$ decays. The measurement gives a result of $174.0\pm 1.8({\rm stat})\pm 2.4({\rm syst})=174.0\pm 3.0$ GeV. The result based on a neutrino weighting procedure~\cite{vwt} that uses the assumed $\eta$ for each neutrino through a range of values at each $m_t$  is $173.7\pm 2.8({\rm stat})\pm 1.5({\rm syst})$. When both results are combined~\cite{d0-dil-new}, the final measurement from the dilepton channel is
$m_t=173.9\pm 1.9({\rm stat})\pm 1.6({\rm syst})$.

\subsection{Combination of Results}
To improve the precision of the world average $m_t$, the two Collaborations have established procedures to combine their measurements regularly. Figure~\ref{fig:mass-combo} at the top summarizes the measurements included in the Tevatron combination~\cite{mtop-combo}. Most recently, the Tevatron experiments, CDF and D0, and the LHC experiments, ATLAS and CMS, combined their $m_t$ measurements using procedures similar to the Tevatron average to produce a world combination~\cite{mtop-combo-world}. Figure~\ref{fig:mass-combo} at the bottom summarizes the measurements included in the world combination.
\begin{figure}[h,t,b,p]
\centerline{\includegraphics[width=8cm]{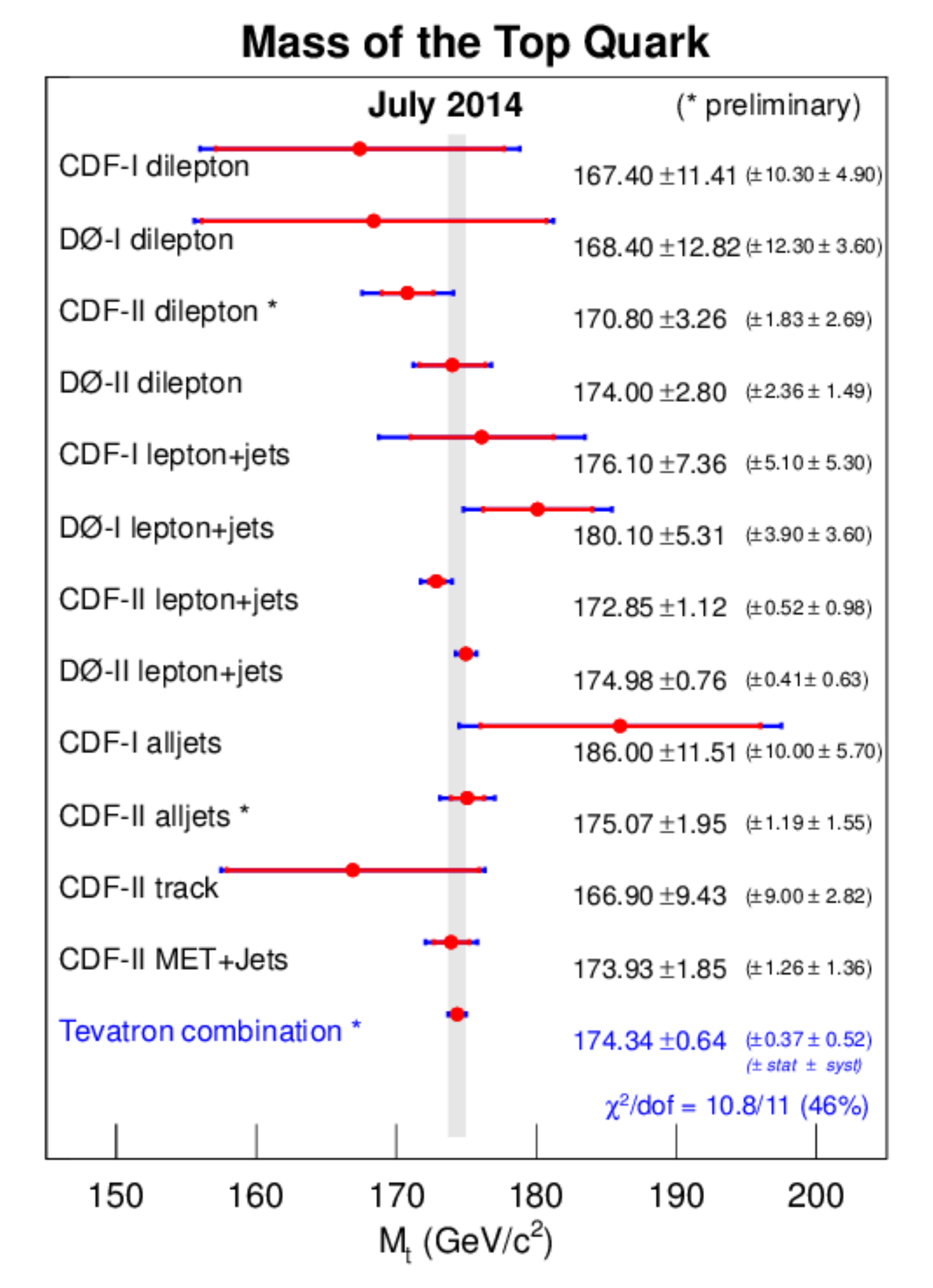}}
\centerline{\includegraphics[width=12cm]{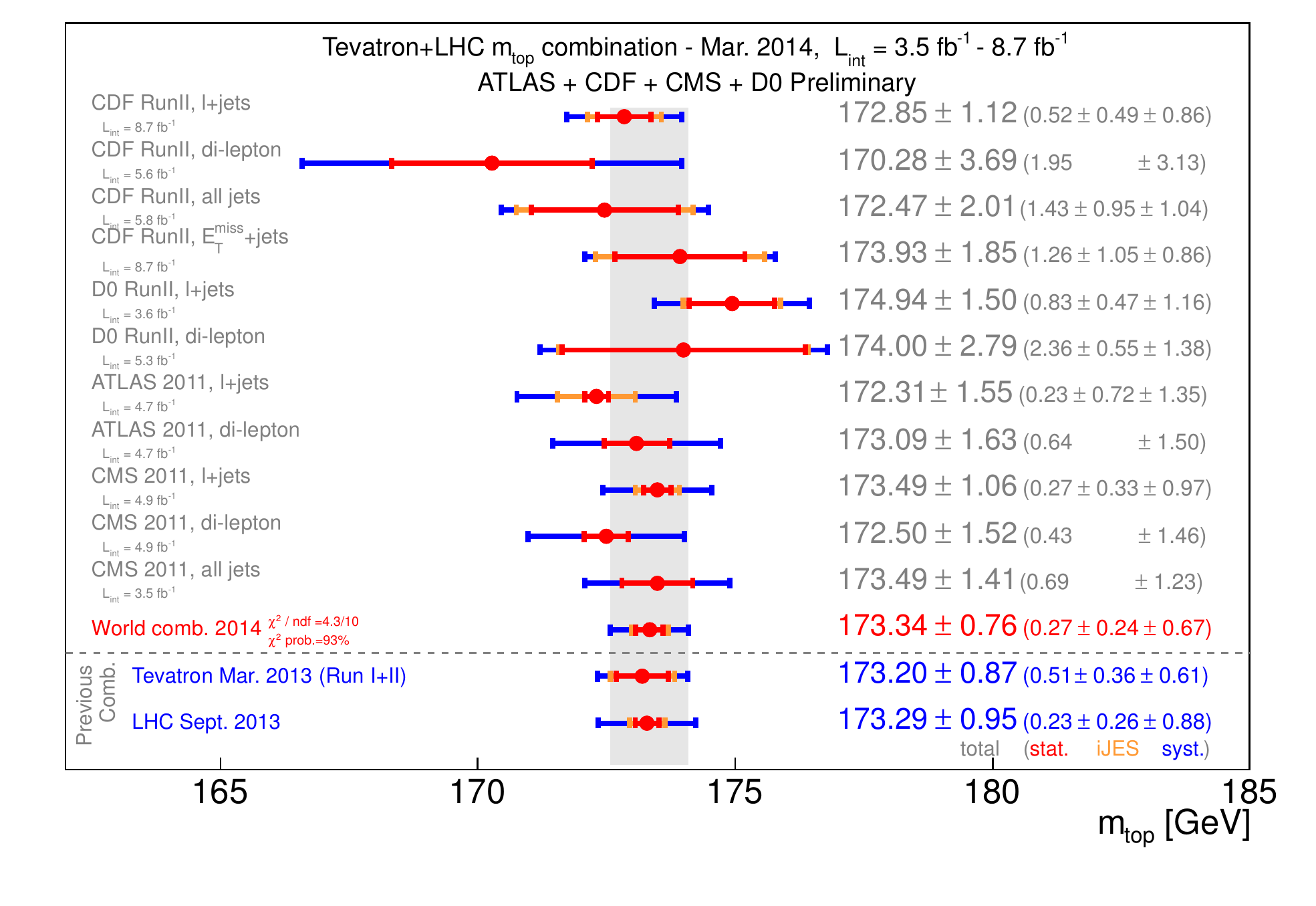}}
\caption{{\bf Top:} The top-quark mass measurements used in the Tevatron combination. {\bf Bottom:} The top-quark mass measurements used in the world combination.
\label{fig:mass-combo}}
\end{figure}

\subsection{Tests of CPT Invariance}
CPT invariance predicts that a particle and its antiparticle partner have the same mass. This was checked by both CDF and D0 Collaboration. CDF tested CPT invariance with a measurement of the mass difference $\Delta m_t=m_t-m_{\bar t}$ in the lepton$+$jets channel using the full luminosity of 8.7 fb$^{-1}$~\cite{cdf-mtop-diff-full-lumi}. The measurement applies a technique identical to the CDF template measurement described previously, but now templates of two mass differences corresponding to the two best $\chi^2$ solutions of the kinematic fit per event are constructed, in place of the two mass templates. The kinematic fit is modified to assume a fixed average top-quark mass $(m_t+m_{\bar t})/2=172.5$ GeV. No {\it in situ} JES calibration is applied in this case. The result of the likelihood fit to the data, $\Delta m_t=-1.95\pm 1.11({\rm stat})\pm 0.59({\rm syst})=-1.95\pm 1.26$ GeV, is consistent with zero within 1.5 standard deviations, and thus consistent with conservation of CPT symmetry. D0 also tested CPT invariance in the lepton$+$jets channel~\cite{d0-mtop-diff}, with a matrix element technique identical to that of the $m_t$ measurement in this channel, except that the matrix element was explictly modified to account for different masses of the top and anti-top quarks. Again, two data sets were used, corresponding to 1 fb$^{-1}$ and 2.6 fb$^{-1}$ of luminosity, respectively, and the results were combined to yield $m_t=0.8\pm 1.8({\rm stat})\pm 0.5({\rm syst})=0.8\pm 1.9$ GeV. This is also consistent with zero, and thus with conservation of CPT symmetry.

\section{Top Quark Properties}

The study of top-quark properties other than its mass, such as the couplings, decay width, charge, and polarization, is important for understanding the nature of this fundamental particle and verifying the standard model (SM) prediction for its identity. This study has been one of the major topics of the Tevatron Run II physics program, using data samples up to two orders of magnitude larger than were available when the top quark was first observed.

\subsection{$W$ Boson Helicity in Top Quark Decays} 
A property characterizing the dynamics of the top-quark decay is the helicity state of the on-shell $W$ boson produced in this decay. The $W$ boson can have three possible helicity states, and the fractions of the helicity states in $W^+$ bosons produced in these states are denoted as $f_0$ (longitudinal), $f_-$ (left-handed), and $f_+$ (right-handed). In the SM, the top quark decays through the $V-A$ weak charged-current interaction, which strongly suppresses right-handed $W^+$ bosons or left-handed $W^-$ bosons. The SM expectation for the helicity fractions depends on the masses of the top quark $m_t$ and of the $W$ boson $M_W$. For the values of $m_t=173.3\pm 1.1$ GeV~\cite{mtop-combo-2010} and $M_W=80.399\pm 0.023$ GeV~\cite{wmass-pdg-2010}, the expected SM values are $f_0=0.688\pm 0.004$, $f_-=0.310\pm 0.004$, and $f_+=0.0017\pm 0.0001$~\cite{whel-sm}. A measurement that deviates significantly from these expectations would provide strong evidence of physics beyond the SM, indicating either a departure from the expected $V-A$ structure of the $tWb$ vertex or the presence of a non-SM contribution to the $t\bar t$ candidate sample. Both the CDF and D0 Collaborations have measured the $f_0$ and $f_+$ helicity fractions using matrix element techniques, and they have combined their results to maximize the precision~\cite{whel-combo}. The results of the combination are $f_0=0.722\pm 0.062({\rm stat})\pm 0.052({\rm syst})=0.722\pm 0.081$ and $f_+=-0.033\pm 0.034({\rm stat})\pm 0.031({\rm syst})=-0.033\pm 0.046$, consistent with the SM expectations. Figure~\ref{fig:whel-combo} shows the 1- and 2-standard deviation $\chi^2$ contours of the combination, together with the individual input measurements and the SM prediction. D0 measured the helicity fractions in jointly analyzed dilepton and lepton$+$jets events using 5.4 fb$^{-1}$ of luminosity, whereas CDF measured them in lepton$+$jets events using 2.7 fb$^{-1}$ and dilepton events using 5.1 fb$^{-1}$. After the combination, CDF updated the helicity fractions measurement using the full luminosity of 8.7 fb$^{-1}$~\cite{cdf-whel-lj-full-lumi}, with the results of $f_0=0.726\pm 0.066({\rm stat})\pm 0.067({\rm syst})=0.726\pm 0.094$ and $f_+=-0.045\pm 0.044({\rm stat})\pm 0.058({\rm syst})=-0.045\pm 0.073$. The updated CDF result is consistent
 with the Tevatron average result.

\begin{figure}[h,t,b,p]
\centerline{\includegraphics[width=12cm]{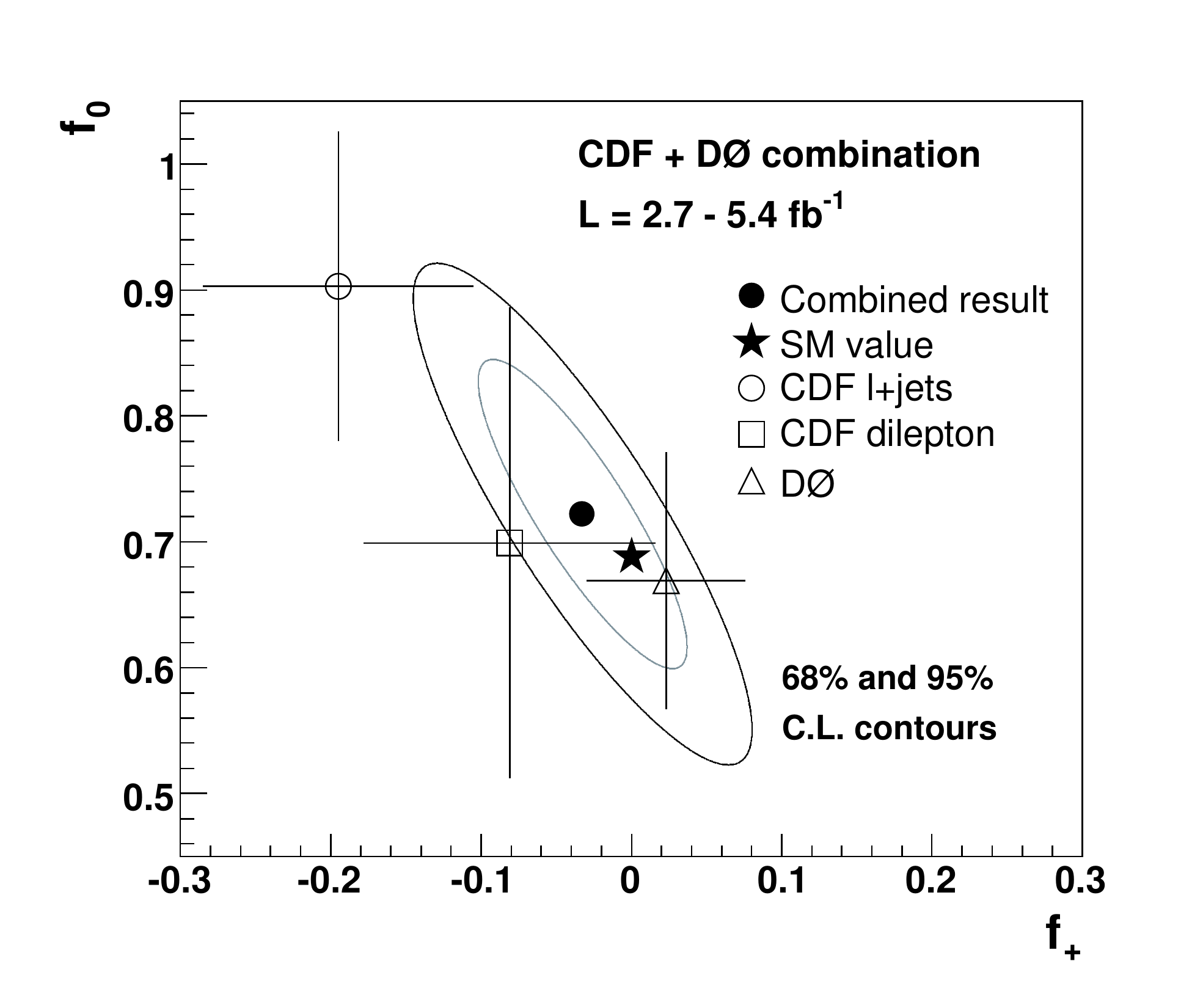}}
\caption{$W$-helicity fractions in top-quark decays measured by CDF and D0, and compared with the SM prediction. The contours show the experimentally observed region from the combination of these measurements.
\label{fig:whel-combo}}
\end{figure}

\subsection{Top Quark Width}
Another decay property of the top quark is its decay width. The large mass of the top quark endows it with a large decay width and, hence, the shortest lifetime of all known fermions~\cite{top-width}. In leading order, the top-quark decay width $\Gamma_t$ depends on the top-quark mass $m_t$, the Fermi coupling constant $G_F$, and the magnitude of the top-to-bottom quark coupling $\vert V_{tb}\vert$ in the CKM quark-mixing matrix~\cite{top-width-lo}. The NLO calculation with QCD and electroweak corrections predicts $\Gamma_t=1.33$ GeV at $m_t=172.5$ GeV~\cite{top-width-nlo}. This is consistent with the NNLO result of $\Gamma_t=1.32$ GeV~\cite{top-width-nnlo}, indicating that higher-order corrections are unimportant. A deviation from the SM prediction could indicate the presence of non-SM decay channels, such as decays through a charged Higgs boson~\cite{top-to-charged-higgs}, the supersymmetric top-quark partner~\cite{top-to-susy}, or a flavor-changing neutral current~\cite{top-fcnc}. A direct measurement of $\Gamma_t$ provides general constraints on such hypotheses. CDF measured $\Gamma_t$ in lepton$+$jets events~\cite{cdf-width-lj-full-lumi} using the full luminosity of 8.7 fb$^{-1}$ and the template method applied in the $m_t$ measurement described previously. The change in this case is that $m_t$ is fixed to 172.5 GeV and instead teamplates of $\Gamma_t$ are constructed. The fit of the data yields $1.10<\Gamma_t<4.05$ GeV at the 68\% C.L., corresponding to a lifetime of $1.6\times 10^{-25}<\tau<6.0\times 10^{-25}$ s. For a typical quark hadronization time scale of $3.3\times 10^{-24}$ s~\cite{q-to-hadron-time-scale}, this result supports the assertion that top-quark decay occurs before hadronization.

D0 extracted indirectly the decay width from the $\vert V_{tb}\vert$ CKM matrix element, which was derived from a single top-quark cross section measurement using data corresponding to 5.4 fb$^{-1}$ of luminosity~\cite{PRD85-091104-2012}. In this analysis, the total width $\Gamma_t$ is extracted from the partial decay width $\Gamma(t\to Wb)$ and the branching fraction $B(t\to Wb)$. $\Gamma(t\to Wb)$ is obtained from the t-channel single top-quark production cross section and $B(t\to Wb)$ is measured in $t\bar t$ events. For a top-quark mass of 172.5 GeV, the result is $\Gamma_t=2.00^{+0.47}_{-0.43}$ GeV, which translates to a top-quark lifetime of $\tau=(3.29^{+0.90}_{-0.63})\times 10^{-25}$ s, in good agreement with the results from the CDF direct measurement and the SM prediction.

\begin{figure}[h,t,b,p]
\centerline{\includegraphics[width=6.5cm]{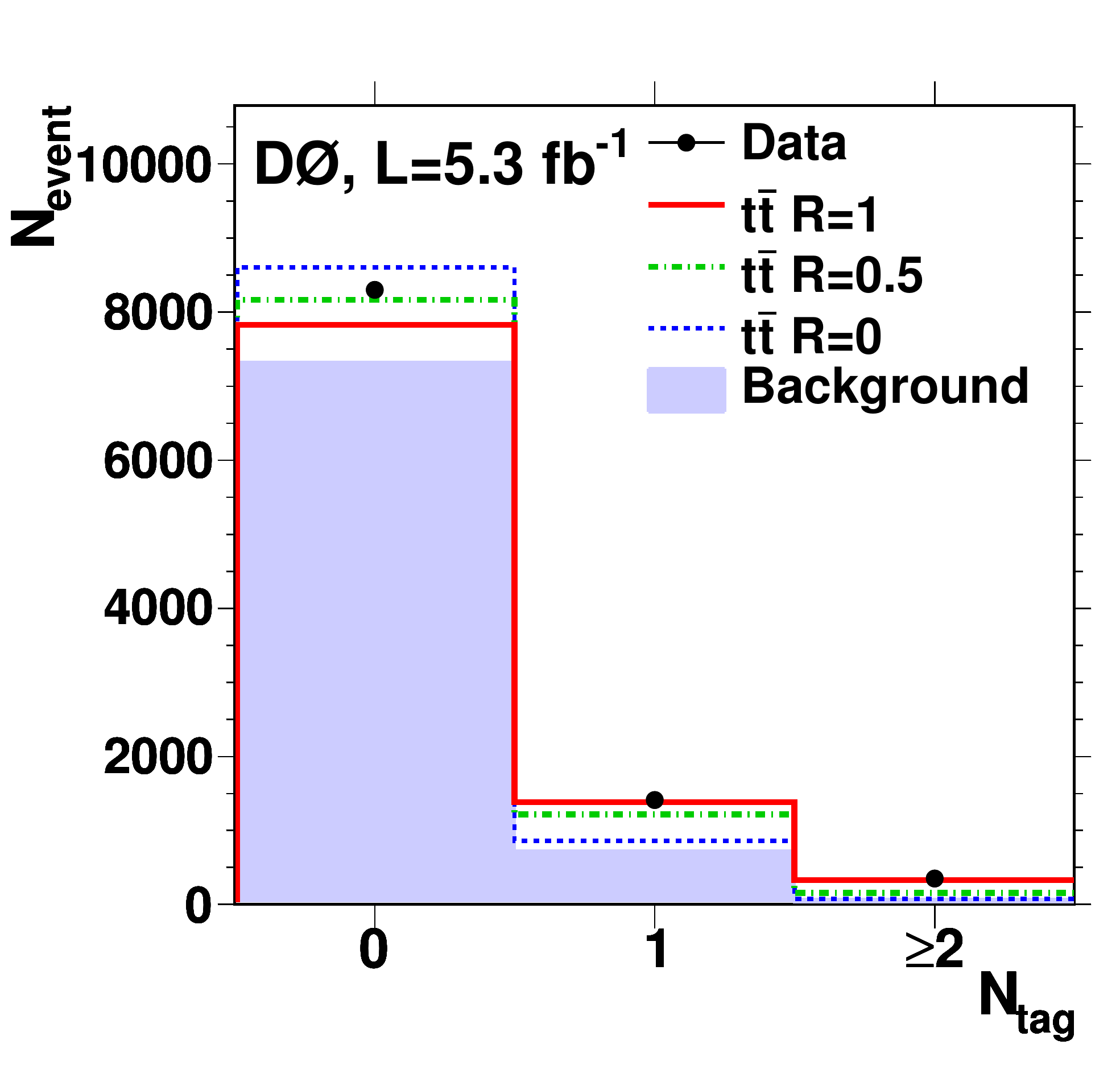}\includegraphics[width=6.5cm]{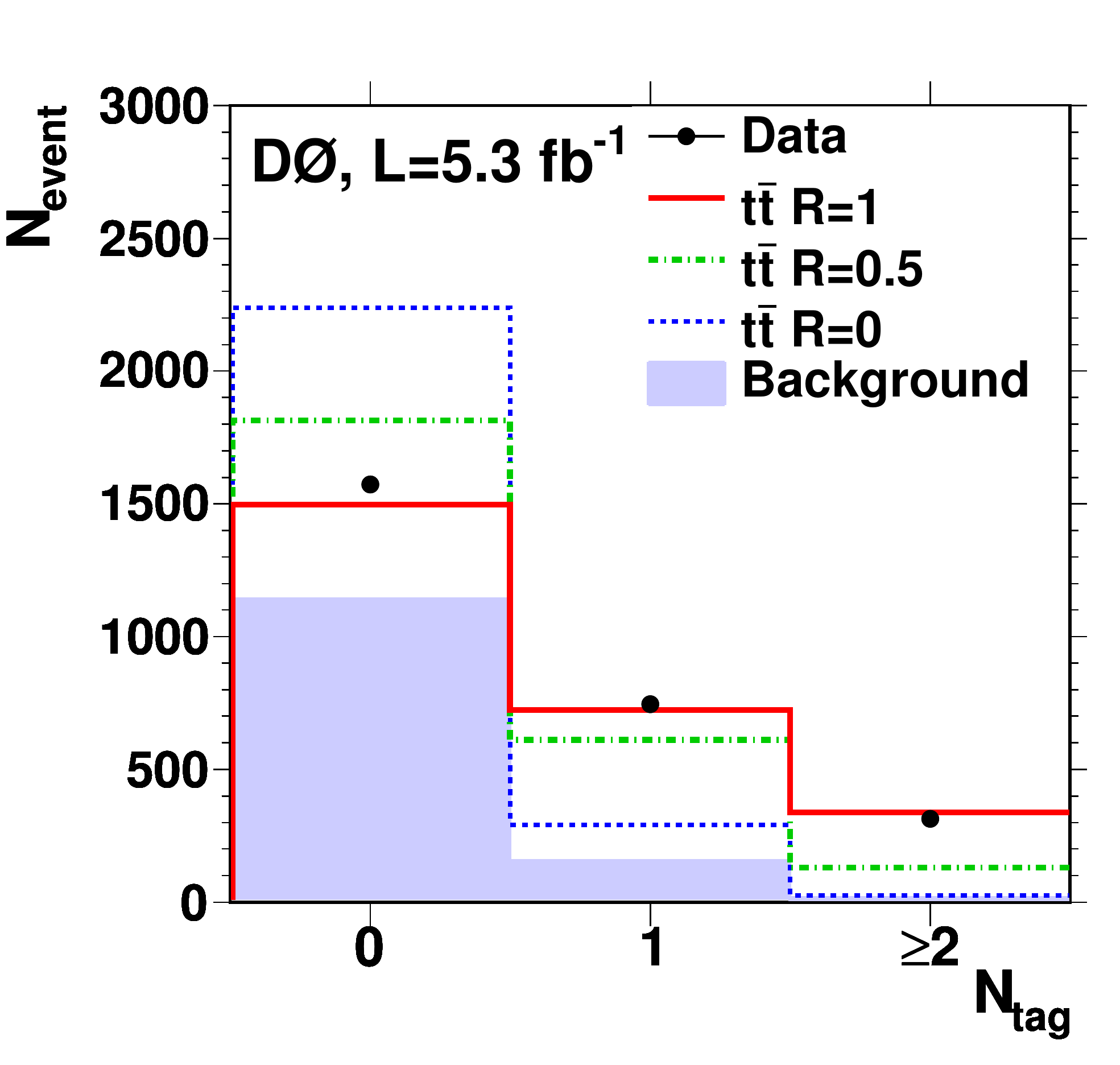}}
\caption{Number of $b$-tagged jets in the D0 data, compared with the signal and background expectations for various values of $R=B(t\to Wb)/B(t\to Wq)$. {\bf Left:} Data and expectations in lepton$+$3 jets events. {\bf Right:} Data and expectations in lepton$+$$\ge$4 jets events.
\label{fig:br-d0-lj}}
\end{figure}

\subsection{Top Quark Couplings}
The $\vert V_{tb}\vert$ CKM matrix element is directly constrained by measurements of the single top-quark production cross section, as discussed in section~\ref{sec:stop-Run2}. However, it can be indirectly constrained by measuring the ratio $R=B(t\to Wb)/B(t\to Wq)$ of the branching fraction for top-to-bottom quark decays to the branching fraction for top-to-any ``down'' quark decays, including $b$, if a hypothesis on the dimension of the CKM mixing matrix (i.e., the number of quark flavors) is made and the unitarity property of the matrix is used. The SM hypothesis involves six flavors, three ``up'' ($u$, $c$, $t$) and three ``down'' ($d$, $s$, $b$), corresponding to a 3$\times$3 CKM matrix. Then $R=\vert V_{tb}\vert^2/\sum_{q=d,s,b}\vert V_{tq}\vert^2$~\cite{top-ckm}. Using the existing constraints on $\vert V_{ts}\vert$ and $\vert V_{td}\vert$, the SM predictions are $\vert V_{tb}\vert=0.99915^{+0.00002}_{-0.00005}$ and $R=0.99830^{+0.00004}_{-0.00009}$~\cite{top-br-sm}. A measured deviation from these predictions would be an indication of new physics: an extra generation of quarks, non-SM top-quark production, or non-SM background to top-quark production. Both CDF and D0 Collaborations have measured $R$ by measuring the number of $b$-tagged jets in the final state of $t\bar t$ candidate events. CDF analyzed events both in the lepton$+$jets channel~\cite{top-br-cdf-lj} and in the dilepton channel~\cite{top-br-cdf-ll} from a sample corresponding to the full luminosity of 8.7 fb$^{-1}$. In the lepton$+$jets channel, a ratio $R=0.94\pm 0.09$ or $R>0.785$ at 95\% C.L. was extracted and a $\vert V_{tb}\vert=0.97\pm 0.05$ or $\vert V_{tb}\vert>0.89$ at 95\% C.L was deduced. In the dilepton channel, a ratio $R=0.87\pm 0.07$ or $R>0.73$ at 95\% C.L. was extracted and a $\vert V_{tb}\vert=0.93\pm 0.04$ or $\vert V_{tb}\vert>0.85$ at 95\% C.L was deduced. D0 analyzed together lepton$+$jets and dilepton events from a sample corresponding to 5.4 fb$^{-1}$ of luminosity~\cite{top-br-d0-lj-dil}, extracted a ratio $R=0.90\pm 0.04$ or $0.82<R<0.98$ at 95\% C.L., and deduced a $\vert V_{tb}\vert=0.95\pm 0.02$ or $0.90<\vert V_{tb}\vert<0.99$ at 95\% C.L. Figure~\ref{fig:br-d0-lj} shows the number of $b$-tagged jets expected for various $R$ values, compared with the D0 lepton$+$jets data. The results from both experiments are consistent with the SM expectation, although the D0 result agrees with the expectation only at the 1.6\% level.

D0 has also explored departures from the $V-A$ form of the electroweak interaction of the top quark predicted by the SM, by combining the information from single top-quark production and $W$ helicity fractions measured with data corresponding to 5.4 fb$^{-1}$ of luminosity~\cite{d0-fromfacs}. In this search, 95\% C.L. upper limits on the magnitudes of anomalous $Wtb$ R-handed vector and L,R-tensor coupling form factors are placed at a new physics scale $\Lambda=1$ TeV.

\begin{figure}[h,t,b,p]
\centerline{\includegraphics[width=6.5cm]{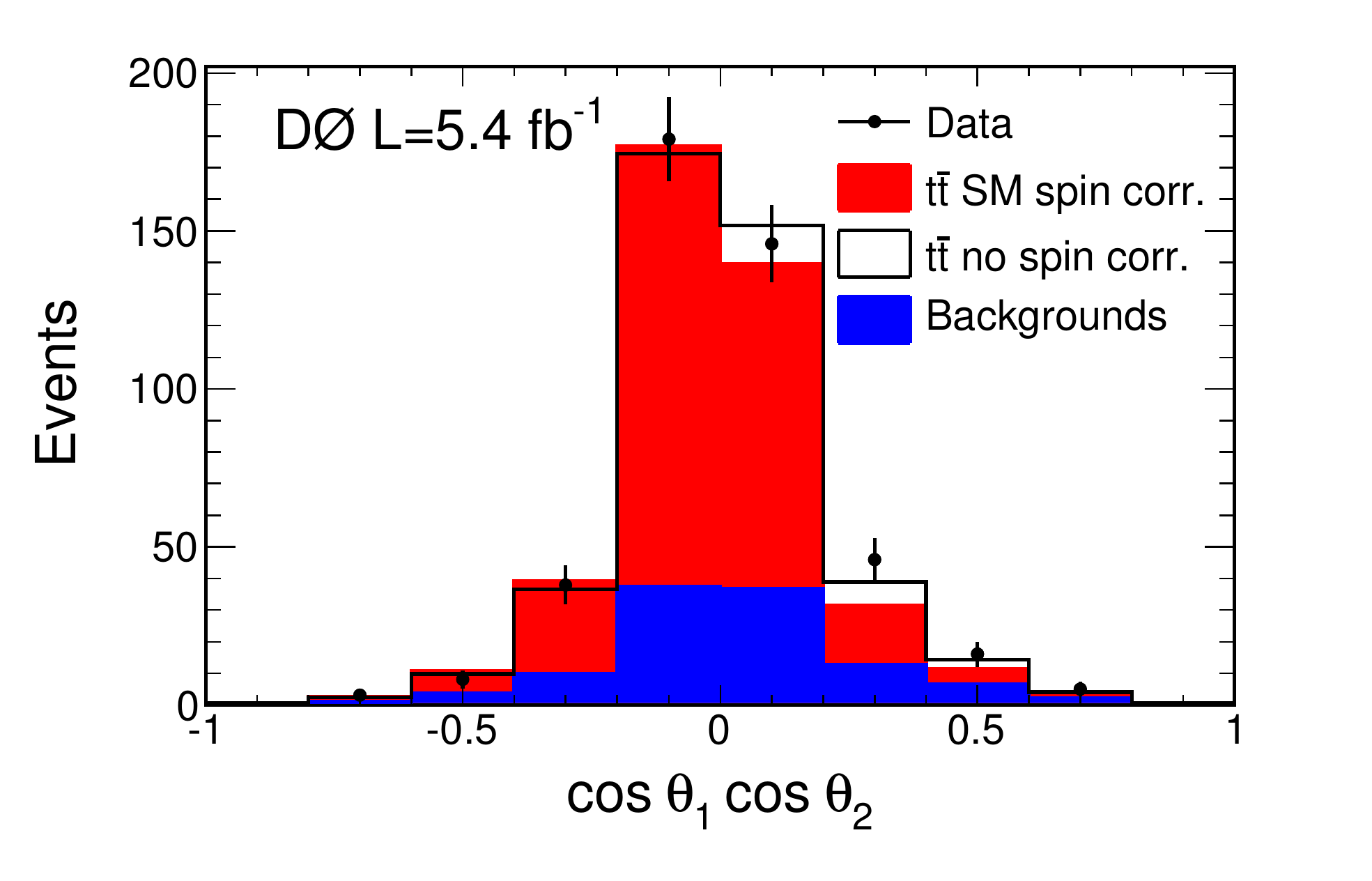}\includegraphics[width=6.5cm]{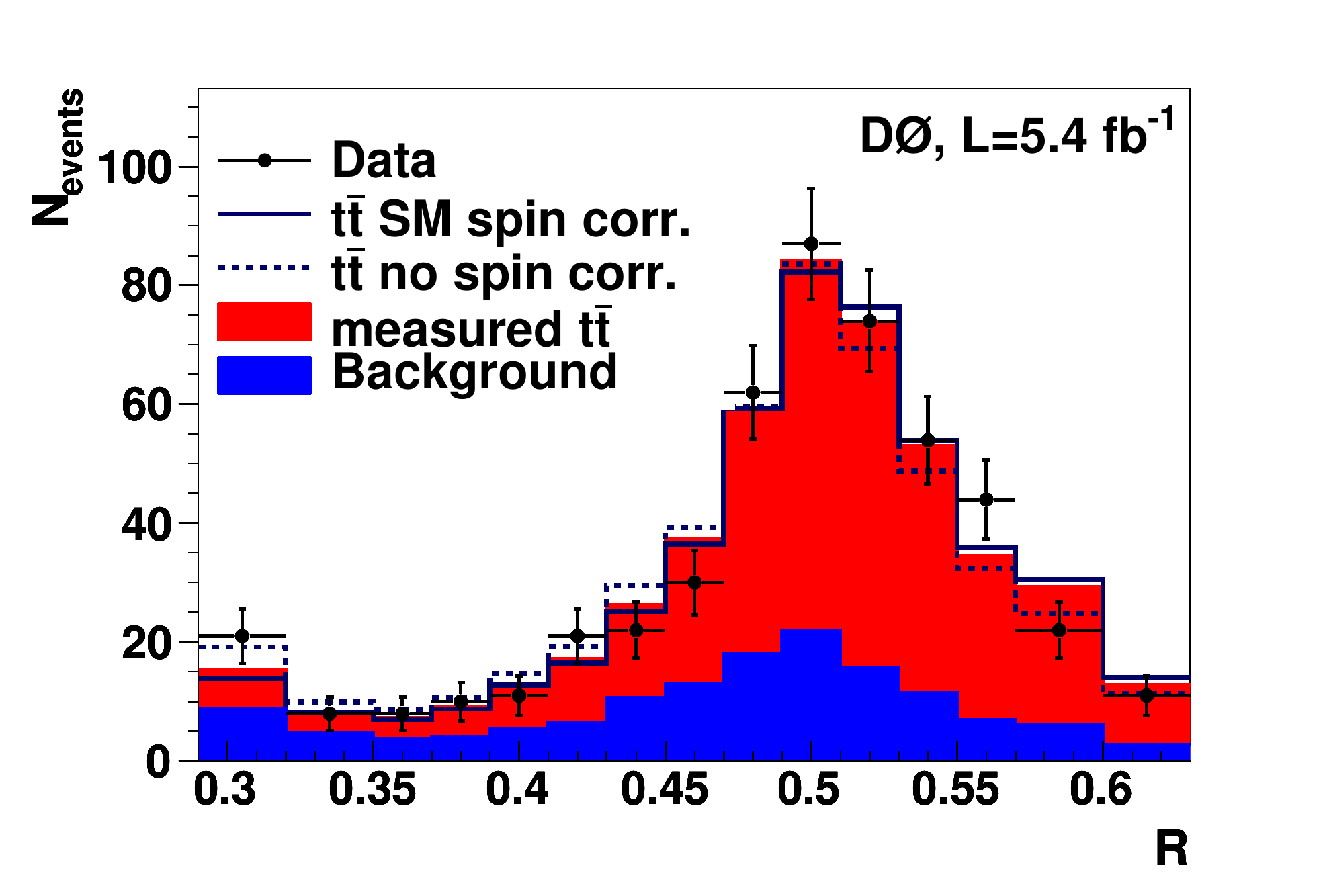}}
\caption{{\bf Left:} The distribution of $\cos\theta_1 \cos\theta_2$ measured by D0 in dilepton events and compared with SM predictions for the total background and for the signal with and without spin correlations. {\bf Right:} The discriminant $R$ distribution fitted to the D0 dilepton data and compared with SM predictions for the total background and for the signal with and without spin correlations.
\label{fig:spin-d0-dil}}
\end{figure}

\subsection {Spin Correlation in \ttbar Events}
On the production side, while in the SM top and anti-top quarks are produced unpolarized in hadron collisions, the orientations of their spins are correlated~\cite{top-spin}. Since the top quarks decay before hadronization, the correlation between the direction of the spin of the top and the anti-top quark cannot be affected by fragmentation~\cite{q-to-hadron-time-scale}. This contrasts with longer-lived quarks, for which the spins become decorrelated by strong interactions before they decay~\cite{quark-spin-decorr}. The orientation of the spin of the top quark is therefore reflected in the angular distributions of its decay products. The strength of spin correlation depends on the production mechanism and differs, for example, for $q\bar q\to t\bar t$ and $gg\to t\bar t$~\cite{top-spin-prod}. The correlation strength is defined $d^2\sigma/d\cos\theta_1 d\cos\theta_2=\sigma(1-C\cos\theta_1 \cos\theta_2)/4$, where $\sigma$ denotes the cross section and $\theta_1$ and $\theta_2$ are the angles between the spin quantization axis and the direction of flight of the decay leptons (for leptonically decaying $W$ bosons) or jets (for hadronically decaying $W$ bosons) in the parent $t$ and $\bar t$ rest frames. The value $C=+1$~$(-1)$ corresponds to fully correlated (anticorrelated) spins and $C=0$ corresponds to uncorrelated spins, while the SM prediction at NLO using the beam momentum vector as the spin quantization axis is $C=0.777^{+0.027}_{-0.042}$~\cite{top-spin-nlo}. The charged leptons from the $t\to W^+b\to l^+\nu_lb$ and $\bar t\to W^-\bar b\to l^-\bar\nu_l\bar b$ decays are the probes with the highest sensitivity to the direction of the $t$ and $\bar t$ quark spin, respectively. Therefore, the dilepton final state is more suitable for measurements of the spin correlation strength~\cite{top-spin-test,top-spin-nlo}.

Both CDF and D0 have measured the spin correlation strength. CDF measured a strength of $C=0.60\pm 0.50({\rm stat})\pm 0.16({\rm syst})=0.60\pm 0.52$ using 1001 lepton$+$jets candidate events selected from a data sample corresponding to 4.3 fb$^{-1}$ of luminosity~\cite{top-spin-cdf}. The measurement was done by fitting a 2-dimensional distribution constructed from the $\cos\theta_1 \cos\theta_2$ bilinear, where in one dimension the angles correspond to the lepton and the $d$-quark jet (identified as the closest to the $b$-quark jet in the $W$ boson rest frame) from the leptonic and hadronic $W$ decay, respectively, and in the other dimension they correspond to the lepton and the $b$-quark associated with the hadronically decaying $W$. D0 initially measured a strength of $C=0.10\pm 0.45({\rm stat+syst})$ by fitting the $\cos\theta_1 \cos\theta_2$ distribution of the leptons in a data sample of dilepton candidate events corresponding to 5.4 fb$^{-1}$ of luminosity~\cite{top-spin-d0-dil}. Next, D0 measured the fraction of strength relative to the SM prediction $f=0.74^{+0.40}_{-0.41}({\rm stat+syst})$ using the same data sample of dilepton candidate events~\cite{top-spin-d0-me}, but this time employing a matrix element technique based on the discriminant $R=P_{c}/(P_{c}+P_{u})$, where $P_{c}$ and $P_{u}$ are the per-event probability densities for the hypotheses of correlated spins as predicted by the SM and of uncorrelated spins, respectively~\cite{top-spin-me}. Figure~\ref{fig:spin-d0-dil} shows the distributions of $\cos\theta_1 \cos\theta_2$ used in the first measurement and of $R$ used in the second one. The second D0 measurement excluded the hypothesis of uncorrelated top-quark spins at the 97.7\% level. Later, D0 measured $f=1.15^{+0.42}_{-0.43}({\rm stat+syst})$ from a sample of lepton$+$jets events corresponding to 5.3 fb$^{-1}$ of luminosity using the same matrix element technique as the measurement in the dilepton channel~\cite{top-spin-d0-combo}. By combining with the result measured in the dilepton channel, the fraction $f=0.85\pm 0.29({\rm stat+syst})$ is determined, which translates to a probability of 0.16\% for the true fraction to be zero (no correlations). Therefore, the combined result provided the first evidence of SM spin correlation at 3.1 standard deviations.

\subsection{Lorentz Invariance}
D0 also investigated the possibility of violation of Lorentz invariance in the framework of the standard-model extension (SME)~\cite{lorentz-sme} in the top-quark sector~\cite{lorentz-top}, by examining the $t\bar t$ production cross section in lepton$+$jets events selected from data corresponding to 5.3 fb$^{-1}$ of luminosity~\cite{lorentz-top-d0}. The violation is quantified using the SME prediction of a dependence of the cross section on sidereal time as the orientation of the detector changes with the rotation of the Earth. Within this framework, D0 measured coefficients used to parametrize violation of Lorentz invariance in the top-quark sector and found them consistent with zero, within uncertainties.

\subsection{Top Quark Charge}
Finally, an interesting investigation performed by both CDF and D0 concerns the top-quark electric charge. Determining that the top quark decays into a $W^+$ boson and a bottom quark while the anti-top quark decays into a $W^-$ boson and an anti-bottom quark would ensure indirectly that the electric charge of the (anti-)top quark is indeed $(-)2/3 e$, as expected by the SM. If events were found to contain decays into a $W^-$ and bottom-quark final state, the charge of the decaying particle would be $-4/3 e$, incompatible with the SM top quark~\cite{top-charge}. In this hypothesis, the quark of mass around the 173 GeV is assumed to be part of a fourth generation of quarks and leptons, while the SM top quark is heavier than 230 GeV. Even though by the time of the CDF and D0 measurements this model was strongly disfavored by the measurements of single top quark production, the charge correlations between $b$-quark jets and $W$ bosons in $t\bar t$ events had not yet been definitively established by that time. Therefore, it was worth of testing directly the hypothesis of an exotic quark decay. Both experiments tested this hypothesis in events with a lepton$+$jets final state, D0 using 2.7 fb$^{-1}$ of luminosity~\cite{top-charge-d0} and CDF using 5.6 fb$^{-1}$ of luminosity~\cite{top-charge-cdf}. Both experiments excluded the exotic quark hypothesis, D0 at the 95\% level and CDF at the 99\% level. The latest result from D0 uses  5.3 fb$^{-1}$ of fully 
reconstructed lepton+jets events~\cite{top-charge-d0-new} and excluded the hypothesis that
the top quark has a charge of $Q=+4/3 e$ at a significance greater than 5 standard deviations. The analysis also places 
an upper limit of 0.46 on the fraction of such quarks that can be present in an admixture
with the standard model top quarks ($Q = +2/3 e$) at a 95\% confidence level.

Figure~\ref{fig:top-charge} shows the top charge in data compared with predictions of the SM and the exotic top-quark model. The SM prediction is clearly preferred by the data.
\begin{figure}[h,t,b,p]
\centerline{\includegraphics[width=7cm]{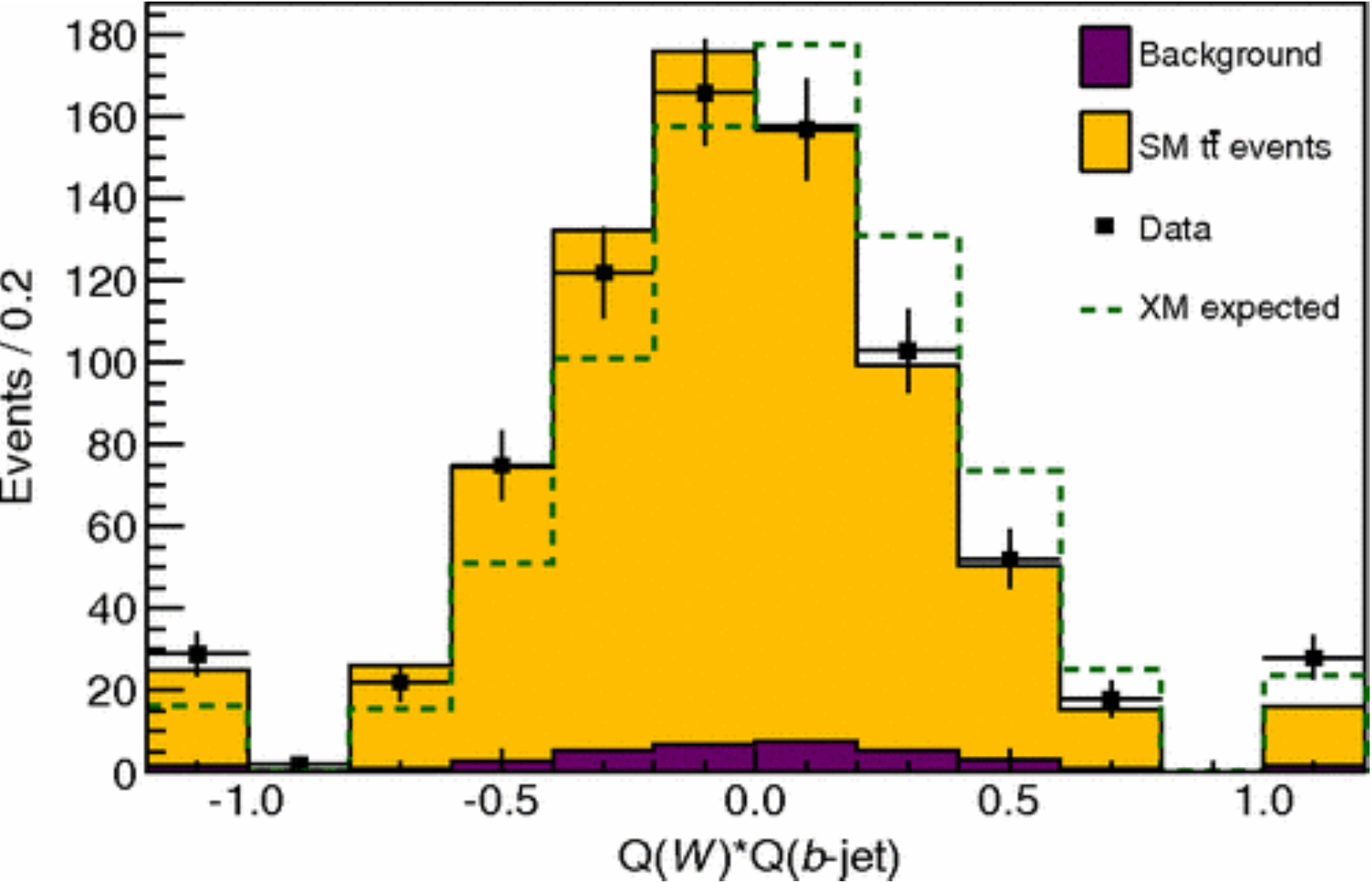}\includegraphics[width=7cm]{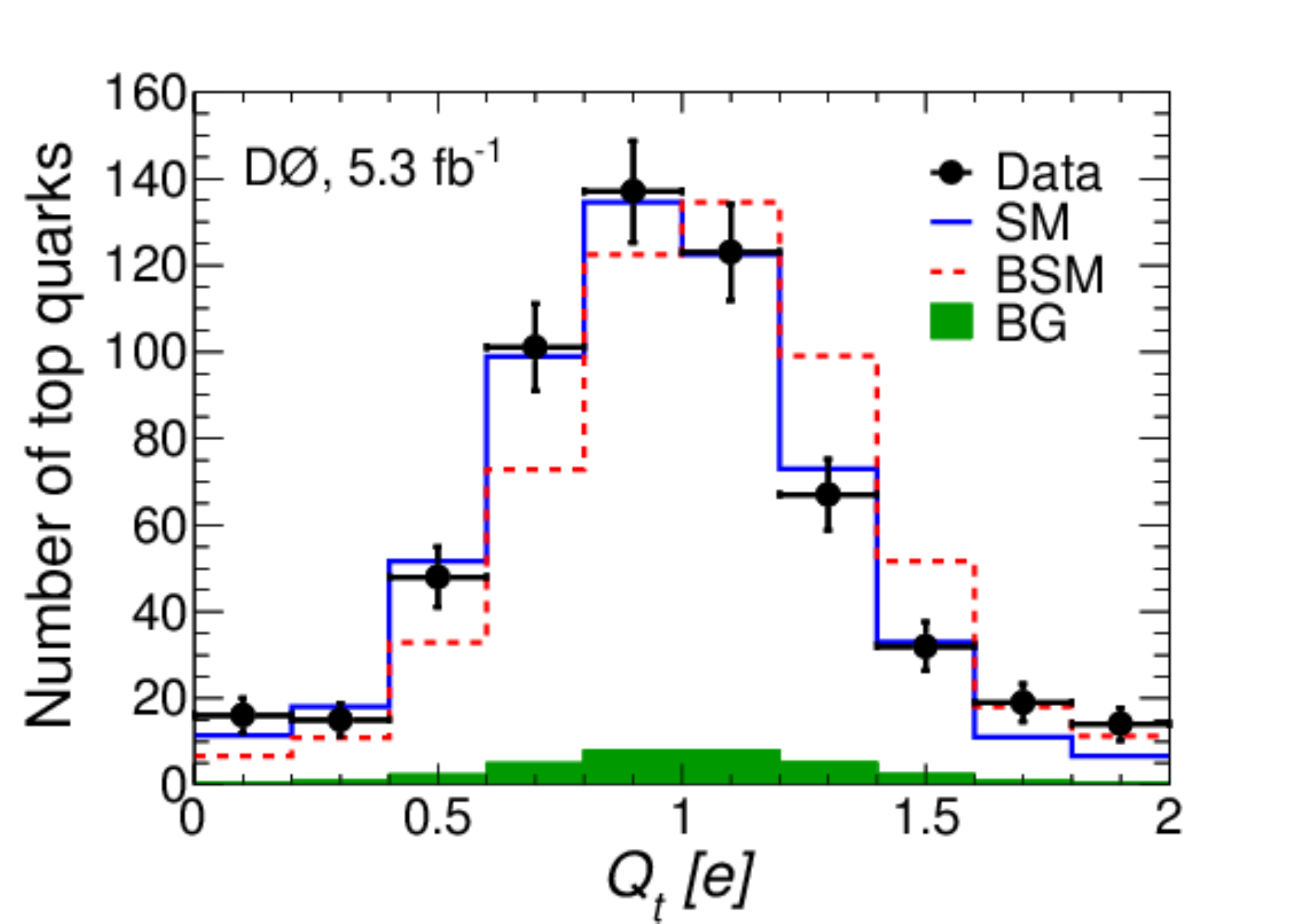}}
\caption{Left: Distribution of the product of the $W$ boson charge times the $b$-jet charge for lepton$+$jets events selected from CDF data corresponding to 5.6 fb$^{-1}$ of integrated luminosity. Shaded histograms show stacked SM signal and total background predictions. The dashed line shows the exotic model (XM) expectation assuming SM backgrounds and a top-quark charge of $-$4/3. SM-like candidates are on the negative side of the plot while XM-like candidates are on the positive side. The outermost bins correspond to $b$-jet charge exactly $\pm$1. 
Right: Combined distribution in the charge for \ttbar candidates
in D0 data compared with expectations from the SM and
the BSM. The background contribution (BG) is represented
by the green-shaded histogram.
\label{fig:top-charge}}
\end{figure}

\section{Summary and Conclusions}

The top-quark discovery in Run I was the major success of the Tevatron. After nearly 30 years from the foundation of the SM, this discovery completed the proof of the SM predictions for the fundamental fields of matter. In Run II, the Tevatron experiments, CDF and D0, studied extensively the properties and interactions of the top quark with the other fields of the SM by using two orders of magnitude larger data samples than those available in Run I. The quest for the top quark in Run I turned into a precision study of top-quark physics in Run II. The top-quark pair production cross section has been measured at a precision of the order of 5\%. Single top-quark production has been observed in both s and t channels. The top-quark mass has been measured with a precision of order 0.4\%. Its properties have been measured and found consistent with SM predictions, in most cases at the level of 5$\sigma$ observation and in some cases at the level of 3$\sigma$ evidence. The top-quark physics extracted from the Tevatron data agrees within 2$\sigma$ of experimental uncertainty, or better in most cases, with SM predictions of NLO and in some cases of NNLO accuracy in the strong and weak coupling constants. Many searches in the top-quark sector have been performed at the Tevatron, testing fundamental symmetries of the SM and setting sensitive exclusion limits on many new physics scenarios. Many new tecnhiques for precision measurements in the top-quark sector have been developed and established at the Tevatron, from multivariate methods for single top-quark measurements where the signal-to-background ratio is very low to sophisticated event reconstruction algorithms and jet energy scale calibration for top-quark mass measurements. The precision of top-quark measurements reached at the Tevatron has been driving the advance of theoretical top-quark physics calculations over the recent years. Top-quark physics, which was established at the Tevatron, is the flagship of its legacy.

\section{Acknowledgment}

We thank Paul Grannis and Andreas Jung for useful comments and discussions.

We thank the Fermilab staff and technical staffs of
the participating institutions for their vital contributions.
We acknowledge support from the DOE and NSF
(USA), ARC (Australia), CNPq, FAPERJ, FAPESP
and FUNDUNESP (Brazil), NSERC (Canada), NSC,
CAS and CNSF (China), Colciencias (Colombia), MSMT
and GACR (Czech Republic), the Academy of Finland,
CEA and CNRS/IN2P3 (France), BMBF and DFG (Germany),
DAE and DST (India), SFI (Ireland), INFN
(Italy), MEXT (Japan), the KoreanWorld Class University
Program and NRF (Korea), CONACyT (Mexico),
FOM (Netherlands), MON, NRC KI and RFBR (Russia),
the Slovak R\&D Agency, the Ministerio de Ciencia
e Innovacion, and Programa Consolider--Ingenio 2010
(Spain), The Swedish Research Council (Sweden), SNSF
(Switzerland), STFC and the Royal Society (United
Kingdom), the A.P. Sloan Foundation (USA), and the
EU community Marie Curie Fellowship contract 302103.



\begin{thebibliography}{0}    

\bibitem{top-obs-1995-cdf}
CDF Collaboration (F.~Abe {\it et al.}), 
Phys.\ Rev.\ Lett.\ {\bf 74}, 2626 (1995).

\bibitem{top-obs-1995-d0}
D0 Collaboration (S.~Abachi {\it et al.}), 
Phys.\ Rev.\ Lett.\ {\bf 74}, 2632 (1995).

\bibitem{toppp}
M. ~Czakon and A. ~Mitov, arXiv:1112.5675.


\bibitem{singletop-xsec-kidonakis}
N.~Kidonakis,
Phys.\ Rev.\ D {\bf 74}, 114012 (2006). 

\bibitem{PRL-103-092001-2009}
D0 Collaboration (V.M.~Abazov {\it et al.}), 
Phys.\ Rev.\ Lett.\ {\bf 103}, 092001 (2009).

\bibitem{PRL-103-092002-2009}
CDF Collaboration (T.~Aaltonen {\it et al.}), 
Phys.\ Rev.\ Lett.\ {\bf 103}, 092002 (2009).

\bibitem{b-discovery}
S.W. Herb {\it et al.},
Phys.\ Rev.\ Lett.\ {\bf 39}, 252 (1977).

\bibitem{PLB147-493-1984}
UA1 Collaboration (G. Arnison {\it et al.}), 
Phys. Lett. B {\bf 147}, 493 (1984).

\bibitem{ZPC37-505-1988} 
UA1 Collaboration (C. Albajar {\it et al.}), 
Z. Phys. C {\bf 37}, 505 (1988).

\bibitem{ZPC46-179-1990}
UA2 Collaboration (T. Akesson {\it et al.}),  
Z. Phys. C 46, {\bf 179} (1990). 

\bibitem{PRL68-447-1992}
CDF Collaboration (F. Abe {\it et al.}),  
Phys.\ Rev.\ Lett.\ {\bf 68}, 447 (1992).

\bibitem{PRL72-2138-1994}
D0 Collaboration (S. Abachi {\it et al.}),  
Phys.\ Rev.\ Lett.\ {\bf 72}, 2138 (1994).

\bibitem{PRL73-225-1994}
CDF Collaboration (F. Abe {\it et al.}),  
Phys.\ Rev.\ Lett.\ {\bf 73}, 225 (1994). 

\bibitem{NPB303-607-1988}
P. Nason, S. Dawson and R. K. Ellis, 
Nuclear Physics B {\bf 303}, 607--633 (1998).

\bibitem{ICHEP94-Grannis}
P. Grannis, 
ICHEP 1994 proceedings, arXiv:hep-ex/9409006 (1994).

\bibitem{PRD64-032002-2001}
CDF Collaboration (T. Affolder {\it et al.}),  
Phys.\ Rev.\ D {\bf 64}, 032002 (2001).

\bibitem{PRD67-012004-2013}
D0 Collaboration (V. M. Abazov {\it et al.}),  
Phys.\ Rev.\ D {\bf 67}, 012004 (2003).

\bibitem{PRL80-2767-1998}
CDF Collaboration (F. Abe {\it et al.}),  
Phys.\ Rev.\ Lett.\ {\bf 80}, 2767 (1998).

\bibitem{PRD58-052001-1998}
D0 Collaboration (B. Abbott {\it et al.}), 
Phys.\ Rev.\ D {\bf 58}, 052001 (1998).

\bibitem{Nature429-638-2004}
D0 Collaboration (M. Abazov {\it et al.}),  
Nature 429, 638-642 (10 June 2004).  

\bibitem{PRL82-271-1999}
CDF Collaboration (F. Abe {\it et al.}),  
Phys.\ Rev.\ Lett.\ {\bf 82}, 271 (1999).

\bibitem{PRD60-052001-1999} 
D0 Collaboration (B. Abbott {\it et al.}), 
Phys.\ Rev.\ D {\bf 60}, 052001 (1999).

\bibitem{PRL79-1992-1997}
CDF Collaboration (F. Abe {\it et al.}),  
Phys.\ Rev.\ Lett.\ {\bf 79}, 1992 (1997).

\bibitem{top-ckm}
N.~Cabibbo, Phys.\ Rev.\ Lett.\ {\bf 10}, 531 (1963);
M.~Kobayashi and T.~Maskawa, Prog.\ Theor.\ Phys.\ {\bf 49}, 652 (1973).

\bibitem{PRL86-3233-2001} 
CDF Collaboration (T. Affolder {\it et al.}),  
Phys.\ Rev.\ Lett.\ {\bf 86}, 3233 (2001).

\bibitem{PRD65-091102-2002}
CDF Collaboration (D. Acosta {\it et al.}),  
Phys.\ Rev.\ D {\bf 65}, 091102 (2002).

\bibitem{PLB517-282-2001} 
D0 Collaboration (V. M. Abazov {\it et al.}),  
Phys.\ Lett.\ B {\bf 517}, 282 {2001}.

\bibitem{PRL79-357-1997}
CDF Collaboration (F. Abe {\it et al.}),  
Phys.\ Rev.\ Lett.\ {\bf 79}, 357 (1997).

\bibitem{PRL82-4975-1999} 
D0 Collaboration (V. M. Abazov {\it et al.}),  
Phys.\ Rev.\ Lett.\ {\bf 82}, 4975 (1999).

\bibitem{PRL80-2525-1998} 
CDF Collaboration (F. Abe {\it et al.}),  
Phys.\ Rev.\ Lett.\ {\bf 80}, 2525 (1998).

\bibitem{PRL84-216-2000} 
CDF Collaboration (T. Affolder {\it et al.}),  
Phys.\ Rev.\ Lett.\ {\bf 84}, 216 (2000).

\bibitem{PRL85-256-2000} 
D0 Collaboration (B. Abbott {\it et al.}), 
Phys.\ Rev.\ Lett.\ {\bf 85}, 256 (2000). 


\bibitem{PRD89-072001-2014}
CDF and D0 Collaborations (T.~Aaltonen {\it et al.}), 
Phys.\ Rev.\ D {\bf 89}, 072001 (2014).

\bibitem{PRD88-091103-2013}
CDF Collaboration (T.~Aaltonen {\it et al.}),  
Phys.\ Rev.\ D {\bf 88}, 091103 (2013).

\bibitem{PRL105-012001-2010}, 
CDF Collaboration (T.~Aaltonen {\it et al.}),  
Phys.\ Rev.\ Lett.\ {\bf 105}, 012001 (2010).

\bibitem{PRD81-052011-2010}
CDF Collaboration (T.~Aaltonen {\it et al.}),  
Phys.\ Rev.\ D {\bf 81}, 052011 (2010).

\bibitem{PLB704-403-2011}
D0 Collaboration (V.M.~Abazov {\it et al.}), 
Phys.\ Lett.\ B {\bf 704}, 403 (2011).

\bibitem{PRD84-012008-2011}
D0 Collaboration (V.M.~Abazov {\it et al.}), 
Phys.\ Rev.\ D {\bf 84}, 012008 (2011).

\bibitem{arXiv:1402.6728}
CDF Collaboration (T.~Aaltonen {\it et al.}),  
Phys.\ Rev.\ D {\bf 89}, 091101 (2014). 


\bibitem{PRD82-071102-2010}
D0 Collaboration (V.M.~Abazov {\it et al.}), 
Phys.\ Rev.\ D {\bf 82}, 071102 (2010).

\bibitem{PRL96-202002-2006}
A.~Abulencia {\it et al.} (CDF Collaboration), 
Phys.\ Rev.\ Lett.\ {\bf 96}, 202002 (2006).

\bibitem{PRL109-132001-2012}
P. Baernreuther, M. Czakon and A. Mitov, Phys.\ Rev.\ Lett.\ {\bf 109}, 132001 (2012).

\bibitem{PRD80-054009-2009}
U. Langenfeld, S. Moch and P. Uwer, Phys.\ Rev.\ D {\bf 80}, 054009 (2009).

\bibitem{PRD78-074005-2008}
N. � and R. Vogt, Phys.\ Rev.\ D {\bf 78}, 074005 (2008).

\bibitem{JHEP09-127-2008}
M. Cacciari, S. Frixione, M. L. Mangano, P. Nason and G. Ridolfi, J.\ High Energy
Phys.\ 09, 127 (2008).

\bibitem{arXiv:1401.5785}
D0 Collaboration (V.M.~Abazov {\it et al.}), 
arXiv:1401.5785, submitted to Phys. Rev. {\bf D}.

\bibitem{JHEP1009-097-2010}
V.Ahrens  {\it et al.}
Journal of High Energy Physics 1009, 097 (2010).

\bibitem{PRD83-112003-2011}
CDF Collaboration (T.~Aaltonen {\it et al.}),  
Phys. Rev. D {\bf 83}, 112003 (2011).

\bibitem{PRD84-112005-2011}
D0 Collaboration (V.M.~Abazov {\it et al.}), 
Phys. Rev. D {\bf 84}, 112005 (2011).

\bibitem{PRD87-092002-2013}
CDF Collaboration (T.~Aaltonen {\it et al.}),  
Phys. Rev. D {\bf 87}, 092002 (2013).

\bibitem{arXiv:1405.0421}
D0 Collaboration (V.M.~Abazov {\it et al.}), 
arXiv:1405.0421

\bibitem{PRD86-034026-2012}
W. Bernreuther and Z.-G. Si
Phys. Rev D {bf 86}, 034026 (2012)

\bibitem{PRD88-112002-2013} 
D0 Collaboration (V.M.~Abazov {\it et al.}), 
Phys. Rev. D {\bf 87}, 011103(R) (2013).

\bibitem{arXiv:1403.1294}
D0 Collaboration (V.M.~Abazov {\it et al.}), 
arXiv:1403.1294, submitted to PRD.

\bibitem{PRD88-072003-2013}
CDF Collaboration (T.~Aaltonen {\it et al.}),  
Phys. Rev. D {\bf 88}, 072003 (2013).


\bibitem{arXiv:1404.3698}
CDF Collaboration (T.~Aaltonen {\it et al.}),  
Phys.\ Rev.\ Lett.\ {\bf 113}, 042001 (2014).


\bibitem{PRL111-182002-2013}
CDF Collaboration (T.~Aaltonen {\it et al.}),  
Phys.\ Rev.\ Lett.\ {\bf 111}, 182002 (2013).

\bibitem{PRD49-4454-1999}
C. T. Hill and S. J. Parke,
Phys. Rev. D {\bf 49}, 4454 (2012).

\bibitem{PLB266-419-1991} 
C. T. Hill
Phys.\ Lett.\ B {\bf 266}, 419 (1991).

\bibitem{PRL83-3370-1999}
L. Randall and R. Sundrum,
Phys.\ Rev.\ Lett.\ {\bf 83}, 3370 (1999).

\bibitem{PRL110-121802-2012}
CDF Collaboration (T.~Aaltonen {\it et al.}),  
Phys.\ Rev.\ Lett.\ {\bf 110}, 121802 (2012).

\bibitem{PRD85-051101-2012}
D0 Collaboration (V.M.~Abazov {\it et al.}), 
Phys. Rev. D {\bf 85}, 051101 (2012).

\bibitem{arXiv:1401.2443}
A Carmona {\it et al.},
arXiv:1401.2443

\bibitem{PRD85-115011-2012}
J.Alwall {\it et al.},
Phys. Rev. D {\bf 85}, 115011 (2012).

\bibitem{PRL112-231802-2014}
CMS Collaboration (S. Chatrchtan {\it et al.}), 
Phys.\ Rev.\ Lett.\ 112, 231802 (2014).

\bibitem{PRD-83-091503-2011}
N.Kidonakis, 
Phys. Rev. D {\bf 83}, 091503 (2011).

\bibitem{PLB690-5-2010}
D0 Collaboration (V.M.~Abazov {\it et al.}), 
Phys.\ Lett.\ B {\bf 690}, 5 (2010).

\bibitem{PRD81-054028-2010}
N.Kidonakis, 
Phys. Rev. D {\bf 81}, 054028 (2010).

\bibitem{PRD-82-112005-2010}
CDF Collaboration (T.~Aaltonen {\it et al.}),  
Phys. Rev. D {\bf 82}, 112005 (2010).

\bibitem{PRD-84-112001-2011}
D0 Collaboration (V.M.~Abazov {\it et al.}), 
Phys. Rev. D {\bf 84}, 112001 (2011).

\bibitem{PLB-705-313-2011}
D0 Collaboration (V.M.~Abazov {\it et al.}), 
Phys.\ Lett.\ B {\bf 705}, 313 (2011).

\bibitem{PLB-726-656-2013}
D0 Collaboration (V.M.~Abazov {\it et al.}), 
Phys.\ Lett.\ B {\bf 726}, 656 (2013).

\bibitem{arXiv:1407.4031}
CDF Collaboration (T.~Aaltonen {\it et al.}), 
arXiv:1407.4031

\bibitem{PRD85-091104-2012}
D0 Collaboration (V.M.~Abazov {\it et al.}), 
Phys.\ Rev.\ D {\bf 85}, 091104 (2012). 

\bibitem{arXiv:1402.0484}
CDF Collaboration (T.~Aaltonen {\it et al.}), 
arXiv:1402.0484.

\bibitem{arXiv:1402.3756}
CDF Collaboration (T.~Aaltonen {\it et al.}), 
arXiv:1402.3756

\bibitem{arXiv:1402.5126}
CDF and D0 Collaborations (T.~Aaltonen {\it et al.}), 
arXiv:1402.5126.

\bibitem{PRD63-014018-2001}
T. Tait and C.-P. Yuan,
Phys.\ Rev.\ D {\bf 63}, 014018 (2001).

\bibitem{PRL102-151801-2008}
CDF Collaboration (T.~Aaltonen {\it et al.}), 
Phys.\ Rev.\ Lett.\ {\bf 102}, 151801 (2008).

\bibitem{PLB693-81-2010}
D0 Collaboration (V.M.~Abazov {\it et al.}), 
Phys.\ Lett.\ B {\bf 693}, 81 (2010).

\bibitem{PRL103-041801-2009}
CDF Collaboration (T.~Aaltonen {\it et al.}), 
Phys.\ Rev.\ Lett.\ {\bf 103}, 041801 (2010).

\bibitem{PLB699-145-2011}
D0 Collaboration (V.M.~Abazov {\it et al.}), 
Phys.\ Lett.\ B {\bf 699}, 145 (2011).

\bibitem{PRL108-201802-2012}
CDF Collaboration (T.~Aaltonen {\it et al.}), 
Phys.\ Rev.\ Lett.\ {\bf 108}, 2021802 (2012).


\bibitem{higgs}
F.~Englert and R.~Brout, Phys.\ Rev.\ Lett.\ {\bf 13}, 321 (1964);
P.W.~Higgs, Phys.\ Lett.\ {\bf 12}, 132 (1964); Phys.\ Rev.\ Lett.\ {\bf 13}, 508 (1964);
G.S.~Guralnik, C.R.~Hagen, and T.W.B.~Kibble, Phys.\ Rev.\ Lett.\ {\bf 13}, 585 (1964);
P.W.~Higgs, Phys.\ Rev.\ {\bf 145}, 1156 (1966).

\bibitem{msbar}
U.~Langenfeld, S.~Moch, and P.~Uwe, Phys.\ Rev.\ D {\bf 80}, 054009 (2009).

\bibitem{pole}
M.C.~Smith and S.S.~Willenbrock,  Phys.\ Rev.\ Lett.\ {\bf 79}, 3825 (1997).

\bibitem{scheme}
A.~Buckley {\it et al.},  Phys.\ Rep.\ {\bf 504}, 145 (2011).

\bibitem{mtop-xsec}
D0 Collaboration (V.M.~Abazov {\it et al.}),  Phys.\ Lett.\ B {\bf 703}, 422 (2011).

\bibitem{mtop-pre95}
C.~Quigg, Phys.\ Today {\bf 50}, No. 5, 20 (1997), and references therein.

\bibitem{mtop-at95}
The LEP Collaborations ALEPH, DELPHI, L3, OPAL,
and the LEP Electroweak Working Group, Report No. CERN-PPE/94-187, 1994,
and references therein.

\bibitem{mtop-review}
A.~Barbaro-Galtieri, F.~Margaroli, and I.~Volobouev, Rep.\ Prog.\ Phys.\ {\bf 75}, 056201 (2012).



\bibitem{lifetime-th}
M.~Jezabek and J.H.~K{\"u}hn, Phys.\ Rev.\ D {\bf 48}, R1910 (1993).

\bibitem{wmass}
K.~Nakamura {\it et al.} (Particle Data Group), J.\ Phys.\ G {\bf 37}, 075021 (2010).

\bibitem{cdf-lj-full-lumi}
CDF Collaboration (T.~Aaltonen {\it et al.}),  Phys.\ Rev.\ Lett.\ {\bf 109}, 152003 (2012).

\bibitem{cdf-vj-full-lumi}
CDF Collaboration (T.~Aaltonen {\it et al.}),  Phys.\ Rev.\ D {\bf 88}, 011101(R) (2013).

\bibitem{d0-lj}
D0 Collaboration (V.M.~Abazov {\it et al.}),  Phys.\ Rev.\ D {\bf 84}, 032004 (2011).

\bibitem{d0-lj-new}
D0 Collaboration (V.M.~Abazov {\it et al.}),  Phys.\ Rev.\ Lett.\ 113, 032002 (2014).

\bibitem{cdf-mtop-tracking}
CDF Collaboration (T.~Aaltonen {\it et al.}),  Phys.\ Rev.\ D {\bf 81}, 032002 (2010).

\bibitem{cdf-mtop-leptons}
CDF Collaboration (T.~Aaltonen {\it et al.}),  Phys.\ Lett.\ B {\bf 698}, 371 (2011).

\bibitem{cdf-allj}
CDF Collaboration (T.~Aaltonen {\it et al.}),  Phys.\ Lett.\ B {\bf 714}, 24 (2012).

\bibitem{cdf-dil}
CDF Collaboration (T.~Aaltonen {\it et al.}),  Phys.\ Lett.\ B {\bf 714}, 24 (2012).

\bibitem{d0-dil}
D0 Collaboration (V.M.~Abazov {\it et al.}),  Phys.\ Rev.\ Lett.\ {\bf 107}, 082004 (2011).

\bibitem{vwt}
D0 Collaboration (S. Abachi  {\it et al.}), 
Phys.\ Rev.\ Lett.\ {\bf 80}, 2063 (1998).

\bibitem{d0-dil-new} 
D0 Collaboration (V.M.~Abazov {\it et al.}), 
Phys.\ Rev.\ D {\bf 86}, 051103(R) (2012).

\bibitem{mtop-combo}
CDF and D0 Collaborations (T.~Aaltonen {\it et al.}), Phys.\ Rev.\ D {\bf 86}, 092003 (2012), 
updated in arXiv:1407.2682

\bibitem{mtop-combo-world}
ATLAS, CDF, CMS, and D0 Collaborations, arXiv:1403.4427.

\bibitem{cdf-mtop-diff-full-lumi}
CDF Collaboration (T.~Aaltonen {\it et al.}),  Phys.\ Rev.\ D {\bf 87}, 052013 (2013).

\bibitem{d0-mtop-diff}
D0 Collaboration (V.M.~Abazov {\it et al.}),  Phys.\ Rev.\ D {\bf 84}, 052005 (2011).


\bibitem{mtop-combo-2010}
Tevatron Electroweak Working Group (CDF and D0 Collaborations), arXiv:1007.3178.

\bibitem{wmass-pdg-2010}
K.~Nakamura {\it et al.} (Particle Data Group), J.\ Phys.\ G {\bf 37}, 075021 (2010).

\bibitem{whel-sm}
A.~Czarnecki, J.G.~K{\"o}rner, and J.H.~Piclum, Phys.\ Rev.\ D {\bf 81}, 111503 (2010),
and private communication.

\bibitem{whel-combo}
CDF and D0 Collaborations (T.~Aaltonen {\it et al.}), Phys.\ Rev.\ D {\bf 85}, 071106(R) (2012).

\bibitem{cdf-whel-lj-full-lumi}
CDF Collaboration (T.~Aaltonen {\it et al.}),  Phys.\ Rev.\ D {\bf 87}, 031104(R) (2013).

\bibitem{top-width}
J.~Beringer {\it et al.} (Particle Data Group), Phys.\ Rev.\ D {\bf 86}, 010001 (2012).

\bibitem{top-width-lo}
A.~Denner and T.~Sack, Nucl.\ Phys.\ B {\bf 358}, 46 (1991).

\bibitem{top-width-nlo}
A.~Czarnecki and K.~Melnikov, Nucl.\ Phys.\ B {\bf 544}, 520 (1999);
K.G.~Chetyrkin, R.~Harlander, T.~Sedensticker, and M.~Steinhauser,
Phys.\ Rev.\ D {\bf 60}, 114015 (1999).

\bibitem{top-width-nnlo}
J.~Gao, C.S.~Li, and H.X.~Zhu, Phys.\ Rev.\ Lett.\ {\bf 110}, 042001 (2013).

\bibitem{top-to-charged-higgs}
V.~Barger, J.L.~Hewett, and R.J.N~Phillips, Phys.\ Rev.\ D {\bf 41}, 3421 (1990).

\bibitem{top-to-susy}
K.I.~Hikasa and M.~Kobayashi, Phys.\ Rev.\ D {\bf 36}, 724 (1987);
C.S.~Li, J.M.~Yang, and B.Q.~Hu, Phys.\ Rev.\ D {\bf 48}, 5425 (1993).

\bibitem{top-fcnc}
J.L.~Diaz-Cruz, M.A.~Perez, G.~Tavares-Velasco, and J.J.~Toscano,
Phys.\ Rev.\ D {\bf 60}, 115014 (1999).

\bibitem{cdf-width-lj-full-lumi}
CDF Collaboration (T.~Aaltonen {\it et al.}),  Phys.\ Rev.\ Lett.\ {\bf 111}, 202001 (2013).

\bibitem{q-to-hadron-time-scale}
I.~Bigi, Y.~Dokshitzer, V.~Khoze, J.~Kuhn, and P.~Zerwas,
Phys.\ Lett.\ B {\bf 181}, 157 (1986); L.H.~Orr and J.L.~Rosner,
Phys.\ Lett.\ B {\bf 246}, 221 (1990).

\bibitem{top-br-sm}
J.~Beringer {\it et al.} (Particle Data Group), Phys.\ Rev.\ D {\bf 86}, 010001 (2012);
CDF Collaboration (A.~Abulencia {\it et al.}), Phys.\ Rev.\ Lett.\ {\bf 97}, 242003 (2006);
D0 Collaboration (V.M.~Abazov {\it et al.}),  Phys.\ Rev.\ Lett.\ {\bf 97}, 021802 (2006);
LHCb Collaboration (R.~Aaij {\it et al.}),  Phys.\ Lett.\ B {\bf 709}, 177 (2012).

\bibitem{top-br-cdf-lj}
CDF Collaboration (T.~Aaltonen {\it et al.}),  Phys.\ Rev.\ D {\bf 87}, 111101(R) (2013).

\bibitem{top-br-cdf-ll}
CDF Collaboration (T.~Aaltonen {\it et al.}),  Phys.\ Rev.\ Lett.\ {\bf 112}, 221801 (2014).

\bibitem{top-br-d0-lj-dil}
D0 Collaboration (V.M.~Abazov {\it et al.}),  Phys.\ Rev.\ Lett.\ {\bf 107}, 121802 (2011).

\bibitem{d0-fromfacs}
D0 Collaboration (V.M.~Abazov {\it et al.}),  Phys.\ Lett.\ B {\bf 713}, 165 (2012).

\bibitem{top-spin}
V.D.~Barger, J.~Ohnemus, and R.J.N.~Phillips, Int.\ J.\ Mod.\ Phys.\ A {\bf 4}, 617 (1989).


\bibitem{quark-spin-decorr}
A.F.~Falk and M.E.~Peskin, Phys.\ Rev.\ D {\bf 49}, 3320 (1994).

\bibitem{top-spin-prod}
G.~Mahlon and S.~Parke, Phys.\ Rev.\ D {\bf 53}, 4886 (1996);
G.~Mahlon and S.~Parke, Phys.\ Lett.\ B {\bf 411}, 173 (1997).

\bibitem{top-spin-nlo}
W.~Bernreuther, Z.G.~Si, and P.~Uwer, Nucl.\ Phys.\ B {\bf 690}, 81 (2004).

\bibitem{top-spin-test}
A.~Brandenburg, Z.G.~Si, and P.~Uwer, Phys.\ Lett.\ B {\bf 539}, 235 (2002);

\bibitem{top-spin-cdf}
CDF Collaboration (T.~Aaltonen {\it et al.}),  Phys.\ Rev.\ D {\bf 83}, 031104(R) (2011).

\bibitem{top-spin-d0-dil}
D0 Collaboration (V.M.~Abazov {\it et al.}),  Phys.\ Lett.\ B {\bf 702}, 16 (2011).

\bibitem{top-spin-d0-me}
D0 Collaboration (V.M.~Abazov {\it et al.}),  Phys.\ Rev.\ Lett.\ {\bf 107}, 032001 (2011).

\bibitem{top-spin-me}
K.~Melnikov and M.~Schulze, Phys.\ Lett.\ B {\bf 700}, 17 (2011).

\bibitem{top-spin-d0-combo}
D0 Collaboration (V.M.~Abazov {\it et al.}),  Phys.\ Rev.\ Lett.\ {\bf 108}, 032004 (2012).

\bibitem{lorentz-sme}
D.~Colladay and V.A.~Kostelecky, Phys.\ Rev.\ D {\bf 58}, 116002 (1998);
V.A.~Kostelecky, Phys.\ Rev.\ D {\bf 69}, 105009 (2004).

\bibitem{lorentz-top}
V.A.~Kostelecky and N.~Russell, Rev.\ Mod.\ Phys.\ {\bf 83}, 11 (2011).

\bibitem{lorentz-top-d0}
D0 Collaboration (V.M.~Abazov {\it et al.}),  Phys.\ Rev.\ Lett.\ {\bf 108}, 261603 (2012).

\bibitem{top-charge}
D.~Chang, W.F.~Chang, and E.~Ma, Phys.\ Rev.\ D {\bf 59}, 091503 (1999).

\bibitem{top-charge-d0}
D0 Collaboration (V.M.~Abazov {\it et al.}),  Phys.\ Rev.\ Lett.\ {\bf 105}, 101801 (2010).

\bibitem{top-charge-cdf}
CDF Collaboration (T.~Aaltonen {\it et al.}),  Phys.\ Rev.\ D {\bf 88}, 032003 (2013).

\bibitem{top-charge-d0-new}
D0 Collaboration (V.M.~Abazov {\it et al.}),  arXiv:1407.4837. 

\end{thebibliography}
\end{document}